\documentclass[11pt, a4paper]{article}
\usepackage{jheppub}
\usepackage{graphicx}
\usepackage{epsfig,amsmath,amsthm}
\usepackage{epstopdf}

\newcommand{\be}{\begin{equation}}
\newcommand{\ee}{\end{equation}}
\newcommand{\bea}{\begin{eqnarray}}
\newcommand{\eea}{\end{eqnarray}}
\def\bse{\begin{subequations}}
\def\ese{\end{subequations}}
\newcommand{\IR}{\mathbb{R}} 

\def\IZ{\relax\ifmmode\hbox{Z\kern-.4em Z}\else{Z\kern-.4em Z}\fi}

\newcommand{\non}{\nonumber \\}

\def\half{\frac{1}{2}} 

\def\del{{\partial}}

\def\hphi{{\hat \phi}}

\def\hS{{\hat S}}  \def\hQ{{\hat Q}} \def\hx{{\hat x}}
 \def\hQ{{\hat Q}}
\def\hrho{{\hat \rho}}

 \def\co{{\cal O}}

\def\tphi{\widetilde{\phi}}

  \def\eps{\epsilon}

 \def\hrho{{\hat \rho}}

\def\presub{\vspace{.5cm} \noindent}

\def\bi{\begin{itemize}} \def\ei{\end{itemize}}

\def\({\left(} \def\){\right)}
\def\[{\left[} \def\]{\right]}

\def\w{\omega}
\def\factell{ }		
\def\PhiS{A_E}			
\def\PhiV{A_M}			
\def\REone{R^+_1}
\def\RMone{R^-_1}
\def\d{\partial}

\title{Theory of post-Newtonian radiation and reaction}

\author{Ofek Birnholtz, Shahar Hadar and Barak Kol\\
{\it Racah Institute of Physics, Hebrew University, Jerusalem 91904, Israel} \\
{\tt ofek.birnholtz,barak.kol@mail.huji.ac.il}, {\tt shaharhadar@phys.huji.ac.il}
}

\abstract{We address issues with extant formulations of dissipative effects in the effective field theory (EFT) which describes the post-Newtonian (PN) inspiral of two gravitating bodies by (re)formulating several parts of the theory.
Novel ingredients include gauge invariant spherical fields in the radiation zone; a system zone which preserves time reversal such that its violation arises not from local odd propagation but rather from interaction with the radiation sector in a way which resembles the balayage method; 2-way multipoles to perform zone matching within the EFT action; and a double-field radiation-reaction action which is the non-quantum version of the Closed Time Path formalism and generalizes to any theory with directed propagators including theories which are defined by equations of motion rather than an action. This formulation unifies the treatment of outgoing radiation and its reaction force.
We demonstrate the method in the scalar, electromagnetic and gravitational cases by economizing the following: the expression for the radiation source multipoles; the derivation of the leading outgoing radiation and associated reaction force such that it is maximally reduced to mere multiplication; and the derivation of the gravitational next to leading PN order. In fact we present a novel expression for the +1PN correction to all mass multipoles. We introduce useful definitions for multi-index summation, for the normalization of Bessel functions and for the normalization of the gravito-magnetic vector potential.}

\begin{document}
\maketitle

\section{Introduction}

Much is known about the post-Newtonian approximation of a binary system and in particular on the effective field theory (EFT) approach to this problem, as will be reviewed shortly. The conservative two-body effective action is known up to order 3PN, see the reviews \cite{BlanchetRev,FutamaseItohRev} and references therein, as well as \cite{Blanchet:2009rw}.
It was reproduced within the EFT approach in \cite{FoffaSturani3PN} and certain sectors of order 4PN were recently determined in the EFT approach  \cite{FoffaSturani4PNb} and in the Hamiltonian ADM (Arnowitt-Deser-Misner) formalism \cite{JaranowskiSchaefer4PNa}.

Less is known about the non-conservative sector, namely the emitted radiation and radiation reaction (RR) force, which are the subject of this paper.
The RR force is the force which guarantees the system's dissipation due to the emitted radiation.
Indeed, the RR force is known only to order +1PN beyond the leading order \cite{IyerWill,BlanchetRev,GalleyLeibovich}.  It is true that this apparent low order can be justified by need, since a +2PN correction appears only at 4.5PN.
Still, the lack of high order results could also be influenced by certain issues with the current formulation, including the EFT approach formulation, as we proceed to discuss.

 \bi \item A split between theories of radiation and radiation reaction force such that conservation laws are not manifest.

Currently there are two lines of work: the radiation EFT was treated by \cite{GoldbergerRothstein1, GoldbergerRothstein2,GoldbergerRoss,RossMultipoles} while the EFT formulation of the RR force was treated in \cite{GalleyTiglio,GalleyLeibovich}  and earlier papers (a similar division exists also in earlier, non-EFT work).
There are seemingly essential differences between these two lines of work.
The RR force EFT is based on the Closed Time Path action formalism \cite{CTP} while the radiation EFT is not;  the radiation EFT uses the Feynman propagator for radiation while the RR force uses the retarded and advanced propagators.
However, the two phenomena are clearly closely related. Hence while there is no contradiction in the current split approach one may wish for a unified theory.
As a concrete example, the energy carried by radiation must deplete the system's energy by the same amount.
Yet, looking at the expression for the acceleration due to the radiation reaction force in eq. (168) of the review \cite{BlanchetRev}  energy balance is not at all manifest.

\item Seemingly overly cumbersome computations.

We note three specific cases. The computation of the leading RR force was reproduced within EFT in Appendix A of \cite{GalleyLeibovich}. We notice that while the final result is very simple the computation is not as simple.  The computation of the correction to the mass quadrupole of a system was reproduced within EFT in \cite{GoldbergerRoss} and involves 5 diagrams, yet the number of actual terms in the end result is smaller. Finally, the general computation of the multipole moments was performed in \cite{RossMultipoles} and includes combinatorial factors which we notice to be suggestively similar to the Taylor coefficients of Bessel functions.

\item NRG fields and real Feynman rules are not incorporated yet.

Non-Relativistic Gravitational (NRG) fields are routinely used in computations of the conservative two-body effective action \cite{GilmoreRoss,FoffaSturani3PN,FoffaSturani4PNb} yet they were not applied yet to the dissipative sector.   In addition, much of the literature uses imaginary Feynman rules where all the factors of $i$ cancel at the end to yield real results, yet these unnecessary cancellations can be avoided through the use of purely real (and $\hbar$ free) Feynman rules appropriate for a classical EFT \cite{CLEFT-caged}.

 \ei

In this paper we shall (re)formulate several parts of the theory to achieve unity and economization. The paper, which was at the center of our research since fall 2011, is organized as follows. We begin in subsection \ref{subset:background} with a detailed review of the background and past work. In section \ref{sec:formulation} we formulate the theory while introducing several new ideas. In section \ref{sec:demonstration} we demonstrate the theory by economizing several computations. Finally in section \ref{sec:summary} we summarize our results and discuss them.

\subsection{Background}
\label{subset:background}

Solving the two-body problem in Einstein's gravity is of interest both theoretically and observationally. Observationally, binary star systems are the expected typical source of steady gravitational waves (GWs).  The computation of the waveform of gravitational waves is essential for the world-wide effort to detect gravitational waves, first in order to design detection filters and later for signal interpretation. Advanced LIGO, an upgrade of the LIGO detector is under construction \cite{aLIGO} according to which completion is scheduled for 2015.  eLISA (evolved LISA) (or the New Gravitational-wave Observatory - NGO) is a proposed space-based detector under consideration by the European Space Agency (ESA)
 for launch in 2022  \cite{eLISA}. Several other detectors worldwide are at some stage of development: either at work, or undergoing construction or upgrade, or under consideration, see \cite{IFOprojects} for a list of links, including advanced VIRGO (Italy) \cite{aVIRGO},  KAGRA (Japan) \cite{KAGRA}, GEO (Germany) and AIGO (Australia).

Theoretically, the two body problem in the post-Newtonian limit is of intrinsic interest in General Relativity (GR), being an essential part of the pre-GR limit. Indeed it occupied Einstein ever since the theory's incubation when misconceptions over the Newtonian limit led him astray in 1913  (see for example \cite{Norton}) and  it continued to occupy him at least 22 years later when he worked to derive the first post-Newtonian correction to the equations of motion \cite{EIH}. In addition this problem turns out to require practically all of the deep tools of perturbative field theory including Feynman diagrams, loop computation, regularization and renormalization.  Chronologically these ideas happened to be discovered in the context of quantum field theory yet we maintain that they could have been discovered also in the purely classical context of the two-body problem in GR, or in similar classical problems. Thus this problem provides a fresh perspective on field theory and especially on its classical limit.

In general the full GR equations for this system cannot be solved analytically. Numerical solutions of a binary system are now readily available after a decades long effort (see  \cite{PretoriusRev} for a 2007 review). In order to reach the analytic domain it is necessary to take certain perturbative limits. In the binary system of two masses ($m1,m2$) two such limits exist: the post-Newtonian limit $\frac{v^2}{c^2} \sim \frac{G\,M}{r_{12} c^2} \ll 1$, where $v$ is a typical velocity and $M=m1+m2$; and the extreme mass ratio limit $m_1 \ll m_2$.  Having both numerical and analytical methods is not redundant but rather they complement each other. Only numerical solutions can address the non-perturbative parameter region, yet analytic methods add insight, in particular into the dependence of the problem on its parameters. In addition numerics and analytics serve for important cross checks.

\presub {\bf Key concepts}. \emph{Gravitational radiation} is one of the central predictions of Einstein's gravity \cite{GWEinstein}\footnote{Einstein's \cite{GWEinstein} also obtained the leading quadrupolar expression, but with a missing factor of 2}. Indeed, Special Relativity requires that changes in the gravitational field caused by changes in the configuration of masses would propagate no faster than the speed of light. These propagating perturbations are gravitational waves. GR tells us that a (planar) gravitational wave travels at the speed of light and is characterized by its wavelength and its (transverse) polarization. The early history of the field was plagued with confusion regarding the reality of gravitational waves including second thoughts by Einstein himself in an unpublished  manuscript (1936) \cite{Kennefick}. An important milestone was the realization that GWs carry energy and other conserved charges, see the sticky bead argument (1957) \cite{sticky-bead}. In the non-relativistic limit GWs are generated by a time varying mass quadrupole moment at order 2.5PN, with subleading contributions from all mass and current multipoles of the source.

\emph{Radiation reaction (RR) force} is a direct consequence of radiation. Conservation of energy (and other conserved quantities) requires that the radiated energy be deducted from the system. This happens through the interaction of the bodies with the reaction fields which accompany radiation. In GR this was understood first by Burke and Thorne as late as 1970 \cite{BurkeThorne}. These authors used an asymptotic matched expansion (perhaps introducing it to the GR literature) between the system zone and the radiation zone. They viewed the RR force as arising from a Newtonian-like reaction potential field which enters the system zone through matching with the radiation zone. A relevant counterpart appears in classical electrodynamics, namely the Abraham-Lorentz-Dirac (ALD) force \cite{ALD,Dirac}. Dirac's relativistic derivation stresses that this force can be understood to result from the time-reversal-odd kernel $G_{odd} := (G_{ret}-G_{adv}) /2$ where $G_{ret}$ and $G_{adv}$ are the retarded and advanced Green's functions.
A closely related concept, that of the self-force, appears in the context of the extreme mass ratio limit of the two-body problem. The self-force is a force caused by field perturbations which were sourced by the small body itself in the past. Considerable research effort was devoted to this topic over the last two decades,  especially to the issue of its regularization -- see the reviews \cite{SFEMR-rev} and references therein.

\emph{The effective field theory (EFT) approach} to GR was introduced in \cite{GoldbergerRothstein1}. They recognized a hierarchy of scales in the problem (the sides of objects are much smaller than the two-body separation, which in turn is much smaller than the radiation wavelength) and boldly applied the ideas of effective field theory developed within the context of quantum field theory. More generally the EFT approach is designed for any field-theoretic problem, classical or quantum, with a hierarchy of scales. Once the scales are separated the EFT approach proceeds to eliminate one of the scales (usually the short distance one), replacing it by effective interactions in the surviving scale. Within quantum field theory this procedure is known as ``integrating out'' fields (in the sense of the Feynman path integral) but field ``elimination'' is more appropriate in the non-quantum context.

\presub {\bf Literature review}.  The Newtonian limit was central to the development of GR by Einstein (see for example \cite{Norton}) and  an essentially post-Newtonian computation appeared already in the paper  introducing GR \cite{GR}, namely the perihelion shift of Mercury.  The Lagrangian for the first post-Newtonian correction (1PN) to the two-body motion was obtained by \cite{LorentzDroste1917}, but was left mostly unnoticed. Einstein predicted gravitational waves and obtained the leading quadrupolar expression \cite{GWEinstein}. Einstein, Infeld and Hoffmann (1938) \cite{EIH} obtained the 1PN correction to the equations of motion. After world war II work resumed and the 1PN correction was promoted to the Lagrangian level in \cite{Fichtenholz}. During the 1950s two major groups pursued and developed the subject: one led by Infeld in Poland and one led by Fock in Russia, and their findings are summarized in the books \cite{FockInfeld}. During the 1960s ``golden age'' in GR research attention shifted to black holes. In 1970 Burke and Thorne understood the RR force and obtained the leading order \cite{BurkeThorne}. By the early 1970s Kimura's group in Japan was able to attempt to proceed to the 2PN correction to the motion \cite{Ohta:1974pq}, but still suffered from errors, noticed and corrected by Damour \cite{Damour1982,Damour1983a} and Sch\"{a}fer \cite{Damour:1985mt}. \cite{WagonerWill} found the +1PN correction to the mass quadrupole and hence to radiation. The multipolar decomposition of GWs was given in \cite{ThorneMultipoles}.

In the early 1980s post-Newtonian study was picked up by T. Damour \& N. Deruelle \cite{Bel:1981be2}, by L. Blanchet (see the reviews \cite{BlanchetRev,Blanchet:2009rw}), and by G. Sch\"{a}fer who introduced the ADM Hamiltonian approach (see review \cite{SchaeferRev}).
The +1PN correction to the RR force was obtained in \cite{IyerWill} (1993-5), see also \cite{Jaranowski:1996nv} within the Hamiltonian ADM approach. The +2PN corrections to the radiation field and energy loss (damping) were obtained in \cite{DBIWW}.

In 2004 the EFT approach was introduced by Goldberger and Rothstein  \cite{GoldbergerRothstein1}, reducing PN computations to Feynman diagrams. Certain diagrammatic calculations appeared before, see for example \cite{prevEFT} , but while they can be considered to be precursors of the EFT approach they did not make direct contact with field theoretic Feynman diagrams and were not part of a complete formulation of the PN theory.  \cite{GoldbergerRothstein1} reproduced both the quadrupole formula and the 1PN correction to the two body motion. Black hole absorption was incorporated in \cite{GoldbergerRothstein2}. \cite{Galley:2005tj}  initiated a study of a field theory description of the RR force in terms of the Closed Time Path (CTP) formalism. Porto \cite{PortoSpin2005} initiated a study to incorporate spinning compact objects within EFT and it was applied in \cite{PortoRothstein2006} to achieve the first determination of the next to leading spin-spin interaction, up to certain missing contributions found in \cite{Steinhoff:2007mb} using Hamiltonian methods and also found later to arise from indirect contributions in the EFT method, see also \cite{Levi:2008nh}. Further spin effects were studied in \cite{PortoSpinMore}. \cite{CLEFT-caged} introduced Non-Relativistic Gravitational (NRG) field redefinition of the Einstein field and real Feynman rules while \cite{NRG} proceeded to apply them and economize the 1PN derivation, presumably optimally. The 2PN correction to the 2-body effective action was reproduced within EFT in \cite{GilmoreRoss} and later 3PN by \cite{FoffaSturani3PN}
 (see \cite{Chu:2008xm} for automated N-body at 2PN).
Certain sectors of the conservative dynamics at order 4PN were recently determined, in the EFT approach  \cite{FoffaSturani4PNb} and in the Hamiltonian ADM formalism \cite{JaranowskiSchaefer4PNa}. \cite{GoldbergerRoss} reproduced corrections to radiation at least up to order +1.5PN, and \cite{FoffaSturani4PNa} reproduced the RR tail term at 4PN.

The progress in determinations of the motion enabled in recent years higher order determinations of the radiation: +3PN for quasi-circular orbits in \cite{Blanchet:2008je}, partial results for +3.5PN in \cite{Faye:2012we}.

\section{Formulation of perturbation theory}
\label{sec:formulation}

In this section we (re)formulate an effective field theory of post-Newtonian radiation and RR force to address the issues mentioned in the introduction. Even though our main interest lies in the gravitational two-body system we study also the cases of scalar and electromagnetic (EM) interactions. Indeed in this section we shall mostly  address the scalar case which exhibits most of the relevant features in a somewhat simplified context, while making some comments on the EM and gravitational cases. We shall sometimes refer to the scalar, electromagnetic and gravitational cases as $s=0,\,  1, \,2$ respectively (where $s$ denotes the spin of the corresponding quantized field). A detailed study of all three cases will be given in section \ref{sec:demonstration}.

Consider then a relativistic ``massless'' 4d scalar field $\phi$ whose action is given by \be
 S[\phi] = \int d^4x \( \frac{1}{8 \pi G} \del_\mu \phi\, \del^\mu \phi - \rho\, \phi - V(\phi) \)  \label{scalar-action} ~,\ee
 where $\rho=\rho(x)$ is a general charge density and $V(\phi)$ is a non-quadratic potential which will be assumed to vanish through most of the discussion.
In the scalar two body problem the bodies' trajectories are denoted by $x^\mu_A=x^\mu_A(\tau)$ where $A=1,2$ indexes the bodies. In this case the source term is given by \be
 \int d^4x\, \rho\, \phi = \sum_{A=1,2} q_A \int d\tau\, \phi(x^\mu_A(\tau)) \, \, , \label{scalar-charge} \ee
  where $q_A$ is the A'th body's scalar charge. Actually this theory is not only simple but it is also directly relevant to gravitation: interpreting $\phi$ as the Newtonian potential this action describes a sector of the gauge-fixed gravitational action.

The corresponding action and source term for all types of interaction are organized in table \ref{tab:fields}.

\begin{table}[t!] \centering
$\begin{array}{l|ccc}
	   &\mbox{{\bf field}} & \mbox{{\bf action}}  & 	\mbox{{\bf two-body source}} \\
	  \hline   \\
 \mbox{scalar} &	\phi  	&  \int d^4x \( \frac{1}{8 \pi G} \del_\mu \phi\, \del^\mu \phi - V(\phi) - \rho\, \phi \) &  -\sum_{A=1,2} q_A \int d\tau\, \phi(x^\mu_A(\tau))  \\ \\
\mbox{EM} 	& A_\mu		& -\frac{1}{16\pi} \int d^4 x\, F_{\mu\nu} F^{\mu\nu}  - \int d^4 x\,  J^{\mu}\, A_{\mu} & ~-\sum_{A=1,2} q_A \int dx^\mu_A\,  A_\mu\(x^\mu_A(\tau)\) \\ \\
\mbox{gravity}	& g_{\mu\nu} & \int \, d^4 x\, \sqrt{-g}\, \left( - \frac{1}{16 \pi G} R[g_{\mu\nu}] + \mathcal{L}_{\mathrm{M}} \right) &~ -\sum_{A=1,2} m_A \int \sqrt{g_{\mu\nu}(x^\mu_A)\, dx^\mu_A\, dx^\nu_A}
\end{array} $
\caption{Summary of field, action and 2-body source term for each type of interaction $s=0,\, 1,\, 2$. $\mathcal{L}_{\mathrm{M}}$ is the matter's Lagrangian density. $q_A$ is the body's scalar charge for $s=0$ and its electric charge for $s=1$, while $m_A$ is the body's mass.}
 \label{tab:fields}
\end{table}

We start in the first two subsections by specifying the equations of motion. In subsection \ref{subsec:zones} we specify the fields and the zones in which they are defined. In subsection \ref{subsec:zoom} we address the time-reversal odd propagation which is responsible for dissipative effects. Next we lift the equations of motion to the action level. In subsection \ref{subsec:multipoles} we incorporate the matching procedure into the EFT action and in \ref{subsec:doubling} we allow for elimination of retarded fields in the action through field doubling. Finally in subsection \ref{subsec:summary} we summarize the formulation.

\subsection{Zones and spherical waves}
\label{subsec:zones}

\begin{figure}[t!]
\centering \noindent
\includegraphics[width=15cm]{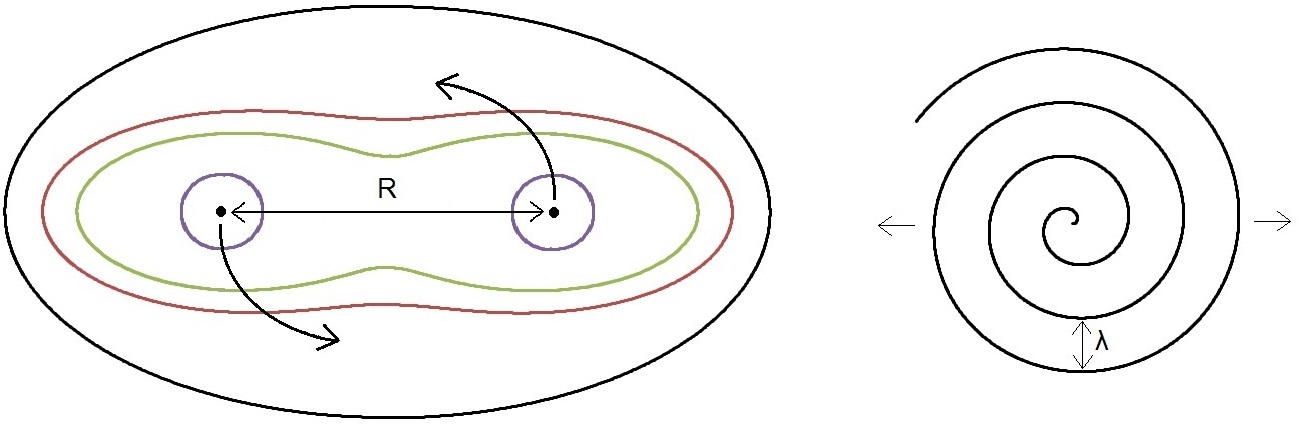}
\caption[]{The two relevant zones. On the left is the system zone with a typical stationary-like field configuration. On the right is the radiation zone with its typical out-spiraling waves.}
 \label{fig:zones}
\end{figure}

The key observation which calls for an EFT approach is that the fields can be decomposed into several parts according to the following hierarchy of 3 length scales: the compact objects' sizes, their separation and the radiation wavelength. In the non-EFT approach one uses Matched Asymptotic Expansion between several zones, starting with \cite{BurkeThorne}. Actually the separation of fields in EFT is completely equivalent to the separation into zones in Matched Asymptotic Expansion (see for example \cite{CLEFT-caged}). We shall start with the zone point of view, and assuming the object size scale was eliminated through the point particle approximation\footnote{This approximation is valid up to a rather high order (5PN in 4d) where finite size effects first enter; see for example \cite{GoldbergerRothstein1}.},we shall concentrate on the two others: the system zone and the radiation zone. Figure \ref{fig:zones} depicts the two zones, each with a typical field configuration. The system zone is defined to keep the two-body separation finite, while the objects are point-like and the radiation wavelength is infinite. It is also known as the induction zone or the near zone (though the latter is liable to be confused with the compact object zone). In the radiation zone  the wavelength is finite while the two-body system shrinks altogether to a point.

One of the main benefits of the division into zones is that each one has an enhanced symmetry (or an approximate one), namely a symmetry absent from the full problem. The symmetry is central to making fitting choices for the formulation of a perturbation theory in each zone including the choice of how to divide the action into a dominant part and a perturbation, the choice of field variables and when relevant the choice of gauge.

In the system zone we follow the standard formulation. This zone is approximately stationary (time independent) since by assumption all velocities are non-relativistic and hence are approximated to vanish. Accordingly the kinetic term of (\ref{scalar-charge}) is written as
\be
S[\phi] \supset \frac{1}{8 \pi G} \int d^3x\, dt\, \[ -\( \vec{\nabla} \phi\)^2 + \frac{1}{c^2} \dot{\phi}^2 \] ~. \label{NR-scalar-action}
\ee
As time derivatives are small here the first term in (\ref{NR-scalar-action}) is dominant and the second term is considered a perturbation \cite{GoldbergerRothstein1}.
Accordingly the propagator is instantaneous
\be
  \parbox{50mm}{\includegraphics[scale=0.3]{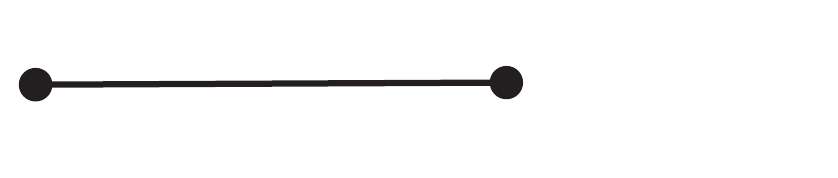}}
	= \frac{4 \pi G}{\vec{k}^2}\, \delta(t_1-t_2) \, \, ,  \label{s0-system-propag} \ee
   while the term with time derivatives contributes a correction 2-vertex \be
  \parbox{55mm}{\includegraphics[scale=0.3]{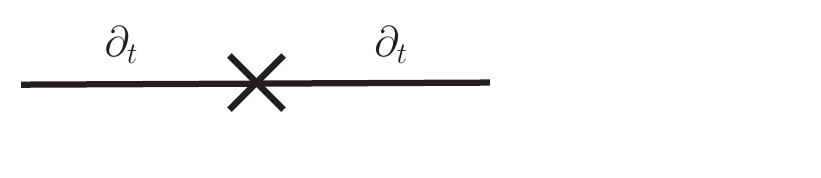}}
	= \frac{1}{4 \pi G\, c^2} ~. \label{s0-propag-corr}
\ee
We are using conventions such that the Feynman rules are real \cite{CLEFT-caged}, as is appropriate for a non-quantum theory.

In the radiation zone we introduce a change. We notice that as the system has shrunk to a point this zone has a spherical symmetry. Hence we insist on using spherical field variables. This is different from extant  EFT work \cite{GoldbergerRothstein2,GoldbergerRoss, GalleyTiglio,GalleyLeibovich, FoffaSturani4PNa}) which characterizes radiation in the standard way by a wavenumber vector $k_\mu$ (and a polarization $h_{\mu\nu}$ for gravity), namely by a planar wave.

We represent the spherical fields by symmetric trace-free tensors $\phi_L(r,t)$ where $L=(i_1 \dots i_\ell)$ is a multi-index and each $i_k=1,2,3$. Separation of variables reads \be
 \phi(r,t,\Omega) = \phi_L(r,t)\, x^L := \sum_{\ell=0}^{\infty} \frac{1}{\ell!} \sum_{i_1 \dots i_\ell=1}^3 \phi_{i_1 \dots i_\ell}(r,t)\,  x^{i_1} \dots x^{i_\ell} \, \, , \label{def:spherical-fields} \ee
where $\Omega$ denotes the angular directions $\theta,\phi$ and we define a convention summation for multi-indices such that $\ell!$ factors are implicit (see Appendix \ref{app:defs}).

We note that any symmetric trace free tensor $\phi_{L_\ell}$ ($\ell$ being given), can be represented also in one of the following equivalent ways
\bi
	\item $\phi_{\ell m}$ -- standard spherical harmonic representation.
	\item $\phi_\ell(\Omega)$ -- a function on the unit sphere.
\ei
The different forms are related by
\be 	
	 \phi_\ell(\Omega) = \phi_{L_\ell}\, \frac{x^{L_\ell}}{r^\ell} = \sum_{m} \phi_{\ell m}\, Y_{\ell m}(\Omega).
\label{basis options}
\ee
There is also a useful complex spinorial representation which we will not discuss here. We found it convenient to use the $\phi_L$ variables for RR force calculations (because they are more analytic around the origin of the radiation zone) and it is our default choice. For calculations of emitted radiation however, we mostly use $\phi_{\ell m}$ which is more convenient asymptotically.

In terms of $\phi_L$ the propagator becomes
\be
  G_{s=0} \equiv
 \parbox{25mm}{\includegraphics[scale=0.7]{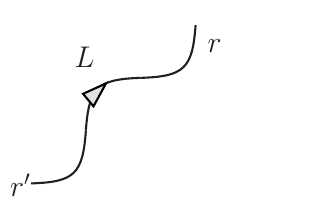}}
 = \frac{(-)^\ell\, (-i \w)^{2\ell+1}}{(2\ell+1)!!} \,G\, \tilde{j}_{\ell}(\w r_1)\, \tilde{h}^{-}_{\ell}(\w r_2) \, \, ,
\label{retarded-prop}
\ee
where $\tilde{j}, \tilde{h}$ are Bessel functions whose normalization is defined in Appendix \ref{app:defs}. This propagator will be derived in detail in  (\ref{Phi propagator scalar}) and the sign convention for $\omega$ will be specified.

Thus we adjust the theory to better exploit the symmetry of each zone separately, thereby economizing computation. For $s=1,2$ these adjustments will include a choice of gauge. The price to pay is in an added complication to the matching between the different fields and respective gauges of the two zones.

\begin{table}[t!]
\centering
$\begin{array}{l|cc}
			   	&  \mbox{\bf System}  	& 	\mbox{\bf Radiation} \\
		\hline \\
~\mbox{symmetry}~ & ~\mbox{stationary}~ & ~\mbox{spherical}~ \\ \\
\phi		& \phi				& \phi_L	\\   \\
A_\mu	& (\phi,\, \vec{A})		& A_L^\eps \\ \\
g_{\mu\nu} & (\phi,\, \vec{A},\, \sigma_{ij}) & h_L^\eps \\
	   \end{array}$ \\
\caption{A summary of the two zones and their main properties. The symmetry applies to the unperturbed equations. For each field $s=0,1,2$ we list the variables used in each one of the zones. Our novelty here is in using spherical fields in the radiation zone, labelled here by a multi-index $L$. $\eps=E,M$ denotes electric or magnetic spherical waves and is a parity label. In EM and gravity these waves are gauge invariant with $\ell \ge 2$. In the system zone we make the standard gauge choice: Lorentz (Feynman) for EM and harmonic for gravity.}
 \label{tab:zones}
\end{table}

\presub ${\bf s=1,2}$. When generalizing spherical waves from the scalar case to electromagnetism and gravity we must account for gauge symmetry and polarization. Spherical symmetry is a particular case of co-homogeneity 1 spaces and hence spherical waves are actually gauge invariant \cite{1dPert,AsninKol}. The detailed analysis is given in the corresponding parts of section \ref{sec:demonstration}, and the outcome is that for both $s=1,2$ there are two families of waves, electric and magnetic, which differ by polarization and are denoted by an index $\eps=E,M \equiv +,-$. In all cases only modes with $\ell \ge s$ are dynamic and represent spherical waves while modes with $0 \le \ell \le s-1$ describe stationary properties of the system rather than radiation. For example for $s=2$  the electric type $\ell=0,1$ modes represent the total mass and the center of mass $\{ M,\, \vec{X}_{cm} \}$, while the magnetic type $\ell=0,1$ modes represent the total momentum and the angular momentum $\{ \vec{P},\, \vec{J}  \}$. The long term variation of these modes is given by an RG flow \cite{GRR}.
 The field variables in each zone (for all $s$) are summarized in table \ref{tab:zones}. The propagator is nearly the same as in the scalar case and the difference is accounted for by the factor $R_s^\eps(\ell)$
\bea
 \parbox{20mm}{\includegraphics[scale=0.5]{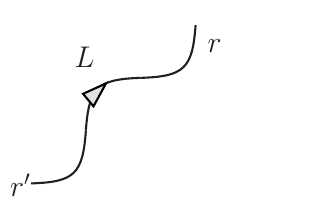}}
 &\equiv& G_s^\eps = G_{s=0}\, R_s^\eps(\ell) \, \, , \non
 R_1^\eps &=& \( \frac{\ell+1}{\ell}\)^\eps \, \, , \non
 R_2^\eps &=& \( \frac{\ell+1}{\ell}\)^\eps \frac{\ell+2}{\ell-1} \, \, ,
 \label{summ:Rs}
 \eea
where in order to avoid a factor of $4$ in $R_2^-$ we find it convenient to redefine the gravito-magnetic vector potential (and consequently the gravito-magnetic spherical waves $h_L^{\eps=M}$) as explained in Appendix \ref{app:defs}.

\subsection{Odd propagation and Balayage}
\label{subsec:zoom}

In the previous subsection we defined our choice of field variables. The equations of motion are contained in the action (\ref{scalar-action}). These must be supplemented by relevant boundary conditions, namely retarded propagation. While in the radiation zone one employs the retarded propagator directly, the system zone needs more care since there one uses the instantaneous propagator (\ref{s0-system-propag}) which does not account for the retarded boundary conditions.

Whereas the action (\ref{NR-scalar-action}) is Lorentz invariant the PN limit and in particular the propagator (\ref{s0-system-propag}) break it. It is interesting to consider the dressed propagator which restores the Lorentz symmetry (see for example \cite{dressed}) \bea
 G_{dr}(k^\mu) &\equiv&
 \parbox{28mm}{\includegraphics[scale=0.3]{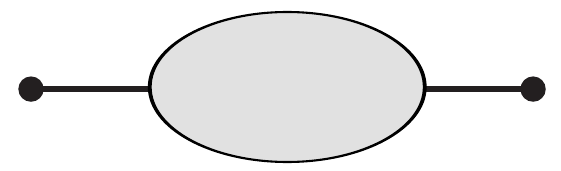}} = \parbox{20mm}{\includegraphics[scale=0.2]{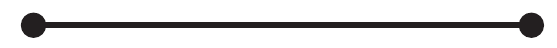}} + \parbox{20mm}{\includegraphics[scale=0.2]{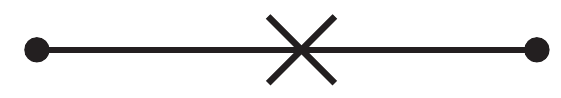}} + \parbox{20mm}{\includegraphics[scale=0.2]{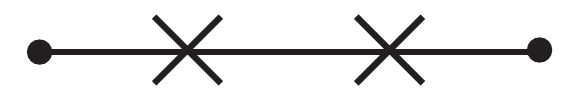}}
 + \dots =
  \non
  &=& \frac{4 \pi G}{\vec{k}^2} \(1 -\frac{\omega^2}{\vec{k}^2} \)^{-1} = \frac {4 \pi G}{\vec{k}^2-\omega^2} \equiv G_{even} \neq G_{ret} \, \, . \label{propagation:dressed} \eea
The dressed propagator is relativistic by construction. However, it is also clearly time-symmetric and therefore cannot equal the retarded propagator but rather it equals the even propagator\footnote{This can be seen by considering the contour around the poles in the complex $\omega$ plane implied by (\ref{propagation:dressed}).} where
 \be
 G_{even/odd} := \half \(G_{ret} \pm G_{adv}\)  \label{even/odd prop} \ee
 defines the even and odd parts of propagation in terms of retarded and advanced propagators.

In order to compensate for the difference we must also account for the odd propagation. That can be done by direct local propagation, namely adding
to the field the consequences of odd propagation \be
 \phi_{odd}(x) := -\int dx'\, G_{odd}(x',x)\, \rho(x') \, \, , \label{local-odd-prop}
 \ee
just like Dirac's relativistic analysis of the self-force of the electron \cite{Dirac}.

Yet in our context there is a better way to obtain $\phi_{odd}$. $G_{odd}$ satisfies the source free wave equation and so does $\phi_{odd}$.
Therefore the data which determines $\phi_{odd}$ can be taken to be the asymptotic boundary conditions (the incoming waves). However, the asymptotic boundary conditions for the 2-body zone are determined by the radiation zone. Thus given our zone structure \emph{there is no need to compute $\phi_{odd}$ through local propagation (\ref{local-odd-prop}) but rather it is already accounted for by matching with the radiation zone}. This is the conclusion of this section. In section \ref{sec:demonstration} we shall confirm by direct perturbative computations that the RR force computed via matching with radiation equals computations through local propagation or relativistic expressions \`{a} la Dirac.

	\begin{figure}[t]
	\begin{center}
	\includegraphics[width=11cm]{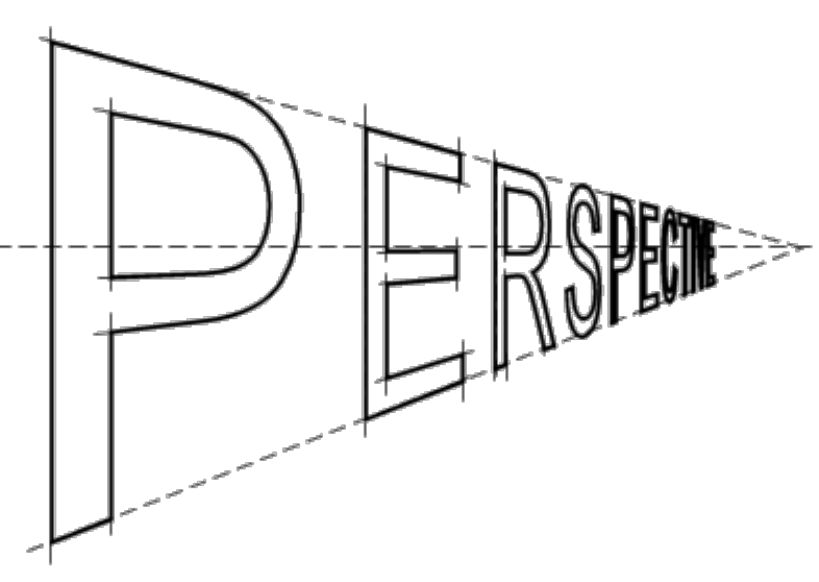}	
	\caption{The idea for zoom balayage is illustrated by the perspective drawing technique which imitates depth. In perspective identical objects appear to be at different depths by drawing them with different sizes. Similarly sources can be shifted while zooming as long as one adjusts their intensity and timing (via a delay or advance in time) such that the observed field is unchanged.}
	\label{fig:perspective}
	\end{center}
	\end{figure}

\presub {\bf Zoom balayage}. We comment that in this way the source for $\phi_{odd}$ is ``combed back'' to infinity and replaced by equivalent asymptotic boundary conditions. This phenomenon resembles the balayage method (French for sweeping, scanning) in potential theory developed by H. Poincar\'e \cite{balayage-encyc,Balayage-Poincare}.
 There one wishes to replace an electrostatic charge distribution within some domain $D$ by an equivalent charge density on the domain's boundary $\del D$ such that the electrostatic potential outside $D$ remains unaltered. One visualizes this process as sweeping the charge from the interior of $D$ into its boundary. In our case the source is swept out to infinity while keeping $\phi_{odd}$ unchanged. It can be thought to arise while zooming in from the radiation zone to the system zone and can be referred to as ``zoom-in balayage''. Later we shall also encounter a complementary case where the charges of the system zone are swept into the origin of the radiation zone in the course of a zoom-out balayage, see (\ref{matching},\ref{I Phi scalar multipoles}).
 See figure \ref{fig:perspective}.

\subsection{Matching lifted to action: 2-way multipoles}
\label{subsec:multipoles}

In the previous subsections we specified our approach for the equations of motion: we chose fields, propagators and gauge for both zones. In the next two subsections we shall lift the equations of motion to the level of an action to benefit from its compactness and from the available field theory tools (Feynman diagrams and the effective action). In this subsection we lift the matching conditions, starting still with a scalar theory (\ref{scalar-action}) and then generalizing to electro-magnetism (EM) and gravity. Within this subsection we shall use the following shorthand notation \be
 \phi \equiv \phi^{sys} \qquad \psi \equiv \phi^{rad} \ee
 to denote the fields in the 2-body zone and in the radiation zone.

It is known how to describe the source for the radiation zone in terms of time-dependent multipole moments at the origin (see for example \cite{GoldbergerRoss}, we use the normalization conventions of \cite{RossMultipoles}) \be
S \supset - \int dt\,  Q^L (t)\, \del_L \psi |_{r=0} \, \, ,  \label{def:Q}
\ee
where $L=(k_1,\dots,k_\ell)$ is a multi-index, $Q^L (t)$ is the charge multipole tensor which is symmetric and trace free, and we are using the multi-index summation convention in Appendix \ref{app:defs}.

\presub {\bf Reaction field multipoles.} In analogy with (\ref{def:Q}) one can parameterize the asymptotic boundary conditions for $\phi \equiv \phi^{sys}$  by multipole quantities $P^L$ as follows \be
 S \supset - \int dt\, P^L (t)\, \del^u_L \tphi |_{u=0} \, \, , \label{def:P}
\ee
 where $u^i=r^i/r^2$ are inverted coordinates such that $u^i=0$ corresponds to $r^i=\infty$, $\del^u_L \equiv (\del/\del u)^L$ and it is convenient to define \be \tphi:= \phi/u ~.\ee
  With these definitions $P_L$ are interpreted as multipoles of the reaction field and $\phi$ is given by  \be
 \phi = -u\, P_L\, (-)^\ell \del^u_L \frac{G}{u}  = -G\, (2\ell-1)!!\, P_L\, x^L \, \, . \ee
The pair of multipole sets $Q_L,\, P_L$ precisely enumerates the necessary boundary conditions in both zones. We refer to this pair of multipoles as 2-way multipoles.

\presub {\bf Matching}. We recovered the correct equations of motion for $\phi,\, \psi$ but it still remains to determine $Q_L,\, P_L$ and that should be done by matching. We know that $Q_L$, the charge multipoles, are determined by the behavior of $\phi$ at $r \to \infty$, while $P_L$ is determined by $\psi$ at $r \to 0$. More precisely \bea
 Q_L &=& -\frac{1}{G\, (2\ell-1)!!} \del^u_L \tphi |_{u=0} \non
 P_L &=& -\frac{1}{G\, (2\ell-1)!!} \del_L \psi |_{r=0} ~. \label{matching} \eea
The first equation can be considered to be a zoom-out balayage while the second one would be zoom-in (see the last paragraph of subsection \ref{subsec:zoom}).

Now we realize that the matching equations can be derived as ordinary equations of motion from the action if we \emph{promote the 2-way multipoles from parameters to action variables} and add a coupling of $P,Q$ as follows \bea
 S &=& \frac{1}{8 \pi} \int d^4 x \[ \del_\mu \phi \del^\mu \phi + \del_\mu \psi \del^\mu \psi  \] \non
    &-& \int dt \sum_{\ell=0}^{\infty}  \[Q^{L_\ell} \del_{L_\ell} \psi |_{r=0} + P^{L_\ell} \del^u_{L_\ell} \tphi |_{u=0} + G\, (2\ell-1)!!\, P^{L_\ell}\, Q^{L_\ell} \] \, \, . \label{matching-action}
 \eea
Indeed variation with respect to $P$ and $Q$ reproduces the matching conditions (\ref{matching}), thereby the addition of the last term accomplishes the \emph{lift of the matching conditions to the level of the action}. Note that it happened to be possible to adjust the single coefficient of the last term such that the two matching conditions are fulfilled.

$P,\, Q$ are algebraic quadratic fields which can be eliminated from the action to yield \bea
 S &=& \frac{1}{8 \pi} \int d^4 x \[ \del_\mu \phi \del^\mu \phi + \del_\mu \psi \del^\mu \psi  \] \non
    &+& \int dt \sum_{\ell=0}^{\infty} \frac{1}{G\, (2\ell-1)!!}  \[  \del_{L_\ell} \psi |_{r=0}  \del^u_{L_\ell} \tphi |_{u=0}  \] ~. \label{direct-matching-action}
 \eea
This form makes evident the coupling of $\phi |_{r=\infty}$ and $\psi |_{r=0}$ which takes place in the overlap region. Yet we shall prefer to keep the $P,\,Q$ fields since they highlight the important physical quantities which are exchanged between the zones, namely the charge multipoles and the reaction field multipoles. It would be interesting to derive this coupling directly from the original action (\ref{scalar-action}).

\presub ${\bf s=1,2}$.
Here there are two wave polarizations: electric and magnetic (alternatively one can use helicity states as a basis). Accordingly we should use two sets of 2-way multipoles \be
 (P,Q) \to (P_\eps,\, Q_\eps), \qquad \eps=E,M \, \, , \ee
 In this case there are also issues of gauge invariance. For the multipoles to be gauge-invariant we couple them to the gauge invariant spherical wave variables. In the radiation zone these are natural variables, while in the 2-body zone they should be understood to be a function of the non-relativistic fields.

\presub {\bf The $\IR^{1,1}$ ring.} We comment that the $(P,Q)$ variables can be thought to belong to the $\IR^{1,1}$ ring \cite{URG} which is closely related to complex numbers, and then the interaction term (\ref{matching-action}) can be expressed in terms of an inner product (in the ring and between multipole sets).

\subsection{Directed propagators, field doubling and action}
\label{subsec:doubling}

\emph{A problem}. In the radiation zone we use the retarded propagator which is clearly asymmetric with respect to exchange of its two endpoints, namely, it is directed. This creates a problem to apply the standard Feynman diagram techniques where the propagator is undirected.  The essentials of this problem appear already in the case of a non-linear scalar theory with source (\ref{scalar-action}) where $V(\phi)$ is a non-quadratic potential.

An undirected propagator is indeed a crucial part of the standard Feynman calculus since the propagator resulting from an action such as (\ref{scalar-action}) is
\bea
 \parbox{30mm}{\includegraphics[scale=0.28]{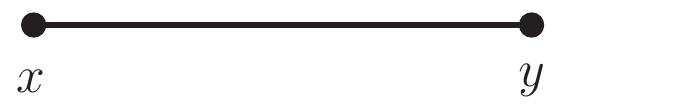}}
= -\( \frac{\delta^2 S}{\delta \phi(x)\, \delta \phi(y)} \)^{-1}
\eea
 (in conventions of classical effective field theory \cite{CLEFT-caged}), and its exchange symmetry (exchanging $x$ and $y$) arises from the commutativity of the second variation of the action (together with exchange-symmetric boundary conditions, which is the element violated by the retarded propagator). Nevertheless, we would like to eliminate the radiation zone, namely to solve for the radiation fields through their equations of motion and only then to pay attention to the other fields. The substitution can certainly be done into the remaining equations of motion, yet in order to employ the powerful diagrammatic techniques of effective field theory we would prefer to perform it at the level of the action. Such elimination at the level of the action is the classical limit of integration out in the quantum path integral, but since the Feynman path integral does not appear in classical physics we refrain from the abuse of language associated with the term ``integration out''  and replace it in this context by ``elimination'' \cite{CLEFT-caged}. In short, \emph{the challenge in this subsection is to define an effective field theory with retarded propagators}.

\emph{An example and a hint}. A diagrammatic representation for some quantities is quite apparent. For concreteness, consider the action
\be
 S[\phi] = \int d^4x \( \frac{1}{8 \pi G} \del_\mu \phi\, \del^\mu \phi - \rho\, \phi - \frac{g}{6} \phi^3 \) \label{phi3-action}
\ee
namely (\ref{scalar-action}) with $V=g\, \phi^3/6$. Suppose we wish to calculate $ \phi= \phi_\rho$, the value of the retarded field in the presence of a source $\rho(x)$. In this case one only needs to refine the standard Feynman rules by attaching a direction to each propagator, which can be thought to represent the direction of time. The vertices remain the same and carry the same value; only the orientation of their propagators needs to be specified, and the correct prescription is that for each vertex exactly one propagator should be chosen as outgoing (and the rest as ingoing). In our example these rules are
\bea
 \parbox{17mm}{\includegraphics[scale=0.4]{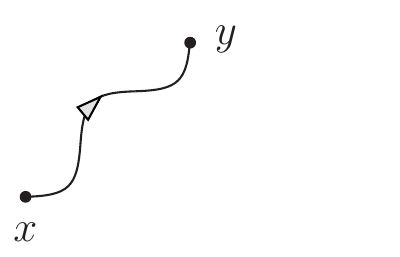}}
	= G_{ret}(x,y) \qquad
 \parbox{17mm}{\includegraphics[scale=0.4]{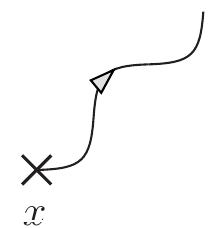}}
	= -\rho(x) \qquad
  \parbox{17mm}{\includegraphics[scale=0.4]{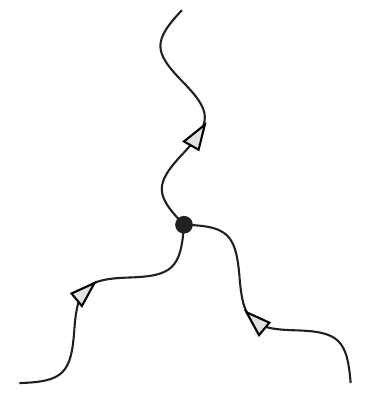}}
=  -g ~, \label{FRex}
\eea
and $\phi_\rho$ is correctly reproduced by
\bea
\phi_\rho =
 \parbox{17mm}{\includegraphics[scale=0.4]{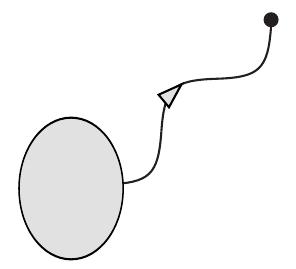}} = \parbox{17mm}{\includegraphics[scale=0.3]{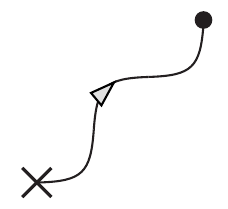}} + \parbox{17mm}{\includegraphics[scale=0.4]{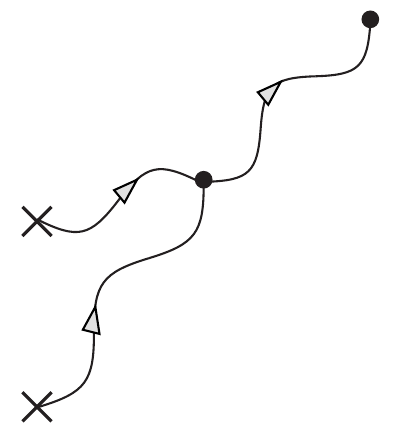}} + \dots \, \, \, \, .
\label{1ptex}
\eea

While field values, namely 1-point functions, can be defined in this way, it does not work for the effective action, namely a 0-point function, which is our goal. The reason is that in order to obtain a 0-point function  the future end of at least one retarded propagator must terminate on a source (a 1-valent vertex), and hence that vertex would not conform with the requirement of having exactly one outgoing (future directed) leg. It is instructive to observe what goes wrong if nevertheless we define an effective action by allowing such a vertex, for instance \be
 \parbox{17mm}{\includegraphics[scale=0.4]{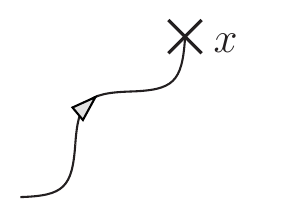}}
= \hspace{-.3cm}^?  -\rho(x) ~. \label{wrongFR4ex} \ee
Then the expansion of the effective action would be \be
 S_{eff} =
\parbox{20mm}{\includegraphics[scale=0.4]{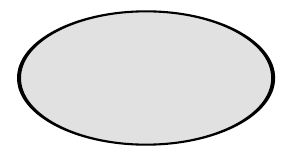}} =  \parbox{10mm}{\includegraphics[scale=0.25]{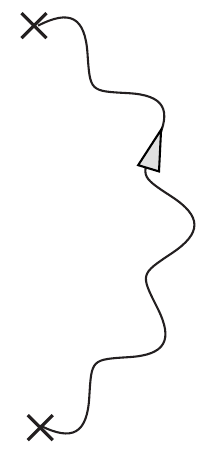}} + \dots
= \int dx dy\, \rho(x) G_{ret}(x,y) \rho(y) + \dots \, \, \, \, . \ee
However, this definition does not satisfy the basic relation $\delta S_{eff} / \delta \rho(x)|_{g=0} =-\phi_{ret}(x)$ but rather one finds $\delta S_{eff} / \delta \rho(x)|_{g=0} =-2\, \phi_{even}(x)$, where only the symmetric part of the propagator appears.

\presub {\bf Theories defined through equations of motion - a generalization}.
 Having identified the essential problem to be the directed nature of the retarded propagator we abstract a more general context for the problem, that of a theory defined through equations of motion (and not through an action). For concreteness we continue to assume a single scalar field $\phi=\phi(x)$ and we write its equation of motion as \be
   0 = EOM(x) = EOM_0 (x)+EOM' (x) - \rho(x) \, \, ,  \label{doubling:eom} ~. \ee
where $EOM_0$ is a dominant linear part, $EOM'$ is a perturbation and $\rho(x)$ is a source. We note that such theories are inherently classical and do not allow a definition of a quantum theory. This class of theories can be naturally described by directed propagators as we proceed to explain.

Even though the theory has no action, it is still possible to define properly generalized Feynman rules which compute the solutions for $\phi$ as follows (see for example \cite{CLFT}) \bea
 \parbox{17mm}{\includegraphics[scale=0.4]{FRex1.pdf}}
&=&    -\( \frac{\delta EOM_0 (x)}{\delta \phi (y)} \)^{-1} \, \, , \non
  \parbox{25mm}{\includegraphics[scale=0.3]{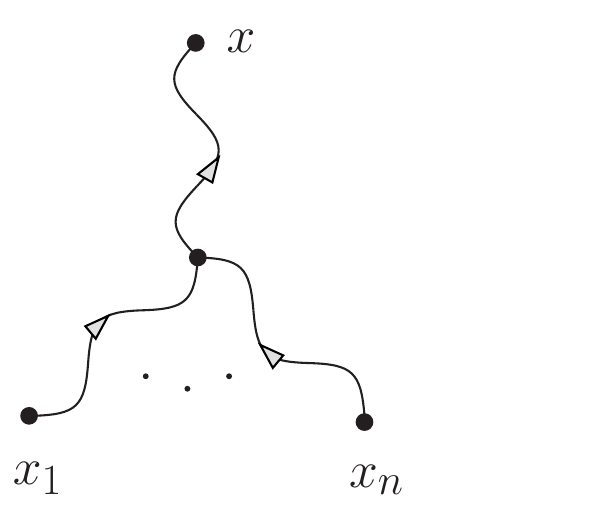}}
&=&   \frac{\delta^n}{\delta \phi(x_1) \dots \delta \phi(x_k) } EOM'(x) \, \, .
\label{EOM-Feynman}
\eea
Note that the propagator is directed because generally
\be
\frac{\delta EOM_0(x)}{\delta \phi (y)} \ne \frac{\delta EOM_0 (y)}{\delta \phi (x)}  \label{doubling:asym1} ~.
\ee
Moreover, each vertex has precisely one outgoing propagator which represents the field which is being solved for, while the ingoing propagators represent a source term.
The proof that these diagrammatic rules produce the correct solution for $\phi$ (\ref{1ptex}) is by induction on the number of vertices.\footnote{A single vertex diagram is equivalent to a term in an equation of motion, and given an arbitrary tree one identifies the vertex attached to the external leg and applies the assumption to the sub-trees which are connected to it.}

The discussion above can be generalized to several fields $\phi^i,~i=1,\dots,N$. In such a case there are even more opportunities for the propagators to be directed, namely asymmetric, since (\ref{doubling:asym1}) becomes
\be
\frac{\delta EOM_{i0}(x)}{\delta \phi^j(y)} \ne \frac{\delta EOM_{j0}(y)}{\delta \phi^i(x)} ~. \label{doubling:asym2}
\ee

The challenge is to find a diagrammatic effective action formulation for the equations of motion (\ref{doubling:eom}).

 \emph{Solution.} First we try to to define an action $\hS$ such as to reproduce the directed Feynman rules (\ref{EOM-Feynman}). We can obtain a direction for the propagator by adding an auxiliary doubling field $\hphi$ such that the propagator connects $\phi$ and $\hphi$ and the exchange symmetry is manifestly broken. To reproduce the vertices of (\ref{EOM-Feynman}) from an action we must take a product of $EOM'_i$ and $\hphi$. Both requirements above are satisfied by the following definition of a double field action \be
 \hS \[ \phi,\hphi \] := \int dx\, EOM_\phi(x)\,\hphi(x) \, \, , \label{doubled-action1} \ee
 where we now added a subscript $\phi$ to the equations of motion in order to stress that they are the equations of motion for $\phi$, not for $\hphi$.
Actually, this definition implies that the new equation of motion with respect to $\hphi$ reproduces the original one for $\phi$, namely $\delta \hS/\delta \hphi(x) = EOM_\phi(x)$.

The definition (\ref{doubled-action1}) reproduces correctly the diagrams for the 1-point function of $\phi$. Yet, in order to eliminate $\phi$ and define an effective action we also need to define a zero point function (no external $\phi$ legs). For that purpose we must have a source (or sink) vertex for $\phi$ -- note that the original source $\rho(x)$ in (\ref{doubling:eom}) is coupled to $\hphi$ in $\hS$.  Now we add to $\hS$ an auxiliary source $\hrho$ for $\phi$ as follows \be
 \hS \[ \phi,\hphi;\, \rho,\hrho \] := \int dx\, \[ EOM_\phi(x)\,\hphi(x) + \frac{\delta EOM_\phi (y)}{\delta \rho(x)}  \hrho(x)\, \phi(y) \] ~. \label{doubled-action2}
 \ee
This is the final form of $\hS$ in the general case. It allows for a general dependence of $EOM_\phi$ on $\rho$, and in the case of (\ref{doubling:eom}) the last term becomes  $S \supset -\int dx\, \hrho (x) \phi(x)$.

It is natural to wonder what is \emph{the meaning of $\hphi$ and $\hrho$}. Let us consider the new equation of motion, namely
\be
 0 = \frac{\delta \hS}{\delta \phi(x)} = \int dx'\, \frac{\delta EOM_\phi(x')}{\delta \phi(x)}\, \hphi(x') -\hrho(x) \, \, .
\ee
This means that $\hphi$ satisfies an equation for linearized fluctuations around the background $\phi$ with a source $\hrho$ and \emph{reversed} propagation (as a source for reversed propagation $\hrho$ can be termed a sink). In particular, in the absence of its source $\hphi$ vanishes, namely
\be
 \hrho = 0 \implies \hphi=0 ~. \label{physical-cond}
\ee
 We view $\phi$ and $\rho$ as being physical observables, unlike  $\hphi$ and $\hrho$. Indeed we set in the equations of motion $\hrho=0$  as a physical condition and hence $\hphi=0$. So $\hphi,\, \hrho$ which describe the hypothetical linearized perturbation of the system as a result of a hypothetical reversed propagation source $\hrho$ are considered auxiliary fields introduced for the purpose of the formalism, and do not reflect a genuine doubling of the system's degrees of freedom.

\presub \emph{\bf Specializing back to retarded propagators}. Now that we have defined an action formulation of a general system with directed propagators, we can specialize back to the case of retarded propagators. In this case \emph{the double field or radiation reaction action} (\ref{doubled-action2}) becomes
\be
 \hS \[ \phi,\hphi;\, \rho,\hrho \] := \int dx\, \[ \frac{\delta S[\phi]}{\delta \phi(x)}\,\hphi(x) + \frac{\delta S[\phi]}{\delta \rho(x)}\, \hrho(x)\,  \] ~. \label{doubled-action3}
\ee
This is the central result of the current  subsection. In this case $\hphi$ describes the hypothetical linearized perturbation  of the system as a result of \emph{advanced} propagation from a hypothetical source $\hrho$.

\presub {\bf Returning to the example}. Given the action (\ref{phi3-action})
 the double field action (\ref{doubled-action3}) is given by
\be
 \hS = \int d^4x \( \frac{1}{4 \pi G} \del_\mu \phi\, \del^\mu \hphi  - \frac{g}{2} \phi^2\, \hphi - \rho\, \hphi  - \hrho\, \phi \) ~.
\ee
The Feynman rules arising from $\hS$ are precisely those anticipated in (\ref{FRex}, namely a directed version of the Feynman rules for the original action (\ref{phi3-action}) where the arrow on each propagator points from $\hphi$ to $\phi$, together with the addition of the new vertex \be
 \parbox{17mm}{\includegraphics[scale=0.4]{FRex4.pdf}}
= -\hrho(x) ~, \label{rightFR4ex}
\ee
which replaces the failed attempt (\ref{wrongFR4ex}).

\presub {\bf  Field doubling and zones}. In our problem there are two zones, and directed propagators appear only in the radiation zone. Therefore we need to specify which fields are to be doubled. The radiation fields should clearly be doubled together with their sources $Q$. The source $Q$, in turn, depends on the fields of the system zone including the trajectories $\vec{x}_A (t)$. This leads us to proceed and double also the fields in the two body zone including $x$. However, there is no need to perform any system zone computation with hatted fields since owing to the undirected nature of the propagator any hatted quantity in the system zone can be readily obtained from the un-hatted ones through linearization. In particular $\hQ$ can be obtained from $Q=Q[x(t)]$  by
\be
 \hQ[x, \hx] = \int dt\, \frac{\delta Q[x]}{\delta x^i_A(t)}\, \hx^i_A(t) \, \, . \label{def:hatQ}
 \ee

Therefore we distinguish between the \emph{essential doubling} of the radiation fields and the inessential doubling, or \emph{cloning} of the system zone fields.

\presub {\bf Discussion: relation with Closed Time Path formalism}. The field doubling theory of this section is closely related to the Closed Time Path (CTP) Formalism \cite{CTP} (also known as the in-in formalism) in quantum field theory. In CTP (see for example the description in \cite{GalleyTiglio})  given a general action $S=S[\phi]$ one doubles the fields and defines the following action for them \be
 \hS_{CTP} [\phi_1,\phi_2] :=  S[\phi_1] - S[\phi_2] \label{CTP-action} ~, \ee
 together with a physical condition to be imposed at the level of the equations of motion \be
  \phi_1=\phi_2 \label{physical-cond-12basis} ~, \ee
 and with the propagators between $\phi_A$ and $\phi_B$, $A,B=1,2$, arranged in a matrix
 \be \( \begin{array}{cc}
 G_F & -G_- \\
 -G_+ & G_D
 \end{array} \) \, \, , \ee
 where $G_F$ is the Feynman propagator, $G_D \equiv G_F*$ is the Dyson propagator and the propagators $G_\pm$ are defined in Appendix C of \cite{GalleyTiglio}.

It is useful to change basis and use the Keldysh representation $\phi,\hphi$ \bea
 \phi_1 &=& \phi + \half \hphi \non
 \phi_2 &=& \phi - \half \hphi \eea
The double field action (\ref{CTP-action}) becomes \be
 \hS = S\[\phi+ \half \hphi \] - S\[\phi - \half \hphi \] = \frac{\delta S[\phi]}{\delta \phi} \hphi + \frac{1}{6}  \frac{\delta^3 S[\phi]}{\delta \phi^3} \hphi^3 +\dots \, \, \, \, . \label{doubled-action-12} \ee
 In the non-quantum limit second and higher powers of $\hphi$ do not contribute to the equations of motion due to the physical condition (\ref{physical-cond-12basis}) which becomes $\hphi=0$ and we reproduce (\ref{doubled-action3}). Here we wish to note a difference between our formulation and CTP. In our formulation we do not impose the physical condition $\hphi=0$, but rather the equation of motion with respect to $\phi$ yields the linear relation $\hphi=\hphi(\hrho)$ and that can be substituted back into the action. Since the physical condition is not imposed the field doubled actions (\ref{doubled-action3},\ref{doubled-action-12}) are inequivalent and we find that in the non-quantum limit the Keldysh basis appears to be the preferred and only basis to be used.

The propagator matrix becomes \ \be \( \begin{array}{cc}
 0 & -iG_{adv} \\
 -i G_{ret} & \half G_H
 \end{array} \) \, \, , \ee
where $G_H \equiv \{ \phi(x), \phi(y) \}$ is the Hadamard function.
 In the non-quantum limit the propagator between $\hphi$ and itself can be taken to vanish (since the action expanded in $\hphi$ is truncated after the first order), and hence we recognize that the propagator matrix reduces to the propagator used here (\ref{FRex}), once the convention for real Feynman rules is accounted for. Now  we recognize the current theory in the context of retarded propagators to be the non-quantum (classical) limit of CTP (however, the general theory founded on equations of motion cannot have a quantum relative).

Normally physical theories are superseded by more general ones which reduce to the original theory in the appropriate limit (the correspondence principle). However, it can sometimes happen that this order is reversed, namely first a more general theory is formulated and only later it is realized that a non-trivial theory can be defined through one of its limits. This is the case at hand where the quantum theory was formulated before its classical limit was known. We may refer to the current theory as the classical ``origin'' of CTP, where origin is in quotes since while it could have been defined in classical physics and then used as an origin for a quantum theory the actual chronological order was reversed.
For a discussion of the extant literature on the classical limit see Appendix \ref{sec:app-Galley}.

\subsection{Summary of formulation}
\label{subsec:summary}

We shall now collect all the ingredients introduced in this section and summarize our formulation. We consider a two-body problem and 3 cases of interactions between them: scalar, electromagnetic and gravitational ($s=0,1,2$ in short). The problem's action for each case is summarized in table \ref{tab:fields}.

\emph{Field variables and zones}. In the post-Newtonian approximation one uses two zones: the system zone and the radiation zone, and accordingly each field is defined over (or has modes in) each zone. We note that each zone has an enhanced symmetry and that guides us in the choice of field variables. In the system zone we use non-relativistic field variables, while in the radiation zone we use spherical wave variables (\ref{def:spherical-fields}) - see table \ref{tab:zones}.

\emph{Action and Feynman rules}. The bulk action is the same in both zones, only expressed in the appropriate field variables. In the system zone the propagator is taken to be instantaneous (\ref{s0-system-propag}).

We define a coupling between the zones through the introduction of 2-way matching multipoles variables $P_L,\, Q_L$. These are algebraic fields which can be thought to reside in the overlap region and their action is given in (\ref{matching-action}). The equations of motion with respect to $P,Q$ reproduce the matching conditions.

In the radiation zone we define the action  \be
 S_{rad}[\phi_L;Q_L] := S\[\phi(x)=\phi_L(r,t)\, x^L\] - \int dt\, Q^L(t)\, \phi_L(t)|_{r=0} \, \, , \ee
 namely we express the scalar action (\ref{tab:fields}) in terms of the spherical fields $\phi_L$ (\ref{def:spherical-fields}) together with the coupling term to the radiation source multipoles $Q_L$ (\ref{def:Q}). For concreteness we present here the scalar case.

In order to use the retarded (and hence directed) propagator (\ref{retarded-prop}) we double the fields and sources by introducing $\hphi,\, \hQ$ and define the radiation reaction action to be (see \ref{doubled-action3}) \be
 \hS \[\phi,\hphi;\, Q,\hQ \] := \int d^4 x\, \[ \frac{\delta S[\phi,Q]}{\delta \phi(x)}\,\hphi(x) +  \frac{\delta S[\phi,Q]}{\delta Q(x)}\,\hQ(x)  \]  \label{sum:doubled-action} ~. \ee
The Feynman rules can be read from these actions.

\presub {\bf Elimination}. At this point the perturbation theory is formulated. In the next section we proceed to demonstrate it via explicit computations, eliminating fields to obtain effective actions involving the bodies and the radiation.

\emph{2-body effective action}. Elimination of system zone fields with no coupling to radiation produces the usual 2-body effective action \be
S_{2bd}[x_1,x_2]:=
 \parbox{17mm}{\includegraphics[scale=0.4]{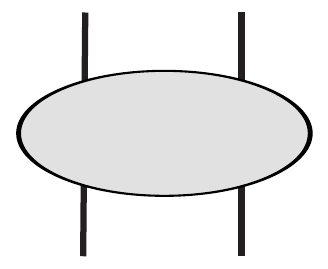}}
\, \, \, \, \, \, \, \, \, \, , \ee
which accounts for conservative interactions. The heavy lines denote the bodies.

\emph{Radiation source multipoles $Q[x]$} are defined through the elimination of near zone fields  in the presence of one external leg of radiation
\bea
-Q[x] \equiv
 \parbox{17mm}{\includegraphics[scale=0.4]{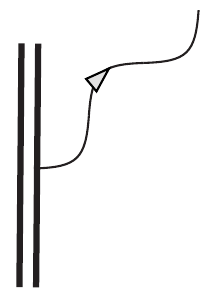}} := \parbox{35mm}{\includegraphics[scale=0.4]{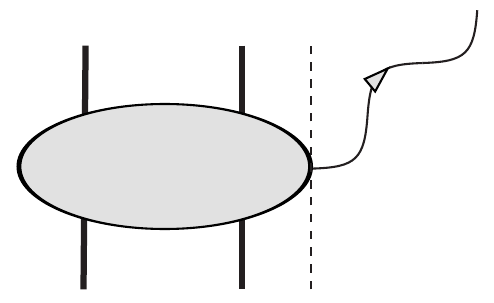}} \, \, ,
\label{def:Qx}
\eea
namely the source vertex in the radiation zone is defined through matching with the system zone. The double heavy line represents the system (in the radiation zone), while a single heavy line represents one of the bodies (in the system zone) and finally the dashed vertical line separating the zones represents their interaction (through the $P,Q$ sector). The doubled source $\hQ$ is defined to be the linearization of $Q$ (\ref{def:hatQ}).
In practice we compute first $Q=Q[\rho]$ for an arbitrary charge distribution $\rho$ (in the system zone) and then substitute the charge distribution of a system of point particles $\rho=\rho[x]$ obtaining altogether \be
 Q[x] := Q\[ \rho[x] \] \, \, . \ee
In computing $Q[\rho]$ we extrapolate the spherical wave from the radiation zone and the overlap region into the system zone and up to the last vertex there.

\emph{Outgoing radiation} is defined through the asymptotic form of the radiation field. Diagrammatically it requires elimination of the radiation fields (except for the asymptotic modes) by
\be
\phi |_{r \to \infty} =
 \parbox{35mm}{\includegraphics[scale=0.4]{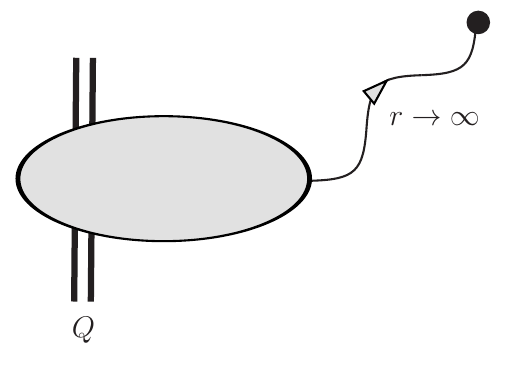}} \equiv \parbox{45mm}{\includegraphics[scale=0.4]{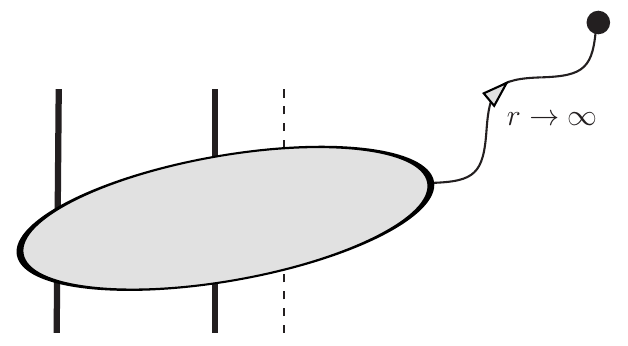}} \, \, ,
\ee
 where the last equality uses the diagrammatic definition of $Q[x]$ (\ref{def:Qx}) \footnote{We remark that the usage of diagonally tilted bubbles in our Feynman diagrams indicates the inclusion of a time-directed propagator.}.
From this quantity we can isolate $\phi^{out}(u,\Omega)$ defined by
\be
\phi(r,t, \Omega) \sim \frac{1}{r}\, \phi^{out}(t-r,\Omega) \mbox{ at } r \to \infty ~. \label{def:phi-out}
\ee
Next the radiated power can be obtained by integrating over the asymptotic radial energy flux (or using the optical theorem).

\emph{The radiation reaction effective action $\hS_{RR}$ and associated RR force}. Elimination of radiation fields through diagrams without external legs yields the radiation reaction effective action
\be
 \hS_{RR}\[Q,\hQ \] :=
 \parbox{25mm}{\includegraphics[scale=0.45]{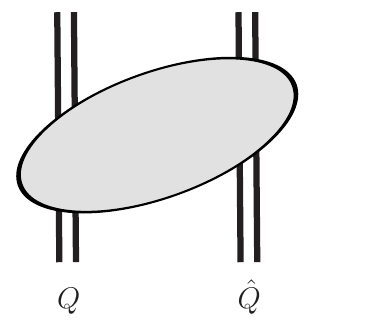}} \equiv \parbox{45mm}{\includegraphics[scale=0.45]{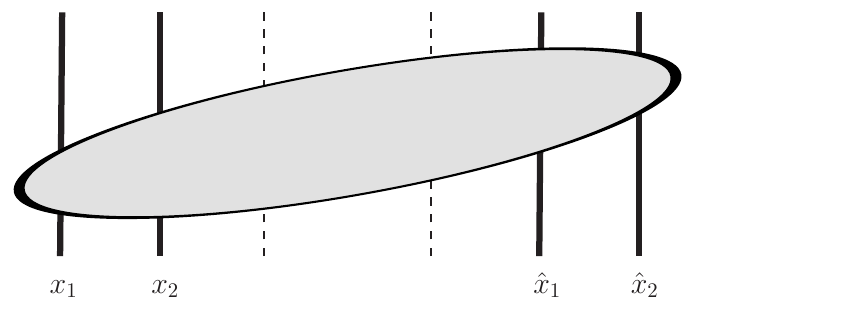}} ~~~~~~~~ .
\label{def:hS-Q}
\ee
 If one wishes one can obtain the reaction field $\phi^{sys}_L |_{r \to \infty}$ and reaction source multipoles $P_L$ as follows \be
 \phi^{rad}_L |_{r=0} \equiv \phi^{sys}_L |_{r \to \infty} \equiv - G\, (2\ell-1)!!\, P_L = - \frac{\delta \hS[Q,\hQ]}{\delta \hQ}  ~. \ee

To obtain $\hS_{RR}$ in terms of $x$ and $\hx$, we substitute $Q[x]$ and $\hQ[x,\hx]$ in (\ref{def:hS-Q}),
\be
 \hS_{RR}[x,\hx] = \hS_{RR}\[Q[x],\, \hQ[x,\hx] \]  \label{def:hS-x}	~. \ee
This quantity encodes in it the RR force through the following contribution to the equations of motion
 \footnote{It is possible to define $\hat{\hS}[x,\hx]:=\int (\delta S/\delta x^i_A) \hx^i_A + \hS[x,\hx]$ such that the equations of motion have a more familiar form $0=(\delta \hat{\hS}/\delta \hx^i_A)$, namely to consist of the variation of a single action functional. However, we find it more convenient not to use $\hat{\hS}$.}
\be
0=\frac{\delta S_{2bd}[x]}{\delta x_A^i} + \frac{\delta \hS_{RR}[x,\hx]}{\delta \hx_A^i} \, \, ,
\label{summ1:eom}
\ee
namely the RR force is
\be
  \(F_{RR}\)^i_A :=  \frac{\delta \hS_{RR}[x,\hx]}{\delta \hx_A^i} \label{def:sf} ~.
\ee

\emph{Dissipated power}. Now one can compute the power dissipated by the RR force (summed over all bodies) \bea
P_{RR} &=&  -\dot{x}^i_A\, \(F_{RR}\)^i_A = -\dot{x}^i_A\, \frac{\delta \hS_{RR}[x,\hx]}{\delta \hx_A^i} = -\dot{x}^i_A\, \int dt'\, \frac{\delta \hS}{\delta \hQ_L(t')}\, \frac{\delta \hQ_L(t')}{\delta \hx_A^i} = \non &=& \dot{x}^i_A\, \int dt'\,  \frac{\delta \hS}{\delta \hQ_L(t')}\, \frac{\delta Q_L(t')}{\delta x_A^i} \, \, ,
\eea
where $P_{RR},\, \dot{x}^i_A$ as well as $\delta \hx_A^i,\, \delta x_A^i$ all occur at time $t$ and we used (\ref{def:sf}) in the second equality and (\ref{def:hatQ}) in the fourth.
In the scalar theory $Q$ depends only on $x$ and not on its time derivatives and hence  \be
\frac{\delta Q_L(t')}{\delta x_A^i(t)}\, \dot{x}^i_A(t) = \frac{dQ}{dt}\, \delta(t-t') \, \, , \label{dQdt}
\ee
and hence \be P_{RR}  = - \frac{d Q_L}{dt}\, \frac{\delta \hS}{\delta \hQ_L} ~. \ee
Actually this holds for any $Q$ of electric type. More generally, namely for $Q$ of magnetic type, (\ref{dQdt}) holds only after averaging over $t$ and hence the time averaged power is given by \be
\left< P_{RR} \right>  = - \left< \frac{d Q_L}{dt}\, \frac{\delta \hS}{\delta \hQ_L} \right> ~. \label{P-sf}
\ee
In the next section we shall confirm that $P_{RR}$ is equal on average to the radiated power.

\presub {\bf Discussion}.

\emph{ Relation between radiation and RR force is made more apparent}. This is due to both quantities depending on the source multipoles $Q$ and moreover, each diagram which contributes to $\hS[Q,\hQ]$ can be associated with one which contributes to $\phi^{out}$ simply by erasing the (double heavy) line for $\hQ$. The equality (on average) of dissipated and radiated power will be seen quite directly, see for example (\ref{radiated energy scalar multipoles}) in the scalar case.

	\begin{figure}[t]
	\begin{center}
 \includegraphics[width=11cm]{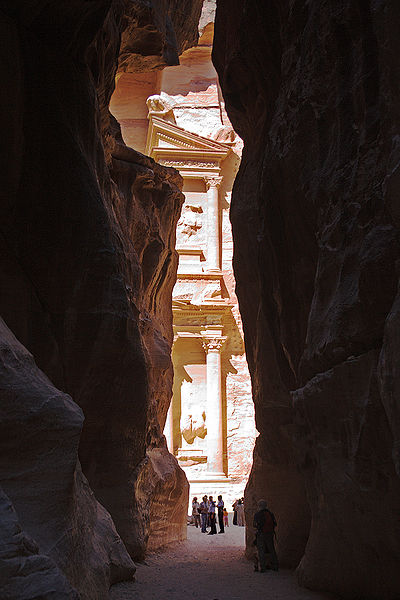}	
	\caption{The narrow passage leading to Petra (Jordan). Similarly the radiation zone is a narrow passage which must be crossed in order to generate a violation of time reversal, which therefore appears at high PN order.}
	\label{fig:petra}
	\end{center}
	\end{figure}

\emph{Bottleneck for T violation}. While time reversal symmetry ($T$) is broken by the retarded propagator, it is still a symmetry of the system zone due to its instantaneous propagation and $S_{2bd}[x]$ is $T$-symmetric. Moreover, $\hS[x,\hx]$ violates $T$ and causes dissipation starting only at a high PN order (2.5PN in gravity, 1.5PN in electromagnetism). We explain that by the observation that $T$ violating interactions in the system zone can only appear from diagrams which communicate with the radiation sector which thereby behaves as a bottleneck that increases the PN order,
see figure \ref{fig:petra}.
In other words the radiation sector is a $T$-breaking sector while $T$ is preserved in the system sector, and the violation of this symmetry (or anomaly) is communicated only though zone coupling (namely the $P,Q$ messengers).

\emph{Characterization of radiation reaction and its relation with the self-force}. In the introduction we loosely defined the RR force  to be the force which guarantees that the system dissipates to ensure overall conservation when radiation is accounted for. The precise definition (\ref{def:sf}) means that the RR force is the one which arises from the elimination of the radiation zone, or equivalently it is caused by system zone fields which originate from the overlap region. This definition is rooted in the post-Newtonian (PN) context as it assumes the existence of the various zones. However, the RR force is not defined by a specific PN order. This suggests that it might be possible to define it in a wider context than PN.

A related issue is to characterize the RR force, namely to distinguish it by expression and not by the way it is computed. The system zone and hence the conserving force preserve $T$, conserve $M, \, \vec{J}, \, \vec{P}$, and finally are independent of the choice of wave boundary conditions (retarded, advanced etc.). It might be that violation of some of these properties can serve as characterization.

The RR force is closely related to the notion of self-force. The self-force is defined to be a force which is affected by a field which was sourced by the very same body in its past. In gravity this notion is natural in the extreme mass ratio (EMR) and it is usually reserved for a force of order ${\cal O} \( m^2 \)$ where $m$ is the smaller mass. The two concepts can be compared and distinguished in a double limit which assumes both PN and EMR,
\footnote{For interesting tests in this double limit see \cite{Blanchet:2010cx} and references therein.}
as can be seen from the following two examples. In this limit the 1PN conserving (and hence non-RR) force which arises from the gravitation of potential energy has the scaling of a self-force, while on the other hand the leading quadrupole RR force has a component proportional to $m M$, where $M$ is the large mass, and hence would not be considered self-force (at least as long as equations of motion are not substituted in the expression). However, in certain cases the two forces coincide, for example in the context of a  single body coupled to the electromagnetic field the ALD force is both RR and self-force.

\section{Demonstration}
\label{sec:demonstration}

In this section we demonstrate the formulation presented in the previous section through explicit perturbative calculations of radiation source multipoles $Q[J]$, the outgoing radiation and the RR force in the cases of a scalar, electromagnetic and gravitational fields. We start with a scalar field in subsection \ref{section:Scalar SF} where we employ spherical waves; we economize the EFT expression for $Q[J]$ by including retardation effects through Bessel functions; the RR force is conveniently formulated in terms of a double field action and is derived through a 1-line evaluation of a Feynman diagram. Next we proceed to the electromagnetic case in subsection \ref{subsection:Electromagnetism} where one must account for the two possible polarizations and we add to the previous ingredients a gauge invariant formulation of waves and sources. Finally we reach the gravitational case including its non-linearities in subsection  \ref{gravitational RR force}. In addition to obtaining an expression for the linearized radiation sources economized by Bessel functions we compute the leading radiation reaction effective action and the associated RR force as well as the next to leading corrections in a way which uses fewer Feynman diagrams and thereby economizes the computation in \cite{GoldbergerRoss}. We moreover compute certain new higher order terms. In many cases we take \cite{RossMultipoles} as a basis for comparison, as it is a thorough treatment of radiation source multipoles and more within the EFT approach.

\subsection{Scalar case: spherical waves and double-field action}
 \label{section:Scalar SF}

We first examine the RR force acting on a charge distribution interacting through a scalar field. We take the action to be \bea
\label{ScalarAction}
S_\Phi= +\frac{1}{8\pi G} \int(\d \Phi)^2 r^2 dr d\Omega dt  - \int \rho\Phi r^2 dr d\Omega dt\, , \,\,\,\,\,\,\,\,
\eea
which is the same as (\ref{scalar-action}) with $V=0$.

{\bf Spherical waves: conventions}. We shall work in the frequency domain and use the basis of spherical multipoles by decomposing the field and the sources as
\bea
\Phi(\vec{r},t)&=&\int\frac{d \w}{2\pi}\sum_L e^{-i \w t} \Phi_{L \w}(r) x^L \, , \nonumber \\
\rho({\vec{r}},t)&=&\int\frac{d \w}{2\pi}\sum_L e^{-i \w t} \rho_{L \w}(r) x^L \, ,
\label{decomposition of scalar field and source}
\eea
where for a multi-index $L=(k_1k_2 \cdots k_\ell)$, $x^L$ is the symmetric-trace-free (STF) multipole
\bea
x^L=(x^{k_1}x^{k_2} \cdots x^{k_\ell})^{STF} \equiv r^\ell n^L.
\label{nL STF}
\eea
The multipole basis satisfies (following (\ref{basis options}))
\bea
&\int& x_{L_\ell}(r,\Omega) x^{L'_{\ell '}}(r,\Omega) d \Omega = \frac{4\pi r^{2\ell}}{(2\ell+1)!!} \delta_{\ell \ell'} \delta_{L_\ell L'_{\ell'}} \, \, \, \, , \nonumber \\
&\int& g^{\Omega \Omega'} \partial_{\Omega} x_{L_\ell} \partial_{\Omega'} x^{L'_{\ell '}} d \Omega = c_s \frac{4\pi r^{2\ell}}{(2\ell+1)!!} \delta_{\ell \ell'} \delta_{L_\ell L'_{\ell'}} \, \, \, \, ,
\label{scalar spherical harmonics normalization}
\eea
where $c_s:=\ell(\ell+1)$, $g^{\Omega \Omega'}$ is the metric on the 2-dimensional unit sphere (for summation convention and definitions, see Appendix \ref{app:Multi-index summation convention}).
We shall also use the inverse transformation\\
\bea
\rho_{L\w}(r)  &=&
\!\int\!\! \rho_\w(\vec{r}) x_L \frac{(2\ell+1)!!d\Omega}{4\pi\factell r^{2\ell}}=
\int\!\!\!\!\int\!\!{dt}e^{i \w t} \rho(\vec{r},t) x_L \frac{(2\ell+1)!!d\Omega}{4\pi\factell r^{2\ell}} \, .
\label{scalar inverse source}
\eea

{\bf Spherical waves: dynamics}. In the new notation, using $\Phi_{L -\w}=\Phi^*_{L \w}$, the action (\ref{ScalarAction}) becomes
\bea
S_\Phi&=&\frac{1}{2}\!\!\int\!\!\frac{d \w}{2\pi}\!\sum_L \!\!\!\int\!\!{dr}
\left[\frac{r^{2\ell+2}\factell }{(2\ell+1)!!\, G}
\Phi^*_{L \w} \left(\w^2+\d_r^2+\frac{2(\ell+1)}{r}\d_r \right) \Phi_{L \w}
-\left(\rho^\Phi_{L \w}\Phi^*_{L \w}+c.c.\right)\right] ~, \non
\label{scalar action spherical}
\eea
with the source term defined as
\bea
\rho_{L \w}^\Phi (r)=\frac{4\pi r^{2\ell+2}\factell }{(2\ell+1)!!}\rho_{L \w}(r)=r^2 \int{d\Omega}\rho_\w(\vec{r}) x_L.
\label{scalar action source}
\eea
From (\ref{scalar action spherical}) we derive the equation of motion
\bea
0=\frac{\delta S}{\delta \Phi_{L\w}^*}=\frac{r^{2\ell+2}\factell }{(2\ell+1)!!\, G} \left(\w^2+\d_r^2+\frac{2(\ell+1)}{r}\d_r \right) \Phi_{L \w}-\rho^\Phi_{L \w} \, \, .
\label{EOM Phi}
\eea
Changing to the dimensionless variable $x:= \omega r$ the homogenous part of this equation becomes
\bea
[ \del_x^2 + \frac{2(\ell+1)}{x}\del_x + 1]\, \tilde{j}_\ell=0,
\label{Modified Bessel equation}
\eea
and its solutions $\tilde{j}_\ell$,$\tilde{h}^+_\ell$  are Bessel functions up to normalization (see Appendix \ref{app:Normalizations of Bessel functions}). Thus \emph{the propagator for spherical waves is}
\bea
 \parbox{20mm}{\includegraphics[scale=0.5]{ScalarProp.pdf}}	
 \equiv G^\Phi_{ret}(r',r)
=\frac{-i\w^{2\ell+1}\, G}{\factell (2\ell+1)!!}\, \tilde{j}_{\ell}(\w r_1)\, \tilde{h}^+_{\ell}(\w r_2) \, \, \, \, ;  \non
r_1:=\min\{r',r\}, ~ r_2:=\max\{r',r\} \, .
\label{Phi propagator scalar}
\eea

We turn to derive the source term in the radiation zone through matching with the system zone according to the diagrammatic definition (see (\ref{def:Qx}) where the diagrammatic notation is explained)  \be
 \includegraphics[scale=0.5]{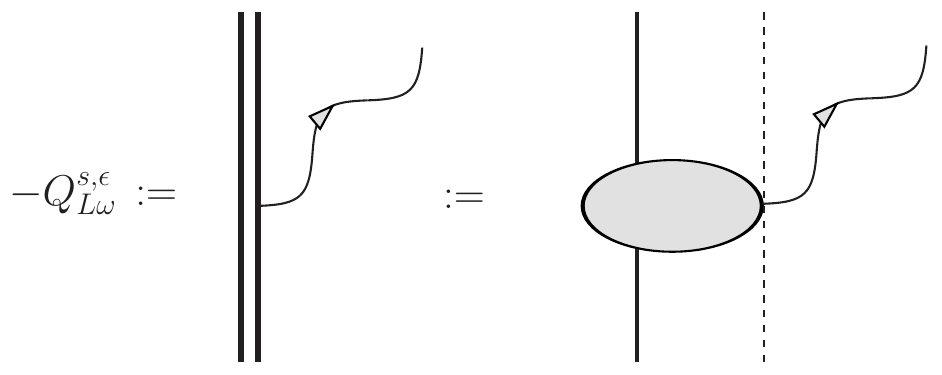} \, \, \, \, .
\label{vertex_definition}
\ee
In the radiation zone we think of the sources $Q_{L \w}$ as located at the origin or $r=0$. Hence the radiation zone field can be written as
\be
\Phi_{L \w}^{EFT}(r) =
\parbox{20mm}
 {\includegraphics[scale=0.5]{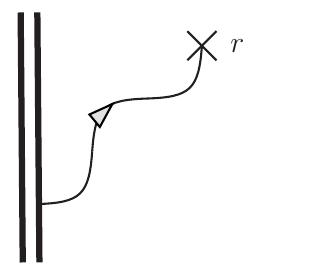}}
= -G\, Q_{L \w} \frac{-i\w^{2\ell+1}}{\factell(2\ell+1)!!} \tilde{h}^+_{\ell}(\w r) \, \, .
\label{scalar wavefunction at radiation zone1}
\ee
On the other hand, in the full theory (or equivalently in the system zone) we can also use spherical waves to obtain the field outside the source as
\bea
\Phi_{L \w}(r) &=& -\int dr' \rho^{\Phi}_{L'\w}(r') G^\Phi_{ret}(r',r)
	= -G\, \left[ \int dr' \tilde{j}_{\ell}(\w r') \rho_{L \w}^\Phi (r')  \right]  \frac{-i\w^{2\ell+1}}{\factell(2\ell+1)!!} \tilde{h}^+_{\ell}(\w r) \nonumber \\
 &=& -G\, \left[ \int d^3 x' \tilde{j}_{\ell}(\w r') \rho_\w(\vec{r} \, ') x'_L  \right]  \frac{-i\w^{2\ell+1}}{\factell(2\ell+1)!!} \tilde{h}^+_{\ell}(\w r) \, \, .
\label{scalar wavefunction at radiation zone2}
\eea

By comparing the expressions for the field (\ref{scalar wavefunction at radiation zone1},\ref{scalar wavefunction at radiation zone2}) and using identity (\ref{scalar inverse source}) to return to the time domain we find that \emph{the radiation source multipoles are}
\be
Q_L = \int d^3 x\, \tilde{j}_\ell(ir\d_t)\, x^{STF}_L\, \rho(\vec{r},t) ~ .
\label{I Phi scalar multipoles}
\ee
Using the series expansion (\ref{Bessel J series2}), we see that the scalar multipoles coincide with Ross's multipoles \cite{RossMultipoles} eqs. (10,11). We note that we can think of this process as a zoom out balayage of the original charge distribution $\rho({\bf r})$ into $Q_L$ carried out through propagation with $\tilde{j}_{\ell}(\w r)$. We also note a useful representation of this result
\bea
Q_L = \int d^3 x \, x^{STF}_L \int_{-1}^{1}dz \delta_\ell(z) \rho(\vec{r},u+z r) \, ~ ,
\label{I Phi scalar multipoles using delta ell}
\eea
where we have used the generating time-weighted function (following \cite{Blanchet:1989ki,Damour:1990gj})
\bea
\delta_\ell(z)&=&\frac{(2\ell+1)!!}{2^{\ell+1}\ell!}(1-z^2)^\ell~,\nonumber\\
\int_{-1}^{1}dz \delta_\ell(z) f(\vec{r},u+z r)&=&\sum_{p=0}^{+\infty}\frac{(2\ell+1)!!}{(2p)!!(2\ell+2p+1)!!} \left( r\d_u \right)^{2p} f(\vec{r},u)~.
\label{generating funtion delta ell}
\eea

Altogether \emph{the Feynman rules in the radiation zone} for the propagator and sources are
\begin{align}
\label{Feynman Rule Vertex}
\parbox{20mm}
 {\includegraphics[scale=0.3]{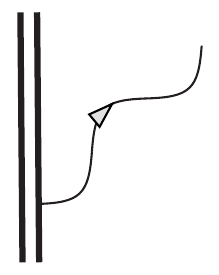}}
=-Q^{s,\epsilon}_{L \omega}\,\,\,\,\,\,\,\,\,\, ,\,\,\,\,\,
\parbox{20mm}
 {\includegraphics[scale=0.3]{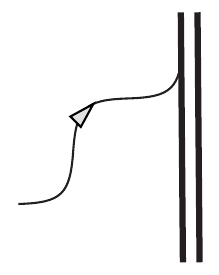}}
=-\hat{Q}^{s,\epsilon *}_{L \omega}\,\,\,\, ,
\end{align}
\begin{align}
\label{Feynman Rule Propagator}
\parbox{20mm}
 {\includegraphics[scale=0.5]{DiagsGeneralRulesProp.pdf}}
=G^{s,\epsilon}_{ret}(r',r)=\frac{-i\omega^{2\ell+1}}{(2\ell+1)!!}\, G\,
\tilde{j}_{\ell}(\w r_1)\, \tilde{h}^+_{\ell}(\w r_2)\,  R_s^\epsilon
\,\,\,\,\,\,\,\,\,\,\,r_1  \leq r_2	 \, \, ,
\end{align}
where for future use we included $\eps$, a possible polarization label  and $R_s$, a rational $\ell$-dependent factor which is absent in the scalar case, namely $R|_{s=0}=1$.

\subsubsection{Outgoing radiation and the RR effective action}
We can now use these Feynman rules to obtain our central quantities.

{\bf Outgoing radiation}. It can now be found diagrammatically as
\begin{align}
\label{Radiation Phi using feynman}
\Phi_{L\w}(r)=&
\parbox{20mm}
 {\includegraphics[scale=0.5]{DiagRadiationScalar.pdf}}
= 	-Q_{L'\w}G^\Phi_{ret}(0,r)
=G\, (-iw)^\ell \frac{Q_{L\w}}{r^{\ell}} \frac{e^{iwr}}{r},
\end{align}
where we used the asymptotic form of $\tilde{h}(x)$ (\ref{Bessel H asymptotic2})  and $\tilde{j}(wr')|_{r'=0}=1$ for the source at $r'=0$. Using (\ref{decomposition of scalar field and source}) we find that \emph{the outgoing radiation is}
\bea
\Phi(\vec{r},t) \sim \frac{G}{r} \sum_L n^L \d_t^\ell Q_L(t-r) ~~ \mbox{as } r \to \infty \, \, ,
\label{radiation Phi}
\eea
which coincides with \cite{RossMultipoles} (equation (21); note a $4\pi$ normalization factor which must be accounted for due to our different normalization of the action).

{\bf Radiation reaction effective action}. It encodes the RR force and is given by \begin{align}
\label{S Phi using feynman}
\hS_\Phi=&
\parbox{20mm}
 {\includegraphics[scale=0.5]{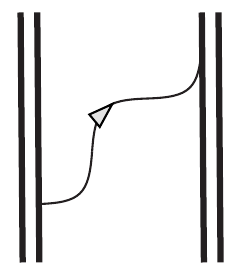}}
= \frac{1}{2}\int\!\frac{d \omega}{2\pi}\sum_{L,L'}
	\left(-Q_{L\w}\right)
	G^\Phi_{ret}(0,0)
	\left(-\hat{Q}^*_{L'\w}\right)+c.c.
	\nonumber\\
=& G \!\int\!\frac{d \omega}{2\pi}\sum_L
	\frac{-i\w^{2\ell+1}}{2\factell (2\ell+1)!!}
	Q^{L\w} \hat{Q}_{L\w}^*+c.c. \, \, \, \, ,
\end{align}
where now both vertices are at $r=r'=0$, and we regularized to zero the divergent (and time symmetric) $\tilde{y}_\ell$ part of $\tilde{h}_\ell^{(1)}$.
Using (\ref{scalar inverse source}) we return to the time-domain finding \emph{the radiation reaction effective action} to be
\bea
\hS_\Phi=G\, \int\!\!{dt}\!\sum_{L}\frac{(-)^{\ell+1}}{\factell (2\ell+1)!!} \hat{Q}^L \d_t^{2\ell+1}Q_L,
\label{S Phi scalar multipoles}
\eea
where $Q_L$ was given by (\ref{I Phi scalar multipoles}) and according to (\ref{def:hatQ})
\bea
\hat{Q}^L&=&\frac{\delta Q^L}{\delta \rho}\hat{\rho}=\frac{\delta Q^L}{\delta x^{i}}\hat{x}^{i}=\frac{\d Q^L}{\d x^{i}}\hat{x}^{i} + \frac{\d Q^L}{\d v^{i}}\hat{v}^{i} + \frac{\d Q^L}{\d a^{i}}\hat{a}^{i} + \cdots
 \, \, .
\label{I Phi scalar multipoles hat}
\eea

Note that the computation reduces to a mere multiplication: vertex -- propagator -- vertex.

\subsubsection{Applications and tests}

{\bf Perturbative expansion of the RR force}. Consider the case of a single charged body with a prescribed trajectory (``being held and waved at the tip of a wand'') interacting with a scalar field.  A fully relativistic force expression is known, analogous to the Abraham-Lorentz-Dirac (ALD) self force \cite{Dirac} familiar from electromagnetism. In this case the notion of RR force and self-force coincide. We compute our perturbative expansion up to order +1PN and compare it with the ALD-like expression.

The ALD self force of radiation reaction on an accelerating scalar charge can be derived in analogy with the electromagnetic case and is found to be
\begin{equation}
\label{F ALD}
F_{ALD}^\mu \equiv \frac{d p^\mu}{d\tau} = \frac{1}{3} G\, q^2\, \( \frac{d^3 x^\mu}{d\tau^3} - \frac{d^3 x^\nu}{d\tau^3} \frac{dx_\nu}{d\tau}\,  \frac{dx^\mu}{d\tau}\).
\end{equation}
The leading and next-to-leading-order terms in the PN expansion of this force are
\bea
\vec{F}_{ALDLO}= \frac{1}{3}G\, q^2\dot{\vec{a}},
\label{F LO ALD}
\eea
\bea
\vec{F}_{ALDNLO} = G\, q^2\left[ \frac{1}{3}v^{2}\dot{\vec{a}} + (\vec{v} \cdot \vec{a})\vec{a} + \frac{1}{3}(\vec{v} \cdot \dot{\vec{a}})\vec{v}\right] \, \, .
\label{F NLO ALD}
\eea
 where all derivatives are now with respect to $t$, namely $v^i:=dx^i/dt,\, a^i:=d^2x^i/dt^2$ and $\dot{a}^i:=d^3 x^i/dt^3$.

We compared our result of the RR force directly to these ALD terms, obtaining a perfect match up to next to leading order. Our method derives the RR force from the action and multipoles (\ref{I Phi scalar multipoles},\ref{S Phi scalar multipoles}) in three stages: by using a source term of a point particle for $\rho$, by matching the appropriate $\hat{\rho}$, and by finally calculating the contribution from generalized Euler-Lagrange equation $\delta S / \delta \hat{x}^i$ (following (\ref{summ1:eom})). The Source term corresponding to a scalar-charged point particle with a trajectory $\vec{x}(t)$ is
\bea
\rho(\vec{x} \, ' ,t) = q \int{\delta^{(4)}(x'-x) d\tau} = \frac{q}{\gamma}\delta^{(3)}(\vec{x} \, ' - \vec{x}) = \sum_{s=0}^{\infty} \frac{-(2s-3)!! v^{2s}}{(2s)!!} \delta^{(3)}(\vec{x} \, ' - \vec{x}) \,  .
\label{scalar point source}
\eea
Thus we find
\bea
Q^L&=&(2\ell+1)!!\sum_{p=0}^\infty\sum_{s=0}^\infty \frac{(2s-3)!!}
	{(2p)!!(2\ell+2p+1)!!(2s)!!}
	\d_t^{2p}(v^{2s}r^{2p}{x}^L_{TF}),	\nonumber\\
\hat{Q}^L&=&(2\ell+1)!!\sum_{\hat{p}=0}^\infty\sum_{\hat{s}=0}^\infty \frac{(2\hat{s}-3)!!}
	{(2\hat{p})!!(2\ell+2\hat{p}+1)!!(2\hat{s})!!}
	\d_t^{2\hat{p}}\frac{\delta}{\delta x^{i}}(v^{2\hat{s}}r^{2\hat{p}}{x}^L_{TF})\hat{x}^{i}
\eea
Accordingly we obtain the Lagrangian, which after moving $2\hat{p}$ time derivatives from the $\hat{x}^L$ multipoles to the ${x}^L$ multipoles by partial integration becomes
\bea
\hat{L}_\Phi=G q^2	\!\sum_{L}	\!\!	{(-)^{\ell+1}(2\ell+1)!!}	\!	
\sum_{p=0}^\infty\sum_{s=0}^\infty\sum_{\hat{p}=0}^\infty\sum_{\hat{s}=0}^\infty
	& \frac{(2s-3)!!}{(2p)!!(2\ell+2p+1)!!(2s)!!}
		\frac{(2\hat{s}-3)!!}{(2\hat{p})!!(2\ell+2\hat{p}+1)!!(2\hat{s})!!}
		\,\,\,\,\,\,\,\,\,\,\,\,\,\,\,\,\,\,\,\,\,\,\,\,\,\,	\nonumber\\
	\times& \hat{x}^{i} \frac{\delta}{\delta x^{i}}(v^{2\hat{s}}r^{2\hat{p}}{x}^L_{TF})
		\d_t^{2\ell+2p+2\hat{p}+1}(v^{2s}r^{2p}{x}_L)
		~	.\,\,\,\,\,\,\,\,\,\,\,\,\,\,\,
\label{scalar lagrangian}
\eea
In the Euler-Lagrange equation, the RR force contribution is given by the variation by $\hat{x}^j$,
\bea
F^j=\frac{\delta \hat{S}}{\delta \hat{x}^j}=
	\left[\frac{\d  \hat{L}}{\d  \hat{x}^j}
			- \frac{d}{dt}\left(\frac{\d \hat{L}}{\d \dot{\hat{x}}^j}\right)
	\right].
\label{scalar EL}
\eea
We thus find the leading force ($\ell=1,p=\hat{p}=s=\hat{s}=0)$, namely the leading dipole term, shown in table \ref{table:scalar multipoles leading} to coincide with
the leading order ALD result (\ref{F LO ALD}). Out of the 15 possible action terms in the next-to-leading order (for different $\ell,p,\hat{p},s,\hat{s}$), we find using the Euler-Lagrange equation 9 non-zero contributions to the force (recorded in table \ref{table:scalar multipoles next to leading scalar}). Adding these contributions we exactly reproduce the ALD result to this order (\ref{F NLO ALD}).\\
\begin{table}[h!]
  \centering \caption{Leading order contribution to the scalar self-force}
\begin{center}
\begin{tabular}{cccccccc}
  \hline
  $\ell$ $p$ $\hat{p}$ $s$ $\hat{s}$ & $\hat{L}/(G\, q^2)$ & $F^j/(G\, q^2)$ \\  \hline
  1 0 0 0 0 & $\frac{1}{3} \hat{x}^{i}\d_t^3x_{i}$ & $\frac{1}{3}\dot{a}^j$\\ \hline
\label{table:scalar multipoles leading}
\end{tabular}
\end{center}
\end{table}
\begin{table}[h!]
  \centering \caption{Next-to-Leading order contribution to the scalar self-force}
\begin{center}
\begin{tabular}{cccccccc}
  \hline
  $\ell$ $p$ $\hat{p}$ $s$ $\hat{s}$ & $\hat{L}/(G\, q^2)$ & $F^j/(G\, q^2)$ \\  \hline
  2 0 0 0 0 & $-\frac{1}{30}\hat{x}^{k} \frac{\delta}{\delta x^{k}}[x^i x^j-\frac{1}{3} x^2 \delta^{ij}]\d_t^5(x_i x_j)$
	& $-\frac{1}{15}[\d_t^5(x^i x^j)x_i-\frac{1}{3}\d_t^5(x^2)x^j]$\\  \hline
  1 0 1 0 0 & $\frac{1}{30} \hat{x}^{k} \frac{\delta}{\delta x^{k}} [x^i x^2] \d_t^5 x_i$
	& $\frac{1}{30}[x^2 \d_t^5 x^j+2x^j x_i \d_t^5 x^i]$\\  \hline
  1 1 0 0 0 & $\frac{1}{30} \hat{x}^i \d_t^5 (x^2 x_i)$
	& $\frac{1}{30} \d_t^5 (x^2 x^j)$\\  \hline
  1 0 0 0 1 & $-\frac{1}{6}\hat{x}^{k} \frac{\delta}{\delta x^{k}}[v^2 x^i] \d_t^3 x_i$
	& $-\frac{1}{6}v^2 \d_t^3 x^j +\frac{1}{3}\d_t[v^j x^i \d_t^3x_i]$\\  \hline
  1 0 0 1 0 & $-\frac{1}{6}\hat{x}^i \d_t^3(v^2 x_i)$
	& $-\frac{1}{6}\d_t^3(v^2 x^j)$\\  \hline
  0 1 1 0 0 & $-\frac{1}{18}\hat{x}^j x_j\d_t^5 x^2$
	& $-\frac{1}{18}x^j \d_t^5 x^2 $\\	\hline
  0 1 0 0 1 & $\frac{1}{6} \hat{v}^j v_j \d_t^3 x^2$
	& $-\frac{1}{6}\d_t [v^j \d_t^3 x^2]$	\\	\hline
  0 0 1 1 0 & $\frac{1}{6} \hat{x}^j x_j \d_t^3  v^2$
	& $\frac{1}{6} x^j \d_t^3 v^2 $\\	\hline
  0 0 0 1 1 & $-\frac{1}{2} \hat{v}^j v_j \d_t v^2 $
	& $\frac{1}{2}\d_t[v^j\d_t v^2]$ \\  \hline
\label{table:scalar multipoles next to leading scalar}
\end{tabular}
\end{center}
\end{table}

{\bf Dissipated power}. We calculate the total power dissipated by the RR force by substituting (\ref{S Phi scalar multipoles}) in (\ref{P-sf})
\be
 P_{RR}  = \sum_{L} \frac{(-)^{\ell}\,G}{\factell (2\ell+1)!!} \del_t Q^L\, \d_t^{2\ell+1}Q_L.
\label{radiated energy scalar definition}
\ee
Using $\ell$ integrations by parts, we find the time-averaged power to be
\bea
 <P_{RR}>&=& \sum_{L} \frac{G}{\factell (2\ell+1)!!} <(\d_t^{\ell+1}Q_L)^2>=\sum_{L} \frac{G}{\ell! (2\ell+1)!!} <(\d_t^{\ell+1}Q_{k_1 k_2 \cdots k_\ell}^{STF})^2> \non
     &=& P_{rad} ~.
\label{radiated energy scalar multipoles}
\eea
This result coincides with the average power emitted by the radiation, $P_{rad}$, which can be calculated directly from the energy flux of the outgoing radiation (\ref{radiation Phi}), or alternatively through eq.(15) of \cite{RossMultipoles}, (accounting again for a $4\pi$ normalization factor, and for comparison we re-introduced the $\frac{1}{\ell!}$ where the notation is non-multi-index, see Appendix \ref{app:Multi-index summation convention}).

\subsection{Electromagnetism: gauge invariant waves}
\label{subsection:Electromagnetism}

The EM action is given by
\bea
\label{MaxwellEMAction}
S=-\frac{1}{16\pi} \int F_{\mu \nu} F^{\mu \nu} r^2 dr d\Omega dt  -\int A_{\mu} J^{\mu}  r^2 dr d\Omega dt\, . \,\,\,\,\,\,\,\,\,\,\,
\eea
 Working in spherical coordinates $(t,r,\Omega)$ and reducing over the sphere as in (\ref{decomposition of scalar field and source}), we decompose the EM field and sources
\bea
A_{t/r}&=&\int\frac{d \w}{2\pi}\sum_L A_{t/r}^{L \w}  x_L e^{-i \w t} \, , \nonumber \\
A_{\Omega}&=&\int\frac{d \w}{2\pi}\sum_L \left( A_{S}^{L \w} \partial_{\Omega} x_L + A_{V}^{L \w} x^L_{\Omega} \right)  e^{-i \w t} \, , \nonumber \\
J^{t/r}&=&\int\frac{d \w}{2\pi}\sum_{L} J^{t/r}_{L \w}  x^L e^{-i \w t} \, , \nonumber \\
J^{\Omega}&=&\int\frac{d \w}{2\pi}\sum_L \left( J^{S}_{L \w} \partial^{\Omega} x^L + J^{V}_{L \w} x_L^{\Omega} \right)  e^{-i \w t} \, ,
\label{decomposition of EM field and sources}
\eea
where the scalar multipoles $x^L$ (\ref{scalar spherical harmonics normalization}) are now supplemented by the divergenceless vector multipoles $x^L_\Omega\!=\!\eps_{\Omega \Omega'}\d^{\Omega'} x^L\!=\!(\vec{r}\! \, \times\!\! \, \vec\nabla x^L)_\Omega$. The complete normalization conditions are
\bea
&\int& x_{L_\ell} x^{L'_{\ell '}} d \Omega
= \frac{4\pi\factell r^{2\ell}}{(2\ell+1)!!} \delta_{\ell \ell'} \delta_{L_\ell L'_{\ell'}} \, , \nonumber \\
&\int& g^{\Omega \Omega'} \partial_{\Omega} x_{L_\ell} \partial_{\Omega'} x^{L'_{\ell '}} d \Omega
= \frac{4\pi c_s \factell r^{2\ell}}{(2\ell+1)!!} \delta_{\ell \ell'} \delta_{L_\ell L'_{\ell'}} \, ,\nonumber \\
&\int& x_{L_\ell}^{\Omega} x^{L'_{\ell'}}_{\Omega} d \Omega
= \frac{4\pi c_s \factell r^{2\ell}}{(2\ell+1)!!} \delta_{\ell \ell'} \delta_{L_\ell L'_{\ell'}} \, ,\nonumber \\
&\int&  g^{\Omega \Omega'} g^{P P'} D_{\Omega}x^{L_\ell}_{P} D_{\Omega'} x^{L'_{\ell'}}_{P'} d \Omega = \frac{8\pi c_s^2 \factell r^{2\ell}}{(2\ell+1)!!} \delta_{\ell \ell'} \delta_{L_\ell L'_{\ell'}} \, ,
\label{vector sphercal harmonics normalization}
\eea
where $c_s:=\ell(\ell+1)$ and $D_{\Omega}$ is the covariant derivative on the sphere. Accordingly we shall also use the inverse transformations:\\
\bea
J^t_{L\w}(r)  &=&
\frac{(2\ell+1)!!}{4\pi\factell r^{2\ell}}\!\int\!\! \rho_\w(\vec{r}) x_L d\Omega=
\frac{(2\ell+1)!!}{4\pi\factell r^{2\ell}}\!\int\!\!\!\!\int\!\!{dt}e^{i \w t} \rho(\vec{r},t) x_L d\Omega \, ,	\nonumber\\
J^r_{L\w}(r) &=&
\frac{(2\ell+1)!!}{4\pi\factell r^{2\ell}}\!\int\!\! \vec{J}_w(\vec{r})\cdot \vec{n} x_L d\Omega=
\frac{(2\ell+1)!!}{4\pi\factell r^{2\ell}}\!\int\!\!\!\!\int\!\!{dt}e^{i\w t} \vec{J}(\vec{r},t)\cdot \vec{n} x_L d\Omega,\nonumber\\
J^V_{L\w}(r) &=&
\frac{(2\ell+1)!!}{4\pi c_s \factell r^{2\ell}}
\!\int\!\! \vec{J}_w(\vec{r})\!\cdot\!\left[\vec{r} \!\times\! \vec\nabla x_L\right] d\Omega=
\frac{(2\ell+1)!!}{4\pi c_s \factell r^{2\ell}}
\!\int\!\!\!\!\int\!\!{dt}e^{i\w t} \vec{J}(\vec{r},t) \!\cdot\! \left[\vec{r} \!\times\! \vec\nabla x_L\right] d\Omega.\,\,\,\,\,\,\,\,\,\,\,\,\,\,\,
\label{EM inverse sources}
\eea
We note that only three inverse transformations are needed, as we henceforth replace $J^S_{L \w}$ using the equation of current conservation
\bea
0=D_\mu J^\mu_{L\w} = iw J^t_{L\w} + (\d_r+\frac{\ell+2}{r})J^r_{L\w}-c_s J^S_{L\w}.
\label{current conservation}
\eea
We plug (\ref{decomposition of EM field and sources}) into Maxwell's action (\ref{MaxwellEMAction}) to obtain
\bea
S=\frac{1}{2}\int\frac{d \w}{2\pi}\sum_L
	 &\!\!\!\!\!\!\!\!\!\!\!\!\!\!\!\!\!\!\!\!\!\!\!\!\!\!\!\!\!\!\!\!\!\!\!\!\!\!\!\!\!\!\!\!\!\!\!\!\!\!\!\!\!\!\!\!\!\!\!\!\!\!\!\!\!\!\!\!\!\!\!\!\!\!\!\!\!\!\!\!\!\!\!\!\!\!\!\!\!\!\!\!\!\!\!\!\!\!\!\!\!\!\!\!\!\!\!\!\!\!\!\!\!\!\!\!\!\!\!\!\!\!\!\!\!\!\!\!\!\!\!\!\!\!\!\!\!\!\!\!\!\!\!\!\!\!\!\!\!\!\!\!\!\!\!\!\!\!\!\!\!\!\!\!
	 \frac{1 \factell }{(2\ell+1)!!} S_{L\w},	\nonumber\\
S_{L\w}=\!\int\!\!dr r^{2\ell+2} &
	\!\!\left\{
	\left[\left| i \w A^{L\w}_r - \frac{1}{r^\ell}(r^\ell A^{L\w}_t)' \right|^2
	+\frac{c_s}{r^2}  \left| i \w A^{L\w}_S - A^{L\w}_t \right|^2
	-\frac{c_s}{r^2} \left| \frac{1}{r^\ell}(r^\ell A^{L\w}_S)' - A^{L\w}_r \right|^2 \right. \right.
\nonumber \\
	 &\!\!\!\!\!\!\!\!\!\!\!\!\!\!\!\!\!\!\!\!\!\!\!\!\!\!\!\!\!\!\!\!\!\!\!\!\!\!\!\!\!\!\!\!\!\!\!\!\!\!\!\!\!\!\!\!\!\!\!\!\!\!\!\!\!\!\!\!\!\!\!\!\!\!
\left.\left.+c_s\left(\frac{\w^2}{r^2}-\frac{c_s}{r^4}\right)\left| A^{L\w}_V \right|^2
	-\frac{c_s}{r^2} \left| \frac{1}{r^\ell}(r^\ell A^{L\w}_V)' \right|^2    \right. \right]
\nonumber \\
	&\!\!\!\!\!\!\!\!\!\!\!\!\!\!\!\!\!\!\!\!\!\!\!\!
- \left. 4\pi \left[ A^{L\w}_r J_{L\w}^{r *} + A^{L\w}_t J_{L\w}^{t *}
						+ c_s A^{L\w}_S J_{L\w}^{S *} + c_s A^{L\w}_V J_{L\w}^{V *} + c.c. \right] \right\}
,
\label{EM action spherical}
\eea
where $':=\frac{d}{dr}$, and we have used $A_{L  -\w} = A^{*}_{L \w}$, $J_{L -\w} = J^{*}_{L \w}$ since $A_{\mu}(x)$, $J^{\mu}(x)$ are real.
We notice that $A^{L\w}_r$ is an auxiliary field, i.e. its derivative $A'_r$ does not appear in (\ref{EM action spherical}). Therefore, its EOM is algebraic and is solved to yield
\bea
A_{L\w}^r = -\frac{1}{\w^2 - \frac{c_s}{r^2}} \left[\frac{i \w}{r^\ell}(r^\ell A_{L\w}^{t})' + \frac{c_s}{r^{\ell+2}}(r^\ell A_{L\w}^{S})' - 4\pi J^r_{L\w} \right] ~.
\label{Ar}
\eea
Substituting the solution into the action, it can be seen that the action depends on only two gauge invariant variables: one is $A^{L\w}_V$, as it appears already in (\ref{EM action spherical}), coupled to the vector source term
\bea
\rho^V_{L\w} := J^V_{L\w} \, \, ,
\label{EM source vector}
\eea
and the other is
\bea
\tilde{A}^{L\w}_S:=A^{L\w}_t - i\w A^{L\w}_S\, ,
\label{astilde def}
\eea
which is coupled to its corresponding source term
\bea
\rho^S_{L\w} := - J^t_{L\w}+\frac{i}{\w r^{\ell+2}} \left( r^{\ell+2} \frac{\Lambda}{\Lambda-1} J^r_{L\w} \right)' \, ,
\label{EM source scalar}
\eea
where $\Lambda:=\frac{\w^2 r^2}{c_s}$, and we have used (\ref{current conservation}).
The action can now be concisely decoupled to a scalar part and a vector part (omitting hereafter the field indices $(L \w)$ for brevity):
\bea
S_{EM} = \frac{1}{2}\int\frac{d \w}{2\pi}\sum_L \,\,\,\,\,\,\,\,& \!\!\!\!\!\!\!\!\!\!\!\!\!\!\!\!\!\!\!\!\!\!\!\!\!\!\!\!\!\!\!\!\!\!\!\!\!\!\!\!\!\!\!\!\!\!\!\!\!\!\!\!\!\!\!\!\!\!\!\!\!\!\!\!\!\!\!\!\!\!\!\!\!\!\!\!\!\!\!\!\!\!\!\!\!\!\!\!\!\!\!\!\!\!\!\!\!\!\!\!\!\!\!\!\!\!\!\!\!\!\!\!\!\!\!\!\!\!\!\!\!\!\!\!\!\!\!\!\!\!\!\!\!\!\!\!\!\!\!\!\!\!\!\!\!\!\!\!\!\!\!\!\!\!
\, \, \, \, \, \, \left[ S^{L\w}_S + S^{L\w}_V \right],
\label{SEM}
\\
S^{L\w}_S = \int{dr} \frac{r^{2\ell+2}\factell }{(2\ell+1)!!}& \!\!\!\!\!\!\!\!\!\!\!\!\!\!\!\!\!\!\!\!\!\!\!\!\!\!\!\!\!\!\!\!\!\!\!
\left[\frac{1}{1-\Lambda} \left| \frac{1}{r^\ell} (r^\ell \tilde{A}_S)' \right|^2 +\frac{c_s}{r^2} \left| \tilde{A}_S \right|^2 +4\pi (\tilde{A}_S \rho^{S*}_{L\w} +c.c.) \right],
\label{SS}
\\
S^{L\w}_V = \int{dr} \frac{r^{2\ell+2}\factell }{(2\ell+1)!!}& c_s
\left[\left( \frac{\w^2}{r^2} - \frac{c_s}{r^4} \right) \left| A_V \right|^2 -\left| \frac{1}{r^{\ell+1}} (r^\ell \tilde{A}_V)' \right|^2 -4\pi (\tilde{A}_V \rho^{V*}_{L\w} +c.c.) \right],\,\,\,\,\,\,\,\,\,\,
\label{SV}
\eea
and we treat them separately. We also note here that for $\ell=0$ we have $\rho^S_{L\w}=0$ (see eq. (\ref{current conservation}),(\ref{EM source scalar})) as well as $S_V=0$, and so only $\ell\geq1$ need be considered.

\subsubsection*{The scalar part of the EM action}
\label{subsubsection:EM scalar}

We derive the equation of motion for the scalar action from $S^{L\w}_S$ (\ref{SS}) by treating $(r^\ell\tilde{A}_S)$ as the field, and finding equations for its conjugate momentum $\left(\frac{r^\ell \Pi_S}{(2\ell+1)!!}\right)$,
\bea
\left(\frac{\factell  r^\ell\Pi_S}{(2\ell+1)!!}\right):=\frac{\partial L}{\partial(r^\ell \tilde{A}_S^{*})'}&=&\frac{r^2\factell (r^\ell\tilde{A}_S)'}{(2\ell+1)!!(1-\Lambda)}, \label{PI S1}\\
\left(\frac{\factell  r^\ell\Pi_S}{(2\ell+1)!!}\right)':=\frac{\partial L}{\partial(r^\ell\tilde{A}_S^{*})}&=&\frac{\ell(\ell+1)\factell }{(2\ell+1)!!}(r^\ell\tilde{A}_S)+\frac{4\pi r^{\ell+2}\rho^S_{L\w}}{(2\ell+1)!!}.
\label{PI S2}
\eea
Differentiating (\ref{PI S2}) w.r.t $r$, substituting (\ref{PI S1}), and renaming the field $\PhiS$ and source term $\rho^{\PhiS}_{L\w}$ (recalling (\ref{EM inverse sources}),(\ref{EM source scalar})) as
\bea
\PhiS&=&\frac{\Pi_S}{\ell r}, \\
\rho^{\PhiS}_{L\w}&=&\frac{4\pi r^{\ell+1}(r^{\ell+2} \rho^S_{L\w})'\factell }{(\ell+1) (2\ell+1)!!}
=\frac{1}{\ell+1}\int\!\! d\Omega\, r x_L \left[
- r^2 \rho_\w(\vec{r})
+\frac{i}{\w} \left( r^2 \frac{\Lambda}{\Lambda-1} \vec{J}_w(\vec{r})\cdot \vec{n} \right)'
\right]'\,\, , \nonumber
\label{EM source scalar Phi}
\eea
we find the equation
\bea
0=\frac{r^{2\ell+2}\factell }{(2\ell+1)!!}\frac{\ell}{\ell+1}\left(\w^2+\d_r^2+\frac{2(\ell+1)}{r}\d_r \right) {\PhiS}-\rho^{\PhiS}_{L \w}.
\label{EOM PhiS}
\eea
This equation is of the same form as (\ref{EOM Phi}), up to replacing $G$ by the factor
\bea
\REone&=&\frac{\ell+1}{\ell},
\label{REone}
\eea
thus from its solution we find a propagator similar to (\ref{Phi propagator scalar}),
\bea
G^{\PhiS}_{ret}(r',r)
=\frac{-i\w^{2\ell+1}}{\factell (2\ell+1)!!} \REone \tilde{j}_{\ell}(\w r_1)\tilde{h}^+_{\ell}(\w r_2)\delta_{LL'} \, \, \, \, ; \nonumber\\
r_1:=\text{min}\{r',r\},\,\,\,r_2:=\text{max}\{r',r\}.
\label{EM propagator scalar}
\eea

We again present the EFT Feynman rules following the steps (\ref{scalar wavefunction at radiation zone1},\ref{scalar wavefunction at radiation zone2},\ref{I Phi scalar multipoles}). In the radiation zone, the field can be written as
\bea
\PhiS^{L \w}(r) ^{EFT}= -Q^{E}_{L \w} \frac{-i\w^{2\ell+1}}{\factell(2\ell+1)!!}\REone \tilde{h}^+_{\ell}(\w r) \, \, ,
\label{EM scalar wavefunction at radiation zone1}
\eea
where $Q^{E}_{L \w}$ are the sources (fig. \ref{vertex_definition}). In the full theory the solution outside the sources is given (see \ref{EOM PhiS}) by
\bea
\PhiS^{L \w}(r) = \int\!\! dr' \rho^{\PhiS}_{L\w}(r') G^{\PhiS}_{ret}(r',r)
	= -\left[ \int\!\! dr' \tilde{j}_{\ell}(\w r') \rho^{\PhiS}_{L\w}(r') \right]  \frac{-i\w^{2\ell+1}}{\factell(2\ell+1)!!}\REone \tilde{h}^+_{\ell}(\w r), \,\,\,\,\,\,\,\,\,\,\,\,\,\,\,\,
\label{EM scalar wavefunction at radiation zone2}
\eea
and the sources can be read off and identified (using (\ref{Modified Bessel equation}, \ref{EM inverse sources}, \ref{EM source scalar Phi}) and integration by parts) to be
\bea
Q^{E}_{L \w}&=&\int\!\!{dr'} \tilde{j}_{\ell}(\w r') \rho^{\PhiS}_{L\w}(r')		\nonumber\\
&=&\frac{1}{\ell+1}\int\!{dr'} \tilde{j}_{\ell}(\w r') \int\!\! d\Omega' \, r' x'_L
	\left[- r'^2 \rho_{\w}(\vec{x} \, ')
	+\frac{i}{\w} \left( r'^2 \frac{\Lambda}{\Lambda-1} \vec{J}_\w(\vec{x} \, ')\cdot \vec{n} \, ' \right)'
	\right]'	\nonumber\\
&=& \frac{1}{\ell+1}
	\!\int\!\!{d^3x'} x'_L
	\left[\frac{1}{r'^\ell}\left(r'^{\ell+1}\tilde{j}_{\ell}(\w r')\right)' \rho_\w(\vec{x} \, ')
	-i\w \tilde{j}_{\ell}(\w r') \vec{J}_\w(\vec{x} \, ') \cdot \vec{x} \, '
	\right].
\eea
Returning to the time domain using (\ref{EM inverse sources}) we find \emph{the electric type radiation source multipoles} (compare (\ref{I Phi scalar multipoles}))
\bea
Q^L_E&=&\frac{1}{\ell+1}\int\!\!{d^3x} x^L_{TF}
	\left[\frac{1}{r^{\ell}}\left(r^{\ell+1}\tilde{j}_{\ell}(ir \d_t)\right)' \rho(\vec{x})
	-\tilde{j}_{\ell}(ir \d_t)\, \d_t \vec{J}(\vec{x} \, ') \cdot \vec{x} \right],
\label{I EM scalar multipoles}
\\
\hat{Q}_E^L&=&\frac{\delta Q_E^L}{\delta J^{i}}\hat{J}^{i}=\frac{\delta Q_E^L}{\delta x^{i}}\hat{x}^{i}=\frac{\d Q_E^L}{\d x^{i}}\hat{x}^{i} + \frac{\d Q_E^L}{\d v^{i}}\hat{v}^{i} + \frac{\d Q_E^L}{\d a^{i}}\hat{a}^{i} + \cdots.
\label{I EM scalar multipoles hat}
\eea

We note here that by expanding $\tilde{j}_\ell$ with (\ref{Bessel J series2}) this PN multipole expansion reproduces eq.(47) of \cite{RossMultipoles} (after using current conservation, eq. (49) there).

\subsubsection*{The vector part of the EM action}
\label{subsubsection:EM vector}

For the vector sector of the action, we rewrite (\ref{SV}) in 4d in a form similar to (\ref{scalar action spherical}),
\bea
S^{L\w}_V = \frac{1}{\RMone}\int{dr}
\left[\frac{r^{2\ell+2}\factell }{(2\ell+1)!!}\PhiV^* \left(\w^2+\d_r^2+\frac{2(\ell+1)}{r}\d_r \right) \PhiV
-\left(\rho^{\PhiV}_{L\w} \PhiV^* +c.c.\right) \right],\,\,\,\,\,\,\,\,\,\,\,\,
\label{S_V in Phi}
\eea
where we have defined the prefactor, field and source term (recalling (\ref{EM inverse sources})) as
\bea
\RMone&=&\frac{\ell}{\ell+1},		\nonumber\\
\PhiV&=&\frac{\ell A_V}{r},		\nonumber\\
\rho^{\PhiV}_{L\w}&=&\frac{4\pi (\ell+1) r^{2\ell+3}\factell }{(2\ell+1)!!}\rho^V_{L\w}(r)
	=\frac{1}{\ell} \, r^3\!\!\int\!\! \vec{J}_w(\vec{r})\!\cdot\!\left[\vec{r} \!\times\! \vec\nabla x_L\right] d\Omega .
\label{EM source vector Phi}
\eea
Again this action (\ref{S_V in Phi}) is identical to (\ref{scalar action spherical}) up to the global prefactor of $\RMone$, with a source similar to (\ref{scalar action source}). Thus the propagator is (compare \ref{Phi propagator scalar}, \ref{EM propagator scalar})
\bea
G^{\PhiV}_{ret}(r',r)
=\frac{-i\w^{2\ell+1}}{\factell (2\ell+1)!!} \RMone \tilde{j}_{\ell}(\w r_1)\tilde{h}^+_{\ell}(\w r_2)\delta_{LL'}~;\nonumber\\
r_1:=\text{min}\{r',r\},\,\,\,r_2:=\text{max}\{r',r\}.
\label{EM propagator vector}
\eea
We find these sources $Q^{M}_{L \w}$ by again matching $\Phi^{\PhiV}_{L \w}(r)$ for large $r$ and from the diagrammatic representation (in analogy with \ref{scalar wavefunction at radiation zone1},\ref{scalar wavefunction at radiation zone2},\ref{I Phi scalar multipoles},\ref{EM scalar wavefunction at radiation zone1},\ref{EM scalar wavefunction at radiation zone2},\ref{I EM scalar multipoles}), to find
\bea
Q^{M}_{L \w}&=&\int\!\!{d^3x} \tilde{j}_{\ell}(\w r) \left(\vec{r}\!\times\! \vec{J}_w(\vec{r})\right)\!_{(k_\ell} x_{L-1)},
\eea
where we have used (\ref{EM inverse sources},\ref{EM source vector}). Returning to the time domain we find \emph{the magnetic radiation source multipoles} (compare (\ref{I Phi scalar multipoles}, \ref{I EM scalar multipoles})),
\bea
Q^L_M&=&\int\!\!{d^3x} \tilde{j}_{\ell}(ir\d_t) \left[(\vec{r}\times \vec{J}(\vec{r}))^{k_\ell} x^{L-1}\right]^{STF},
\label{J EM vector multipoles}
\\
\hat{Q}_M^L&=&\frac{\delta Q_M^L}{\delta J^{i}}\hat{J}^{i}=\frac{\delta Q_M^L}{\delta x^{i}}\hat{x}^{i}=\frac{\d Q_M^L}{\d x^{i}}\hat{x}^{i} + \frac{\d Q_M^L}{\d v^{i}}\hat{v}^{i} + \frac{\d Q_M^L}{\d a^{i}}\hat{a}^{i} + \cdots.
\label{J EM vector multipoles hat}
\eea

Presenting $\tilde{j}_\ell$ as a series expansion using (\ref{Bessel J series2}), these coincide with eq.(48) of \cite{RossMultipoles}.

\subsubsection{Outgoing EM radiation and the RR effective action}

Both polarizations of outgoing EM radiation can now be found diagrammatically (compare \ref{Radiation Phi using feynman}) as
\begin{align}
\label{Radiation EM using feynman scalar}
\PhiS^{L\w}(r)=&
\parbox{20mm}
 {\includegraphics[scale=0.5]{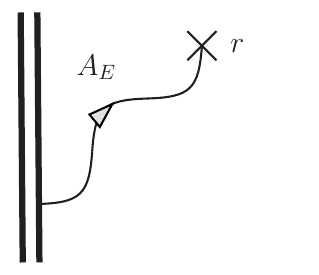}}
= 	-Q_E^{L'\w}G^{\PhiS}_{ret}(0,r)
=\REone (-iw)^\ell \frac{Q_E^{L\w}}{r^{\ell}} \frac{e^{iwr}}{r}\,\, , \,\,\,\,\,\,
\nonumber\\ 
\PhiV^{L\w}(r)=&
\parbox{20mm}
 {\includegraphics[scale=0.5]{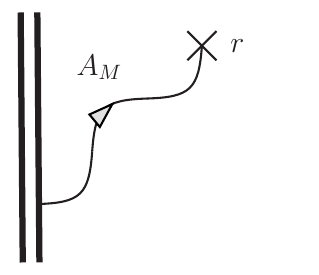}}
= 	-Q_M^{L'\w}G^{\PhiV}_{ret}(0,r)
=\RMone (-iw)^\ell \frac{Q_M^{L\w}}{r^{\ell}} \frac{e^{iwr}}{r}\,\, . \,\,\,\,\,\,
\end{align}
In the time domain, we find for either polarization $\eps$
\bea
A_\eps(\vec{r},t)=\frac{1}{r} R^\eps_1\, n_L \d_t^\ell Q^L_\eps(t-r) ~.
\label{radiation A}
\eea

The EM double field effective action can be written using our Feynman rules as a sum of the scalar and vector action diagrams (again at $r=r'=0$, and without $\tilde{y}_\ell$), finding
\begin{align}
\label{S EM multipoles}
\hS_{EM}=&\,\,\,\,\,\,\,\,\,\,\,\,\,\,\,\,\,\,\,\,\,\,\,\,\,\,\,\,\,\,\,\,\,\,\,\,\,\,\,\,\,\,
\parbox{20mm}
 {\includegraphics[scale=0.5]{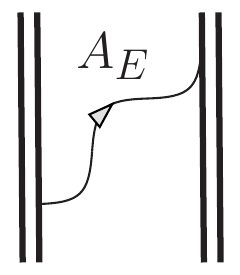}}
\,\,\,\,\,\,\,\,\,\,\,\,\,\,\,\,\,\,\,\,\,\,\,\,\,\,\,\,\,\,\,\,\,+\,\,\,\,\,\,\,\,\,\,\,\,\,\,\,\,\,\,\,\,\,\,\,\,\,\,\,\,
\parbox{20mm}
 {\includegraphics[scale=0.5]{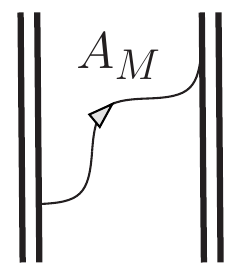}}
\nonumber\\
=& \frac{1}{2}\int\!\frac{d \omega}{2\pi}\sum_{L,L'}
	\left[
		\left(-Q^{E}_{L\w}\right)
		G^{\PhiS}_{ret}(0,0)
		\left(-\hat{Q}^{E*}_{L'\w}\right)
	+
		\left(-Q^{M}_{L\w}\right)
		G^{\PhiV}_{ret}(0,0)
		\left(-\hat{Q}^{M*}_{L'\w}\right)
	\right]
	+c.c.\,\,\,\,\,\,\,\,\,\,
	\nonumber\\
=& \!\int\!\frac{d \omega}{2\pi}\sum_L
	\frac{-i\w^{2\ell+1}}{2\factell (2\ell+1)!!}
	\left[
		\REone	Q^{L\w}_E \hat{Q}^{E*}_{L\w}
	+
		\RMone 	Q^{L\w}_M \hat{Q}^{M*}_{L\w}
	\right]
	+c.c.
	\nonumber\\
=&\int\!\!{dt}\!\sum_{L}\frac{(-)^{\ell+1}}{\factell (2\ell+1)!!}
	\left[
		\frac{\ell+1}{\ell} \hat{Q}^E_L \d_t^{2\ell+1}Q_E^L
	+
		\frac{\ell}{\ell+1} \hat{Q}^M_L \d_t^{2\ell+1}Q_M^L
	\right] \, \, ,
\end{align}
where $Q^L_E,\hat{Q}^L_E,Q^L_M,\hat{Q}^L_M$ were given by (\ref{I EM scalar multipoles},\ref{I EM scalar multipoles hat},\ref{J EM vector multipoles},\ref{J EM vector multipoles hat}).

\subsubsection{Applications and tests}	

{\bf Perturbative expansion of the RR force and comparison with ALD}. For the radiation reaction force on a single accelerating electric charge we have the Abraham-Lorentz-Dirac (ALD) formula \cite{Dirac},
\begin{equation}
\label{EM ALD}
F_{ALD}^\mu \equiv \frac{d p^\mu}{d\tau} = \frac{2}{3} q^2\, \( \frac{d^3 x^\mu}{d\tau^3} - \frac{d^3 x^\nu}{d\tau^3} \frac{dx_\nu}{d\tau}\,  \frac{dx^\mu}{d\tau}\).
\end{equation}
This expression is the same as in the scalar (\ref{F ALD}) case up to a prefactor. Expanding in PN orders, the leading order and next-to-leading-order are
\bea
\vec{F}_{ALDLO}= \frac{2}{3}q^2\dot{\vec{a}},
\label{F LO EM ALD}
\eea
\bea
\vec{F}_{ALDNLO} = q^2\left[ \frac{2}{3}v^{2}\dot{\vec{a}} + 2(\vec{v} \cdot \vec{a})\vec{a} + \frac{2}{3}(\vec{v} \cdot \dot{\vec{a}})\vec{v}\right].
\label{F NLO EM ALD}
\eea
 We also successfully compared this expansion with a method of local odd propagation described in Appendix \ref{app:ALD local propagator}.\\
For our gauge-invariant RR force calculation on a point charge $q$ along a path $\vec{x}(t)$, we rewrite the action
\bea
\hS_{EM}=\int\!dt\,\hat{L}_{EM}, \,\,\,\,\,\,\,\,\,\,\,\,\,\,\hat{L}_{EM}=\hat{L}_{EM}^S+\hat{L}_{EM}^V,
\eea
as a PN series expansion. With (\ref{I EM scalar multipoles},\ref{J EM vector multipoles},\ref{S EM multipoles}) we find
\bea
\hat{L}_{EM}^S=\,q^2\sum_{L}\frac{(-)^{\ell+1}(2\ell+1)!!}{\ell(\ell+1)\factell } 	\nonumber\\
	 &\!\!\!\!\!\!\!\!\!\!\!\!\!\!\!\!\!\!\!\!\!\!\!\!\!\!\!\!\!\!\!\!\!\!\!\!\!\!\!\!\!\!\!\!\!\!\!\!\!\!\!\!\!\!\!\!\!\!\!\!
	\cdot\,
	\sum_{\hat{p}=0}^\infty  \frac{\d_t^{2\hat{p}}}{(2\hat{p})!!(2\ell+2\hat{p}+1)!!}
		 \frac{\delta}{\delta x^{i}}\left[(2\hat{p}+\ell+1)r^{2\hat{p}}x_L
		-\d_t\left(r^{2\hat{p}}x_L \vec{v} \cdot \vec{r} \right) \right]\hat{x}^{i}  \,\,\,\,\,\,\,\,\,\,\,\,\,\,\,\,\,\,		 \nonumber\\
	&\!\!\!\!\!\!\!\!\!\!\!\!\!\!\!\!\!\!\!\!\!\!\!\!\!\!\!\!\!\!\!\!\!\!\!\!\!\!\!\!\!\!\!\!\!\!\!\!\!\!
	\cdot\,
	\d_t^{2\ell+1}\!
	\sum_{p=0}^\infty \frac{ \d_t^{2p}}{(2p)!!(2\ell+2p+1)!!}
		\left[(2p+\ell+1)r^{2p} x^L
		-\d_t\left(r^{2p} x^L \vec{v} \cdot \vec{r} \right) \right]^{STF} ,
		\,\,\,\,\,\,\,\,\,\,\,\,\,\,\,\,
	\nonumber\\
\eea
for the scalar part and
\bea
\hat{L}_{EM}^V=\,q^2\sum_{L}\frac{(-)^{\ell+1} \ell (2\ell+1)!!}{(\ell+1)\factell }
	\sum_{\hat{p}=0}^\infty &	 \frac{\d_t^{2\hat{p}}}{(2\hat{p})!!(2\ell+2\hat{p}+1)!!}
	\frac{\delta}{\delta x^{i}}\left[r^{2\hat{p}}(\vec{r} \times \vec{v})^{k_\ell} x^{L-1}\right]\hat{x}^{i}
	\,\,\,\,\,\,\,\,\,\,\,\,\,\,\,\,\,\,\,\,\,\,\,\,\,\,\,\,\,	\nonumber\\
	\cdot\,
	\d_t^{2\ell+1}
	\sum_{p=0}^\infty &	 \frac{\d_t^{2p}}{(2p)!!(2\ell+2p+1)!!}
		\left[r^{2p}(\vec{r}\times \vec{v})^{k_\ell} x^{L-1}\right]^{STF},
		\,\,\,\,\,\,\,\,\,\,\,\,\,\,\,\,\,\,\,\,\,\,\,\,\,\,\,\,
\label{EM lagrangian}
\eea
for the vector part. Similarly to the scalar RR calculation, we move the $2p$ (or $2p+1$) time derivatives from the $\hat{x}^L$ multipoles to the $x^L$ multipoles by partial integration, and use the Euler-Lagrange equation (\ref{scalar EL}) for $\hat{x}^j$.
We thus find the leading self-force, arising from the electric dipole term ($\ell=1,p=\hat{p}=0$, sources $\rho,\hat{\rho}$ as recorded in table \ref{table:EM leading scalar}), to be as expected from the ALD result  (\ref{F LO EM ALD}).

\begin{table}[h!]
  \centering \caption{Leading order contribution to the EM self-force (only electric $\eps=+$)}
\begin{center}
\begin{tabular}{cccccccc}
  \hline
  $\ell$ $p$ $\hat{p}$ & src & $\hat{L}/q^2$ & $F^j/q^2$ \\  \hline
  1 0 0 & $\rho$ $\hat\rho$
			& $\frac{2}{3} \hat{x}^{i}\d_t^3x_{i}$
			& $\frac{2}{3}\dot{a}^j$\\  \hline
  \label{table:EM leading scalar}
\end{tabular}
\end{center}
\end{table}

The next-to-leading-order includes 5 contributions to the scalar sector, summarized in table \ref{table:EM next to leading scalar}, as well as the leading vector contribution (table \ref{table:EM next to leading vector}). Their sum is identical with the ALD result (\ref{F NLO EM ALD}).
 Note that while this term is the same as in the scalar case (\ref{F NLO ALD}) up to a constant pre-factor the individual terms in the EM case (table \ref{table:EM next to leading scalar}) are completely different from the scalar case (table \ref{table:scalar multipoles next to leading scalar}).

\begin{table}[h!]
  \centering \caption{Next-to-Leading order contribution to the EM self-force, scalar (electric $\eps=+$) sector}
\begin{center}
\begin{tabular}{cccccccc}
  \hline
  $\ell$ $p$ $\hat{p}$ & src & $\hat{L}/q^2$ & $F^j/q^2$ \\  \hline
  2 0 0 & $\rho$ $\hat\rho$
	& $-\frac{1}{20}\hat{x}^{j} \frac{\delta}{\delta x^{j}}[x_i x_k]\d_t^5[x^i x^k-\frac{1}{3} x^2 \delta^{ik}]$
	& $-\frac{1}{10}[x_i \d_t^5(x^i x^j)-\frac{1}{3}x^j \d_t^5 x^2]$\\  \hline
  1 1 0 & $\rho$ $\hat\rho$
	& $\frac{2}{15}\hat{x}^{j} \frac{\delta}{\delta x^{j}}[x^2 x^i] \d_t^5 x_i$
	& $\frac{2}{15}[x^2 \d_t^5 x^j+2x^j x_i \d_t^5 x^i]$\\  \hline
  1 0 1 & $\rho$ $\hat\rho$
	& $\frac{2}{15}\hat{x}_i \d_t^5(x^i x^2)$
	& $\frac{2}{15}\d_t^5(x^2 x^j)$\\  \hline
  1 0 0 & $j_r$ $\hat\rho$
	& $\frac{1}{3}\hat{x}^{j} \frac{\delta}{\delta x^{j}}[v_k x^k x^i]\d_t^4x_i$
	& $\frac{1}{3}[x_i v^i \d_t^4x^j + v^j x_i \d_t^4  x^i
		-\frac{d}{dt}(x^j x^i \d_t^4x_i)]$\\  \hline
  1 0 0 & $\rho$ $\hat{j}_r$
	& $-\frac{1}{3} \hat{x}^i \d_t^4[v_k x^k x_i]$
	& $-\frac{1}{3}\d_t^4(x^j x_i v^i)$\\  \hline
  \label{table:EM next to leading scalar}
\end{tabular}
\end{center}
\end{table}

\begin{table}[h!]
  \centering \caption{Next-to-Leading order contribution, from vector (magnetic $\eps=-$) sector}
\begin{center}
\begin{tabular}{cccccc}
  \hline
  $\ell$ $p$ $\hat{p}$ & $\hat{L}/q^2$ & $F^j/q^2$ \\  \hline
  1 0 0 & $\frac{1}{6}\hat{x}^{i} \frac{\delta}{\delta x^{i}}[\vec{r}\times\vec{v}] \cdot\d_t^3(\vec{r}\times\vec{v})$
		& $[\frac{2}{3}v^j(\vec{v}\cdot \dot{\vec{a}})
			-\frac{2}{3}\dot{a}^j v^2
			-\frac{5}{6}\ddot{a}^j(\vec{x}\cdot\vec{v})$\\
	   & &  $+\frac{1}{3}x^j(\vec{v}\cdot\ddot{\vec{a}})
			+\frac{1}{2}v^j(\vec{x}\cdot\ddot{\vec{a}})
			-\frac{1}{3}\dot{a}^j(\vec{x}\cdot\vec{a})$\\
	 & & $+\frac{1}{3}a^j(\vec{x}\cdot \dot{\vec{a}})
			-\frac{1}{6} x^2\, \d_t^3a^j
			+\frac{1}{6}x^j(\vec{x}\cdot\d_t^3\vec{a})]$\\  \hline
  \label{table:EM next to leading vector}
\end{tabular}
\end{center}
\end{table}

{\bf Dissipated power}. Similarly to (\ref{radiated energy scalar definition}), we calculate the power of the RR force on the accelerating charge, now using (\ref{S EM multipoles}):
\bea
P_{RR} &=&-\vec{v}\cdot \vec{F}=-\frac{d x^{i}}{dt} \left.\frac{\delta \hat{L}}{\delta \hat{x}^{i}}\right|_{\hat{\vec{x}} \to \vec{x}}\nonumber\\
&=&\sum_{L}\frac{(-)^{\ell}}{\factell (2\ell+1)!!}
	\left.\left[
		\frac{\ell+1}{\ell} \frac{\delta Q_E^L}{\delta x^{i}}\frac{d x^{i}}{dt} \d_t^{2\ell+1}Q^E_L
		+ \frac{\ell}{\ell+1} \frac{\delta Q_M^L}{\delta x^{i}}\frac{d x^{i}}{dt} \d_t^{2\ell+1}Q^M_L
	\right] \right|_{\hat{\vec{x}} \to \vec{x}}\!\!\!	.~
\label{radiated energy EM definition}
\eea
The time-averaged power is found using
\bea
\int\!\!{dt} \frac{d x^{i}}{dt} \frac{\del Q_E^L}{\del x^{i}} = \int\!\!{dt} \frac{d Q_E^L}{dt}
	\,\,\,\,,\,\,\,\,
\int\!\!{dt} \frac{d x^{i}}{dt} \frac{\delta Q_M^L}{\delta x^{i}} = \int\!\!{dt} \frac{d Q_M^L}{dt} \, \, ,
\label{EM EL as full derivative}
\eea
followed by $\ell$ integrations by parts, to be
\bea
\label{radiated energy EM multipoles}
<\!\!P_{RR}\!\!>&=& \sum_{L}\frac{1}{\factell (2\ell+1)!!}
	\left<
		\REone (\d_t^{\ell+1}Q_E^L)^2
		+ \RMone  (\d_t^{\ell+1}Q_M^L)^2
	\right>	\nonumber\\
&\!\!\!\!\!\!\!\!\!\!\!\!\!\!\!\!\!\!\!\!\!\!\!\!\!\!\!=&\!\!\!\!\!\!\!\!\!\!\!\!\!\!\!\!
	\!\sum_{L}\frac{(\ell+1)}{\ell\factell (2\ell+1)!!}
		\left<\left(\frac{d^{\ell+1}}{dt^{\ell+1}}Q_E^L	\right)^2 \right>
	+ \sum_{L}\frac{\ell}{(\ell+1)(2\ell+1)!!}
		\left<\left(\frac{d^{\ell+1}}{dt^{\ell+1}}Q_M^L	\right)^2 \right>	\,\,\,\,\,\,\,\,\,\,\,\,\,\,\,\, \\
&\!\!\!\!\!\!\!\!\!\!\!\!\!\!\!\!\!\!\!\!\!\!\!\!\!\!\!=&\!\!\!\!\!\!\!\!\!\!\!\!\!\!\!\!
	\!\sum_{\ell}\!\frac{(\ell+1)}{\ell\, \ell! (2\ell+1)!!}\!
		\left<\left(\frac{d^{\ell+1}}{dt^{\ell+1}}Q^{k_1 k_2 \cdots k_\ell}_{E,STF}	\right)^{\!\!\!2} \right>
	\!\!+\!\! \sum_{\ell}\!\frac{\ell}{(\ell+1)!(2\ell+1)!!}\!
		\left<\left(\frac{d^{\ell+1}}{dt^{\ell+1}}Q^{k_1 k_2 \cdots k_\ell}_{M,STF}	\right)^{\!\!\!2} \right>	\nonumber \\
&\!\!\!\!\!\!\!\!\!\!\!\!\!\!\!\!\!\!\!\!\!\!\!\!\!\!\!=&\!\!\!\!\!\!\!\!\!\!\!\!\!\!\!\!
P_{rad} ~. \nonumber
\eea
We recognize this result as Ross' eq.(52) \cite{RossMultipoles} (with a $4\pi$ normalization factor, re-introducing the $\frac{1}{\ell!}$ factor for comparison, where the notation is non-multi-index, see Appendix \ref{app:Multi-index summation convention}).

\subsection{Gravity: non-linearity}
\label{gravitational RR force}

As in the electromagnetic case, linearized gravitational perturbations in four dimensions are described by two gauge invariant master-functions which describe even-parity (or scalar) and odd-parity (or vector) perturbations  \footnote{In $d>4$ there is an additional tensor sector which is absent in 4d, see for example \cite{AsninKol}.}.
  In this section we will not derive explicitly the 1D reduced action, rather we will use a convenient shortcut: we use the 1D reduced equations of motion derived by Martel and Poisson in (\cite{Martel:2005ir}) valid for gravitational perturbations of the Schwarzschild geometry. Taking the BH mass to zero ($M = 0$) gives the correct equations of motion for gravitational perturbations of \emph{flat} spacetimes with sources. In this section we will use mostly plus signature ($- + + +$) to conform with the notation of (\cite{Martel:2005ir}).

The action for general relativity with a linearized source term is
\bea
S = \frac{1}{16 \pi G} \int \sqrt{-g} R d^4 x \, \, - \, \, \frac{1}{2} \int h_{\mu \nu} T^{\mu \nu} d^4 x \, \, .
\label{GR action}
\eea
The energy momentum tensor obeys
\bea
\nabla_{\mu} T^{\mu \nu} = 0 \, \, .
\label{energy momentum conservation}
\eea

We write the full metric as a perturbation around Minkowski spacetime
\bea
g_{\mu \nu} = \eta_{\mu \nu} + h_{\mu \nu} \, \, ,
\eea
where $\eta_{\mu \nu}$ is the flat space metric. Next, we work in spherical coordinates and decompose the perturbation into tensor spherical harmonics according to
\begin{eqnarray}
h_{ab} &=& \sum_{L} \int \frac{d \omega}{2 \pi} \, \, H^{L \omega}_{ab} x_L e^{- i \omega t} \, \, , \nonumber \\
h_{a \Omega} &=& \sum_{L} \int \frac{d \omega}{2 \pi} \left[ H^{L \omega}_{(E) \, \, a} \partial_{\Omega} x_L + H^{L \omega}_{(M) \, \, a} \epsilon_{\Omega \Theta} \partial^{\Theta} x_L \right] e^{- i \omega t} \, \, , \nonumber \\
h_{\Omega \Theta} &=& r^2 \sum_{L} \int \frac{d \omega}{2 \pi} \left[ H^{L \omega}_{(E)} \tilde{g}_{\Omega \Theta} x_L \right. \nonumber \\
&+& \left. \tilde{H}^{L \omega}_{(E)} \left( D_{\Omega} D_{\Theta} + \frac{1}{2} \ell(\ell+1)\, \tilde{g}_{\Omega \Theta} \right)  x_L + H^{L \omega}_{(M)} \epsilon_{\Omega \Omega'} \epsilon_{\Theta \Theta'} D^{\Omega'} D^{\Theta'} x_L \right] e^{- i \omega t} \, \, , \nonumber \\
\label{h decomposition}
\end{eqnarray}
where lower case Latin indices stand for non-sphere coordinates $a = (t,r)$ and upper case Greek indices stand for coordinates on the sphere $\Omega = (\theta,\phi)$. $\tilde{g}_{\Omega \Theta}$ and $\epsilon_{\Omega \Theta}$ are the metric and Levi-Civita tensor, respectively, on the $2$-sphere. We decompose the energy-momentum tensor in a similar way
\begin{eqnarray}
T^{ab} &=& \sum_{L} \int \frac{d \omega}{2 \pi} \, \, T^{L \omega}_{ab} x_L e^{- i \omega t} \, \, , \nonumber \\
T^{a \Omega} &=& \sum_{L} \int \frac{d \omega}{2 \pi} \left[ T^{L \omega}_{(E) \, \, a} \partial^{\Omega} x_L + T^{L \omega}_{(M) \, \, a} \epsilon^{\Omega \Theta} \partial_{\Theta} x_L \right] e^{- i \omega t} \, \, , \nonumber \\
T^{\Omega \Theta} &=& \sum_{L} \int \frac{d \omega}{2 \pi} \left[ T^{L \omega}_{(E)} \tilde{g}^{\Omega \Theta} x_L \right. \nonumber \\
&+& \left. \tilde{T}^{L \omega}_{(E)} \left( D^{\Omega} D^{\Theta} + \frac{1}{2} \ell(\ell+1) \tilde{g}^{\Omega \Theta} \right) x_L + T^{L \omega}_{(M)} \epsilon^{\Omega \Omega'} \epsilon^{\Theta \Theta'} D_{\Omega'} D_{\Theta'} x_L \right] e^{- i \omega t} \, \, . \nonumber \\
\label{T decomposition}
\end{eqnarray}

In terms of these variables, the $1D$-reduced action describing linearized GR with material sources is
\begin{align}
S_{(E/M)} &= \frac{1}{2} \int \frac{d \omega}{2 \pi} \sum_{L} \int \, \, dr \left[ \frac{r^{2 \ell + 2} \, (R^{\epsilon}_2)^{-1}}{(2 \ell+1)!!}   h_{(E/M)}^{*} \left( \omega^2 + \partial^{2}_{r} + \frac{2(\ell+1)}{r} \partial_{r} \right) h_{(E/M)} \right. \nonumber \\
&- \left. \left( h_{(E/M)}^{*} \mathcal{T}_{(E/M)} + c.c. \right) \right] \, \, ,
\label{1DS}
\end{align}
where
\bea
R^{\epsilon}_2 := \frac{\ell+2}{\ell-1} \left(\frac{\ell+1}{\ell}\right)^{\epsilon} \, \, , \nonumber \\
\label{R2def}
\eea
and we will suppress at times the $(L \omega)$ indices or part of them. The even-parity master-function is given by
\bea
h^{L}_{(E)} := \frac{1}{\ell} \biggl[ \frac{\ell+2}{2}\tilde{K}^{L}
+ \frac{1}{(\ell-1)} \bigl( \tilde{H}^{L}_{rr}
- r \left( \partial_r + \frac{\ell}{r} \right) \tilde{K}^{L} \bigr) \biggr] \, \, , \nonumber \\
\label{even parity master-function}
\eea
where
\bea
\tilde{K}^{L} &:=& H^{L}_{(E)}+\frac{1}{2} \ell (\ell + 1) \tilde{H}^{L}_{(E)} - \frac{2}{r} \left( H^{L}_{(E) \, r} - \frac{1}{2} r^2 \left( \partial_r + \frac{\ell}{r} \right) \tilde{H}^{L}_{(E)} \right) \, \, , \nonumber \\
\tilde{H}^{L}_{rr} &:=& H^{L}_{rr} - 2 \left( \partial_r + \frac{\ell}{r} \right) \left( H^{L}_{(E) \, r} - \frac{1}{2} r^2 \left( \partial_r + \frac{\ell}{r} \right) \tilde{H}^{L}_{(E)} \right) \, \, .
\label{even parity master-function definitions}
\eea
The even parity source term is given by
\bea
\mathcal{T}_{(E)} &:=& \frac{2 \pi \ell}{(\ell+2)} \frac{r^{2 \ell + 2}}{(2 \ell+1)!!} \left\{ 8r T_{(E)}^{r} - 2r^2 (\ell-1)(\ell+2) \tilde{T}_{(E)} \right. \nonumber \\
&+& \left.  \frac{2}{\ell(\ell+1)}\left[ -2r \left( \partial_r + \frac{\ell}{r} \right) \left( T^{rr} - T^{tt} \right) + 4r^2 T_{(E)} + (\ell(\ell+1)-4)\left( T^{rr} - T^{tt} \right) \right] \right\} \, \, . \nonumber \\
\label{even parity source term}
\eea
The odd-parity master-function is given by
\bea
h^{L}_{(M)} := \frac{\ell}{2 (\ell-1)} \left[ \left( \partial_r + \frac{\ell}{r} \right) \tilde{H}_t - \partial_{t} \tilde{H}_r -\frac{2}{r} \tilde{H}_r \right] \, \, ,
\label{odd parity master-function}
\eea
where
\bea
\tilde{H}^{L}_r &=& H^{L}_{(M) \, \, r} - \frac{1}{2} \left( \partial_r + \frac{\ell}{r} \right) H^{\ell m}_{(M)} + \frac{1}{r} H^{L}_{(M)} \, \, , \nonumber \\
\tilde{H}^{L}_t &=& H^{L}_{(M) \, \, t} - \frac{1}{2} \partial_t H^{L}_{(M)} \, \, .
\label{odd parity master-function definitions}
\eea
The odd parity source term is given by
\bea
\mathcal{T}_{(M)} := \frac{8 \pi (\ell+1)}{(\ell + 2)} \frac{r^{2 \ell + 2}}{(2 \ell+1)!!} \left[ i \omega r^2 T_{(M)}^{r} + \left( \partial_r + \frac{\ell}{r} \right)\left( r^2 T_{(M)}^{t} \right) \right] \, \, .
\label{odd parity source term}
\eea
The action (\ref{1DS}) is identical to (\ref{scalar action spherical}) apart for a factor of $(R^{\epsilon}_2)^{-1}$ instead of $G$ in front of the part which is quadratic in $h_{(E/M)}$. The Feynman rules are derived as in the scalar case. The propagator is identical (up to a factor of $R^{\epsilon}_2$) to the scalar propagator, and the vertices, as in subsection \ref{section:Scalar SF}, are read off by matching the full theory with the EFT in the radiation zone. After reinserting cartesian components of the stress-energy tensor in place of the components defined in (\ref{T decomposition}) the multipoles read
\begin{align}
Q^L_{(E)} & =\! \frac{1}{(\ell+1)(\ell+2)}\!\!\int\!\! d^{3}x \, x^{L}\! \left[  r^{-\ell} \left( r^{\ell+2} \, \tilde{j}_{\ell} (r i \partial_t) \right)'' \left( T^{0 0} + T^{a a} \right) \right.\notag \\
&\,\,\,\,\,\,\, \left.
- 4 r^{-\ell-1} \!\left( r^{\ell+2} \, \tilde{j}_{\ell} (r i \partial_t) \right)' \!\!\partial_t T^{0 a} x^{a}
+ 2 \, \tilde{j}_{\ell} (r i \partial_t) \, \partial_t^2 T^{a b} x^{a} x^{b}
+ r^{2} \, \tilde{j}_{\ell} (r i \partial_t) \, \partial_t^2 \left( T^{0 0} - T^{a a} \right) \! \right] \notag \\
& = \sum_{p=0}^\infty \frac{(2 \ell + 1)!!}{(2 p)!! (2 \ell + 2p + 1)!!} \left(1 + \frac{8 p \hspace*{1pt} (\ell + p + 1)}{(\ell + 1)(\ell + 2)}\right) \left[\int d^3 x \, \partial_t^{2p} T^{00} r^{2 p} x^L\right]^{\text{STF}} \notag \\
& + \sum_{p=0}^\infty \frac{(2 \ell + 1)!!}{(2 p)!! (2 \ell + 2p + 1)!!} \left(1 + \frac{4 p}{(\ell + 1)(\ell + 2)}\right) \left[\int d^3 x \, \partial_t^{2p} T^{aa} r^{2 p} x^L\right]^{\text{STF}} \notag \\
& - \sum_{p=0}^\infty \frac{(2 \ell + 1)!!}{(2 p)!! (2 \ell + 2p + 1)!!} \, \frac{4}{\ell + 1} \left( 1 + \frac{2 p}{\ell + 2} \right) \left[\int d^3 x \, \partial_t^{2p+1} T^{0a} r^{2 p} x^{aL}\right]^{\text{STF}} \notag \\
& + \sum_{p=0}^\infty \frac{(2 \ell + 1)!!}{(2 p)!! (2 \ell + 2p + 1)!!} \, \frac{2}{(\ell + 1)(\ell + 2)} \left[\int d^3 x \, \partial_t^{2p+2} T^{ab} r^{2 p} x^{abL}\right]^{\text{STF}} \label{eq_resultIL} \\
Q^L_{(M)} & = \frac{1}{\ell+2} \int d^{3}x \left\{ r^{-\ell-1} \left( r^{\ell + 2} \, \tilde{j}_{\ell} (r i \partial_t) \right)' \left( 2 \, \vec{x} \times \left(\vec{T}^{0 a}\right) \right)^{k_\ell} x^{L-1} \right. \notag \\
&\,\,\,\,\,\,\,\,\,\,\,\,\,\,\,\,\,\,\,\,\,\,\,\,\,\,\,\,\,\,\,\,\,\,\,\,\,\,\,\,\,\,\left.
- \tilde{j}_{\ell} (r i \partial_t) \, 2 \epsilon^{k_\ell ba} \partial_t T^{ac} x^{bcL-1}  \right\}  \notag \\
& = \sum_{p=0}^\infty \frac{(2 \ell + 1)!!}{(2 p)!! (2 \ell + 2p + 1)!!} \left(1 + \frac{2 p}{\ell + 2}\right) \left[ \int d^3 x \, \partial_t^{2p} r^{2p} \left( 2 \, \vec{x} \times \left(\vec{T}^{0 a}\right) \right)^{k_\ell} x^{L-1}\right]^\text{STF}  \notag \\
 & - \sum_{p=0}^\infty \frac{(2 \ell + 1)!!}{(2 p)!! (2 \ell + 2p + 1)!!} \,\frac{1}{\ell + 2} \left[ \int d^3 x \, 2 \epsilon^{k_\ell ba} \partial_t^{2p+1} T^{ac} r^{2p} x^{bcL-1}\right]^\text{STF}  \, . \label{eq_resultJL}
\end{align}
These expressions are a compact form of those in \cite{RossMultipoles} eq. (5.15-16) and are very close to \cite{Blanchet:1989ki} eq. (2.27).
The gravitational Feynman rules are summarized in equations (\ref{Feynman Rule Vertex},\ref{Feynman Rule Propagator}) with $s=2$ together with the definition of the $R_{2}^\eps(\ell)$ factor (\ref{R2def}).

Our definition of the master-functions (\ref{even parity master-function},\ref{odd parity master-function}) and the sources (\ref{even parity source term},\ref{odd parity source term}) is the same as that of \cite{Martel:2005ir} apart from an $r$-independent factor which we determined such that the source will coincide with the standard radiation source multipoles such as in \cite{RossMultipoles}. This affects the form of the propagator and results in the appearance of $R_s^\eps(\ell)$.

\subsubsection{Leading radiation and RR force}

The \emph{linearized} gravitational radiation at infinity is given by
\begin{align}
\label{Radiation gravity using feynman}
h^{(E/M)}_{L\omega}(r)=&
\parbox{20mm}
 {\includegraphics[scale=0.5]{DiagRadiationScalar.pdf}}
= 	-Q_{L'\omega}G^\Phi_{ret}(0,r)	\nonumber\\
=&(-i \omega)^\ell R^{\epsilon}_2 \frac{Q_{L\omega}}{r^{\ell}} \frac{e^{i \omega r}}{r} \, \, ,
\end{align}
where we used (\ref{Bessel H asymptotic2}) and $\tilde{j}(wr')|_{r'=0}=1$.
The radiated energy is given by
\bea
\label{gravitational radiated energy}
P_{rad} &=& \sum_{L} \frac{G \, R^{\epsilon}_2}{(2\ell+1)!!} <(\d_t^{\ell+1}Q_L)^2>  \\
&=& \sum_{L} \frac{G}{\ell! (2\ell+1)!!} \frac{\ell+2}{\ell-1} \left[ \frac{\ell+1}{\ell} <(\d_t^{\ell+1}Q_{k_1 k_2 \cdots k_\ell}^{(E) \, STF})^2> + \frac{\ell}{\ell+1} <(\d_t^{\ell+1}Q_{k_1 k_2 \cdots k_\ell}^{(M) \, STF})^2> \right] \, \, . \nonumber
\eea
The \emph{linearized} double field effective action for a source in a gravitational field
\begin{align}
\hat{S}_{linear}=&
\label{S gravity using feynman}
\parbox{20mm}
 {\includegraphics[scale=0.5]{ActionDiagScalarL.pdf}} \\
 =& \int dt \sum_{\ell} \frac{ G \, (-)^{\ell + 1} \,  ( \ell + 2 ) }{(2 \ell + 1)!! \, ( \ell - 1 ) } \left[ \frac{( \ell + 1 )}{ \ell} \hat{Q}^{L}_{(E)} \partial^{2 \ell + 1}_t Q^{(E)}_{L} + \frac{\ell }{ ( \ell + 1 ) } \hat{Q}^{L}_{(M)} \partial^{2 \ell + 1}_t Q^{(M)}_{L} \right] ~,
 \nonumber
\end{align}
where as in the scalar and EM cases we take $r=r'=0$ in the propagator since we are in the radiation zone and regularize by taking only the $\tilde{j}$ part out of the $\tilde{h}^+$. The balance of energy \be
 <P_{RR}> = P_{rad} \, \, , \label{grav:Pbalance} \ee
 is seen to hold from (\ref{gravitational radiated energy},\ref{S gravity using feynman}) just as in the previous cases (\ref{radiated energy scalar multipoles},\ref{radiated energy EM multipoles}).

{\bf Leading RR force}. The leading order RR force is computed from (\ref{S gravity using feynman}) where as in the scalar and electromagnetic cases the leading contribution comes from the lowest $\ell$ electric channel, which for gravity is the mass quadrupole, namely $(\ell \eps)=(2+)$.  One obtains \bea
  \hS_{LO} &=& -\frac{2\, G}{5} \int dt\, \hat{Q}_E^{I_2}\, \partial_t^5 Q_E^{I_2} \non
   &\equiv& -\frac{G}{5} \int dt\, \hat{Q}_E^{ij}\, \partial_t^5 Q_E^{ij} \, \, ,
\label{LO:effective action}
\eea where for the leading part of the mass quadrupole it suffices to use
\be
Q_E^{ij} \to Q_{E(0)}^{i j} := \sum_{A=1}^2 m_A \left( x^i x^j -\frac{1}{3} \delta^{i j} x^2 \right)_A  ~.
\label{leading order quadrupole}
\ee

The resulting RR force is \be
F_{SF}^i = \frac{\delta \hS}{\delta \hx^i(t)} = -\frac{G}{5} \frac{\delta Q_E^{ij}}{\delta \hx^i(t)} \partial_t^5 Q_E^{ij} \, \, ,
\label{SF-grav-L}
\ee
which is identical to the force derived from the Burke-Thorne potential \cite{BurkeThorne} \be
 V_{BT} (x,t)= \frac{G\, m}{5}\, \del_t^5 Q_E^{ij}(t)\,  x^i x^j \, \, . \ee

We would like to discuss the following points regarding the leading RR force.

 {\it Gauge invariance}. The leading RR effective action (\ref{LO:effective action}) is gauge invariant in the following sense. In the Newtonian limit space-time is flat and coordinates are absolute. As corrections are accounted for we must allow for generators of coordinate redefinition which are small in the PN expansion, namely $\xi^\mu(\vec{x},t) = {\cal O}(1PN)$. Clearly any such gauge variation of terms in the conservative part $S_{2bd}$ cannot generate terms in $\hS$ (since it cannot generate doubled, or hatted, fields) while a variation of (\ref{LO:effective action}) will be of a higher PN order. Therefore (\ref{LO:effective action}) is gauge invariant.

Analyzing the components of (\ref{LO:effective action}) we observe that the leading quadrupole mass moment $Q_{E2}$ is gauge invariant, and so are the spherical waves in the radiation zone. Hence so is $P_{E2}$, the source of reaction fields. Still, the reaction fields themselves are gauge dependent. Indeed, at least two gauges are discussed in the literature, the Burke-Thorne gauge where only the $h_{00}$ component is excited, and the harmonic gauge \cite{MTW} (corrected by \cite{Schaefer:1983hv})\footnote{These are the prominent gauges discussed in the Lagrangian formulations; the Hamiltonian formulation features the ADM gauge, by which the potential and the force are found in \cite{SchaeferRev,Schaefer:1986rd,Buonanno:1998gg}. We followed and compared with Lagrangian methods.}. The force is found to be gauge-dependent and can be thought to include terms which arise from gauge variation (of order 2.5PN) of the Newtonian force. Our observation that $\hS_{RR}$ is gauge invariant at leading order suggests that there might be a way to naturally decompose the standard RR force into an ``essentially RR part'' which would be gauge invariant and a gauge variation of an essentially conservative force.

 {\it Comparison with \cite{GalleyTiglio,GalleyLeibovich}}. The derivation of the gravitational RR force within the EFT approach was first derived in \cite{GalleyTiglio} and later derived again in \cite{GalleyLeibovich}, Appendix A. Our derivation is significantly shorter, consisting essentially of a mere multiplication vertex -- propagator -- vertex according to the Feynman rules (\ref{Feynman Rule Vertex},\ref{Feynman Rule Propagator}).
  Differences in method include: a somewhat different diagram which distinguishes the $Q$ and $\hQ$ sources unlike fig. 1 of \cite{GalleyLeibovich} or fig. 2 of  \cite{GalleyTiglio} ; gauge invariant spherical waves and the associated propagators involving Bessel functions;  no need to compute potentially contributing diagrams at order 0.5PN and 1.5PN as in fig. 1 of \cite{GalleyTiglio} since we are not reproducing the odd propagation locally in the system zone, but rather from interaction with the radiation zone.

 {\it Higher order terms}. Eq. (\ref{S gravity using feynman})  which produces the leading RR force through the E2 channel computes higher terms when other channels are considered. In particular, the current quadrupole M2 and the mass octupole E3 contribute +1PN corrections as follows
\bea
  \hS_{NL}  &\supset& G \int dt\, \left[ -\frac{8\, G}{45} \hat{Q}_M^{I_2}\, \partial_t^5 Q_M^{I_2} + \frac{2\, G}{63} \hat{Q}_E^{I_3}\, \partial_t^7 Q_E^{I_3} \right] \equiv \non
  &\equiv& G \int dt\, \left[ -\frac{4\, G}{45} \hat{Q}_M^{ij}\, \partial_t^5 Q_M^{ij} + \frac{G}{189} \hat{Q}_E^{ijk}\, \partial_t^7 Q_E^{ijk} \right] \, \, ,
\label{NL:channels}
\eea where the leading part of the current quadrupole and mass octupole are
\bea
Q_M^{ij} \to Q_{M(0)}^{i j}  &:=& 2 \sum_{A=1}^2 m_A \left[ \left( \vec{x} \times \vec{v} \right)^i x^j \right]^{TF}_A \, \, ,  \non
Q_E^{ijk} \to Q_{E(0)}^{i j k}  &:=& \sum_{A=1}^2 m_A \left( x^i x^j x^k \right)^{TF}_A  ~.
\eea

\subsubsection{Next to leading correction}

\begin{center}
\begin{figure*}[b!]
        \begin{center}
        \begin{tabular}{ccc}
            (a) & (b) & (c) \\
            \includegraphics[width=4cm,height=5cm]{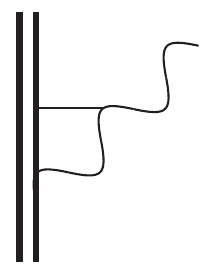} &
            \includegraphics[width=3cm,height=5cm]{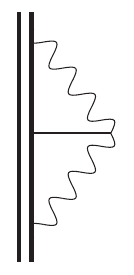} &
            \includegraphics[width=6cm,height=5cm]{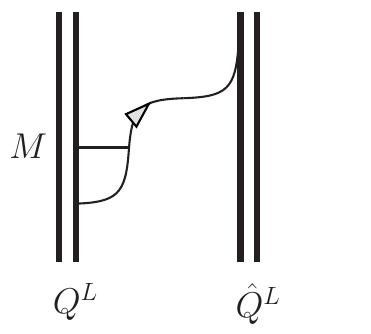} \\
        \end{tabular}
        \end{center}
\caption{Computation of the tail effect in EFT methods: (a) correction to the radiation in \cite{GoldbergerRoss} ; (b) correction to the radiation reaction in \cite{Foffa:2011np} ; (c) correction to the radiation reaction in our formalism.}
\label{fg:tail}
\end{figure*}
\end{center}

Corrections of the RR force can arise either from corrections to $Q[x]$ at the system zone or from corrections to $\hS$ at the radiation zone. System zone corrections are suppressed at least by $\frac{G\,M}{R} \sim v^2$, namely by +1PN, while non-linear interactions in the radiation zone are suppressed at least by $\frac{G\,M}{\lambda}$, namely +1.5 PN (the $GM$ factor comes from an additional vertex and it must be divided by $\lambda$ which is the length scale of the radiation). Therefore the leading +1PN corrections arise from the system zone only and there is no need to consider radiation zone corrections.

The leading nonlinear effect arising from interactions in the radiation zone is the above mentioned +1.5 PN correction, also known as the \emph{tail} effect as it induces propagation inside the light cone. The leading tail effect was first computed in \cite{Blanchet:1987wq,Blanchet:1993ng}. In EFT methods, the leading correction to radiation was computed in \cite{GoldbergerRoss}, and to radiation reaction in \cite{Foffa:2011np}. The tail effect arises from scattering of the emitted wave off the background curvature generated by the mass of the entire system. In order to account for it, we must supplement the spherical wave variables $h_{(E/M)}$ by the stationary modes (mentioned in subsection \ref{subsec:zones}) and take account of the corresponding (non-quadratic) interaction terms (beyond \ref{1DS}). The diagrams in figure \ref{fg:tail} represent the leading tail correction as treated in previous EFT works and as could be treated within (an extension of) our formalism. The diagram should reproduce nothing but the first term in the long distance expansion (in $\frac{G\,M}{r}$) of the Zerilli/Regge-Wheeler equations, as we are using spherical wave variables. Higher order effects which originate from interactions in the radiation zone can be treated in an analogous manner and should reproduce higher order terms in the Zerilli/Regge-Wheeler equations as well as non-linear wave interaction.

Corrections due to the M2 and E3 channels were already discussed in (\ref{NL:channels}). It remains to study the leading order corrections to the mass quadrupole $Q_{E2}=Q_{E2}[T_{\mu\nu}]$ (\ref{eq_resultIL}) and substitute back into (\ref{LO:effective action}). For that purpose we need to clarify the definition of the source energy-momentum tensor $T^{\mu\nu}$ for the case at hand. Considering the standard coupling of matter to weak gravity
\bea
 S_{h-T} = -\half h_{\mu\nu}\, T^{\mu\nu} \label{def:S_ht}
\eea
 we define
\bea
  T^{\mu\nu}(x) := -2 \frac{\delta S_{eff}}{\delta h_{\mu\nu}(x)} \equiv -2\,
 \( \parbox{30mm}{\includegraphics[scale=0.3]{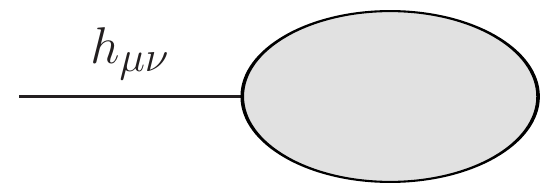}} \)	
  \eea
namely $T^{\mu\nu}$ is defined as the source for $h_{\mu\nu}$ or equivalently as a 1-pt diagram.This definition is equivalent to the one given in \cite{Blanchet:1998in} ($\tau^{\mu\nu}$ in eq. (2.16) there). Considering moreover that in the system zone we use the NRG fields (\ref{def:NRG-fields}) rather than $h_{\mu\nu}$, we find it useful to define a corresponding change of source variables \be
T^{\mu\nu} \longleftrightarrow T^{NRG}:=\(\rho_\phi,\, \vec{J},\, \Sigma^{ij} \) \, \, , \ee
defined by \be
 S_{h-T} = -\rho_\phi\, \phi + \vec{J} \cdot \vec{A} - \half \Sigma^{ij} \sigma_{ij} \, \, , \ee
 which altogether implies \bea
 T^{00} + T^{ii} &=& \rho_\phi =  -
  \( \parbox{30mm}{\includegraphics[scale=0.3]{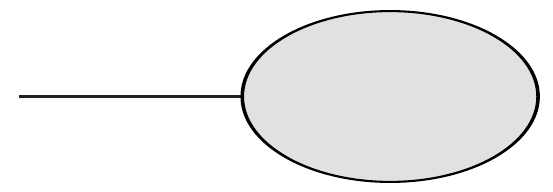}} \) \, \, ,
\non
 T^{0i} &=& J^i =
 \parbox{30mm}{\includegraphics[scale=0.3]{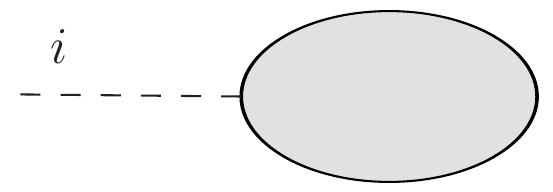}} \, \, ,
\non
 T^{ij} &=& -\Sigma^{ij} = 2\,
 \parbox{30mm}{\includegraphics[scale=0.3]{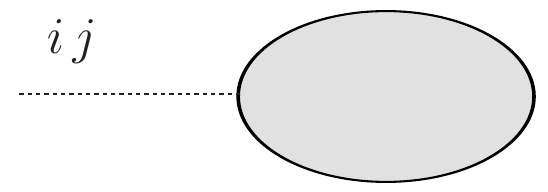}}  ~.
 \eea
 We call $\rho_\phi$ the gravitational mass density, $\vec{J}$ the gravitational source current, and $\Sigma^{ij}$ the stress.

In terms of $T^{NRG}$ we re-write the expression for the radiation source multipoles (\ref{eq_resultJL}) as follows \bea
 Q^L_E & =&  \frac{1}{(\ell+1)(\ell+2)} \!\!\int\!\! d^{3}x \, x^{L} \left[  r^{-\ell} \left( r^{\ell+2} \, \tilde{j}_{\ell} (r i \partial_t) \right)''\, \rho_\phi - 4 r^{-\ell-1} \left( r^{\ell+2} \, \tilde{j}_{\ell} (r i \partial_t) \right)' \!\vec{x} \cdot \dot{\vec{J}} \right. \notag \\
& & \,\,\,\,\,\,\,\,\,\,\,\,\,\,\,\,\,\,\,\,\,\,\,\,\,\,\,\,\,\,\,\,\,\,\,\,\,\,\,\,\,\,\,\,\,\,\,\,\,\,\,\,\,\,\,\,\,\,\,\,
\left. - 2 \, \tilde{j}_{\ell} (r i \partial_t) \, x^{a} x^{b} \ddot{\Sigma}^{a b} + r^{2} \, \tilde{j}_{\ell} (r i \partial_t) \, \partial_t^2 \left( \rho_\phi+2 \Sigma \right)^{\,} \right] \, \, , \label{QT} \\
Q^L_M & =&  \frac{1}{\ell+2} \int d^{3}x \left\{ r^{-\ell-1} \left( r^{\ell + 2} \, \tilde{j}_{\ell} (r i \partial_t) \right)' \epsilon^{k_\ell ab}  x^{aL-1} J^b + \tilde{j}_{\ell} (r i \partial_t) \, \epsilon^{k_\ell ab}   x^{acL-1}  \dot{\Sigma}^{bc} \right\} \, \, . \nonumber
\eea

Now we can proceed to compute the sources for the two body problem \be
 T^{NRG}= T^{NRG}[\vec{x}^A(t)] \, \, , \ee
and the next to leading corrections to the mass quadrupole. The gravitational mass up to NL order is given by \bea
 \rho_\phi &=&
  - \( \parbox{25mm}{\includegraphics[scale=0.5]{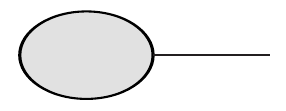}} \) =
  - \( \parbox{18mm}{\includegraphics[scale=0.5]{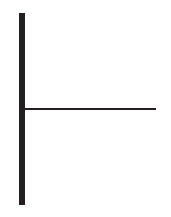}} +  \parbox{18mm}{\includegraphics[scale=0.5]{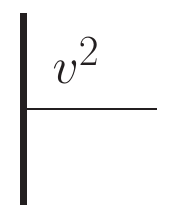}}
 + \parbox{25mm} {\includegraphics[scale=0.5]{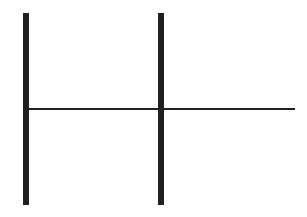}} \)
\non
 &=&  \sum_A m_A\, \delta\(x-x_A\)\, \(1 + \frac{3}{2} v_A^2 - \sum_{B\neq A} \frac{G\, m_B}{r_{AB}} \) ~.
 \label{rhophi-correction}
\eea
For $\vec{J},\, \Sigma^{ij}$ it suffices to use the leading order expressions \bea
 \vec{J} &=& 2 \sum_A m_A\, \vec{v}_A\, \delta\(x-x_A\) \non
 \Sigma^{ij} &=& \sum_A m_A\, \vec{v}_A^i\, \vec{v}_A^j\, \delta\(x-x_A\) ~. \eea
Substituting back into (\ref{QT}) we find \be
 \delta Q_E^L =\sum_A \[ \frac{3}{2} m\, v^2 - \frac{m_1\, m_2}{r} - \frac{4}{l+1} m\, \del_t\, \vec{v} \cdot \vec{x}  + \frac{l+9}{2(l+1)(2l+3)} m\, \del_t^2\, x^2 \]_A x_A^L ~. \label{deltaQE} \ee
In particular the required correction to the quadrupole is \be
 \delta Q_E^{L_2} =\sum_A \[ \frac{3}{2} m\, v^2 - \frac{m_1\, m_2}{r} - \frac{4}{3} m\, \del_t\, \vec{v} \cdot \vec{x} + \frac{11}{42} m\, \del_t^2\, x^2 \]_A x_A^{L_2} ~, \label{deltaQE2} \ee
 where the time derivatives act on everything to their right including the $x^L$ factor. This result was obtained in \cite{Blanchet:1989cu}  and was reproduced in \cite{GoldbergerRoss} within the EFT approach.

In summary, collecting (\ref{LO:effective action},\ref{leading order quadrupole},\ref{NL:channels},\ref{deltaQE2}) the next to leading RR force is encoded by the next to leading part of the body-radiation effective action \be
 \hS = G \int dt\, \left[ -\frac{1}{5} \hat{Q}_E^{ij}\, \partial_t^5 Q_E^{ij} -\frac{4}{45} \hat{Q}_M^{ij}\, \partial_t^5 Q_M^{ij} + \frac{1}{189} \hat{Q}_E^{ijk}\, \partial_t^7 Q_E^{ijk} \right] \, \, ,
 \label{summ:NL-grav}
 \ee
where the radiation source multipoles up to this order are given by \bea
Q_E^{ij}  &=& \sum_{A=1}^2 m_A  \[ 1+ \frac{3}{2}  v^2 - \frac{m_B}{r} - \frac{4}{3}  \del_t\(\vec{v} \cdot \vec{x}\) + \frac{11}{42}\del_t^2 x^2 \]_A  \left( x^i x^j -\frac{1}{3} \delta^{i j} x^2 \right)_A  \, \, ,   \non
Q_M^{i j}  &=& 2 \sum_{A=1}^2  \left[m\, \left( \vec{x} \times \vec{v} \right)^i x^j \right]^{TF}_A \, \, , \non
Q_E^{i j k}  &=& \sum_{A=1}^2  \left(m\, x^i x^j x^k \right)^{TF}_A \, \, ,
\eea
 where given $A$, $m_B$ stands for the other mass.

\presub {\bf Discussion}.
It is interesting to compare this +1PN relativistic correction with the +1PN relativistic correction to the (conservative) two-body effective action, known also as the Einstein-Infeld-Hoffmann (EIH) action. In the case of EIH it was possible to identify each term in the action with one of the following physical effects: kinetic contribution to gravitational mass, potential contribution to gravitational mass, retardation and an additional channel of current-current interaction \cite{NRG}. In this case a very similar interpretation is possible: the first term in (\ref{deltaQE2}) represents a kinetic contribution to the gravitational mass, the second term a contribution from the potential energy (both terms originate in the diagram in (\ref{rhophi-correction})), the last two terms can be thought to represent retardation effects, and finally one must also account for additional M2 and E3 channels.

Comparing our derivation with the derivation of Goldberger and Ross \cite{GoldbergerRoss} who first obtained this correction within the EFT approach we see that our derivation requires fewer diagrams and is more general. In \cite{GoldbergerRoss} the first two terms in (\ref{deltaQE2})  are obtained as a sum of 5 diagrams as follows \bea
\delta T^{00} &\supset&
 \parbox{30mm}{\includegraphics[scale=0.5]{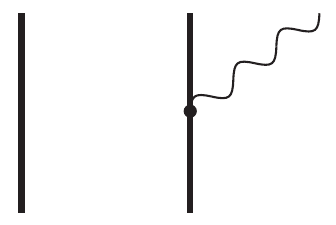}} + \parbox{30mm}{\includegraphics[scale=0.5]{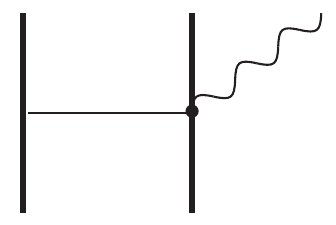}}
 + \parbox{30mm}{\includegraphics[scale=0.5]{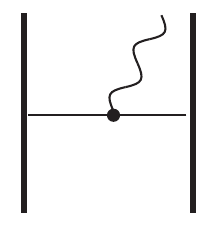}} \, \, ,
 \non	
 \delta T^{ij} &\supset&
  \parbox{30mm}{\includegraphics[scale=0.5]{fig5GR1.pdf}}
 + \parbox{30mm}{\includegraphics[scale=0.5]{fig5GR3.pdf}} \, \, ,		
\eea
 while we used only 2 diagrams  in (\ref{rhophi-correction}) through the use of NRG fields. In addition our expression (\ref{deltaQE}) gives not only the NL correction to the quadrupole, but also for any $\ell$-pole, and therefore constitutes a new result, rather than an economized derivation of a known result. In particular we can write explicitly the correction to the mass octupole which would contribute to radiation and RR force at order +2PN \be
\delta Q_E^{L_3} =\sum_A \[ \frac{3}{2} m\, v^2 - \frac{m_1\, m_2}{r} -  m\, \del_t\, \vec{v} \cdot \vec{x} + \frac{1}{6} m\, \del_t^2\, x^2 \]_A x_A^{L_3} ~. \label{deltaQE3} \ee

\section{Summary and discussion}
\label{sec:summary}

We summarized our formulation and its novel features in subsection \ref{subsec:summary} and here we summarize the results from section \ref{sec:demonstration} and discuss the paper as a whole.  We demonstrated the formulation though application to scalar, electromagnetic and gravitational theories ($s=0,1,2$). We showed how to obtain the Feynman rules (\ref{Feynman Rule Vertex},\ref{Feynman Rule Propagator}). Our definitions are such that the propagator in the radiation zone is the same for all cases $s=0,1,2$ apart for the factor $R^\eps_s(\ell)$ which is summarized in (\ref{summ:Rs}) and was derived in (\ref{Phi propagator scalar},\ref{REone},\ref{EM source vector Phi},\ref{R2def}). The expressions for the radiation source multipoles $Q_L$ for all cases were found in (\ref{I Phi scalar multipoles},\ref{I EM scalar multipoles},\ref{J EM vector multipoles},\ref{eq_resultIL}-\ref{eq_resultJL}). Our expressions for $Q_L$ simplify extant expressions by incorporating Bessel functions to account for retardation (which were missing from \cite{RossMultipoles} but present in \cite{ThorneMultipoles}) while significantly economizing the expressions in \cite{ThorneMultipoles}.

In our formulation the propagator in the two-body sector is restricted to the even part under time reversal and the odd propagation was argued to be accounted for indirectly through interaction with radiation. To confirm this we computed perturbatively the RR force in our formulation and found it to coincide with known non-perturbative results such as ALD in electromagnetism, see tables (\ref{table:scalar multipoles leading},\ref{table:scalar multipoles next to leading scalar},\ref{table:EM leading scalar},\ref{table:EM next to leading scalar},\ref{table:EM next to leading vector}). In addition we further confirmed it with an independent computation using direct local propagation \`{a} la Dirac (Appendix \ref{app:ALD local propagator}).

The scalar sector already demonstrates some of the important features of our formulation, namely the spherical waves in the radiation sector, the above mentioned form of odd propagation, 2-way matching multipoles and field doubling. In the electromagnetic sector the novel features are gauge invariance of the radiation fields and matching multipoles, as well as the polarization of radiation into both electric and magnetic modes. The $R_1^\eps$ factor appears in expressions for radiation and RR force.

Our method was demonstrated further in the gravitational case (for which it was constructed). Here we demonstrated how at leading order our formulation reproduces the known results while economizing the computation to a mere vertex-propagator-vertex multiplication (\ref{S gravity using feynman},\ref{LO:effective action}). We proceeded to reproduce and economize the next to leading order where 5 diagrams which represent relativistic corrections to $T^{\mu\nu}$ are replaced by two diagrams with a clear physical interpretation, namely the kinetic and potential energy contributions to the gravitational mass (\ref{rhophi-correction},\ref{summ:NL-grav}). We obtained new results for the +1PN correction to $Q^L_E$ for all $\ell$ (\ref{deltaQE}, \ref{deltaQE3}) and we expect that our formulation will facilitate additional higher order computations.

Finally we introduced several useful definitions and conventions in Appendix \ref{app:defs}: a multi-index summation convention which takes care of factors of $\ell !$, a normalization of Bessel functions which is useful for our radiation zone propagators and a normalization of the gravito-magnetic vector potential which avoids unnecessary factors of 2 which appear in spin interactions.

\subsection*{Discussion}

Our work addressed the issues mentioned in the introduction. The current formulation renders manifest the close ties between radiation and radiation reaction force: both of them appear in the same theory which uses uniformly the retarded propagator and doubled fields; both of them use the same radiation fields, the same Feynman rules and in particular the same radiation source multipoles $Q_L$; finally the energy balance is rather manifest and was demonstrated in (\ref{radiated energy scalar multipoles},\ref{radiated energy EM multipoles},\ref{grav:Pbalance}) (see also the discussion part in section \ref{subsec:summary}).
Our formulation resulted in several economizations including in the leading and next to leading gravitational RR force and in the expressions for the radiation source multipoles $Q_L$. Finally we incorporated NRG fields and real Feynman rules, and these were some of the reasons which allowed for the previously mentioned economization.

It would be interesting to further refine the computations by substituting in the bodies' unperturbed equations of motion as in \cite{BlanchetRev}, only here at the level of the action. It also seems that our method can be naturally extended to treat the problems of radiation and radiation reaction in generic spacetime dimensions. This will be explored in \cite{HDPNRR}.

{\bf Discussion summary}. We presented several novel ingredients to the EFT formulation of radiation and RR force. These include gauge-invariant spherical radiation fields instead of plane waves, matching lifted to the level of the action through the introduction of 2-way multipoles; novel insights into double-field action (classical origin of the Closed Time Path formalism) including its applicability to arbitrary directed propagators, and the special role of the Keldysh basis in the classical theory. We confirmed our formulation through several tests, demonstrated its utility by performing several economized computations, and obtained a new result (\ref{deltaQE}).

\subsection*{Acknowledgments}

Some of the results were presented in the Israeli joint high energy seminar (Neve-Shalom) on 18.12.2012, and some more in the meeting ``Equations of Motion in Relativistic Gravity'', 17-23.2.2013 Bad Honnef, Germany, and we thank the respective organizers for invitations. BK thanks the organizers of the conference ``Effective Field Theory and Gravitational Physics'', November 28 - 30, 2011 at the Perimeter Institute which contributed to this work.
We thank Avraham Schiller for a discussion on the Keldysh formalism and Luc Blanchet for very useful comments.

This research was supported by the Israel Science Foundation grant no. 812/11. OB was partly supported by an ERC Advanced Grant to T. Piran.

\appendix

\section{Useful definitions and conventions}
\label{app:defs}

In this appendix we collect several definitions and conventions used in this paper.

\subsection{Multi-index summation convention}
\label{app:Multi-index summation convention}

Multi-indices are denoted by capital letters \be
 I \equiv I_\ell :=(i_1 \dots i_\ell) \, \, , \ee
 where each $i_k=1,2,3$ is an ordinary spatial index, and $\ell$ is the number of indices.
 We define a multi-index summation convention by \be
 P_I\, Q_I := \sum_\ell P_{I_\ell}\, Q_{I_\ell} := \sum_\ell \frac{1}{\ell!} P_{i_1 \dots i_\ell}\, Q_{i_1 \dots i_\ell} \, \, .   \ee
This means that not only are repeated multi-indices summed over as in the standard summation convention, but moreover a division by $\ell!$ is implied. When $\ell$ is unspecified the summation is over all $\ell$.

In addition a multi-index delta function is defined through \be
\delta_{I_\ell J_\ell} := \ell!\, \delta_{i_1 j_1} \dots \delta_{i_\ell j_\ell} \, \, . \ee

The definitions are such that factors of $\ell!$ are accounted for automatically.

\subsection{Normalizations of Bessel functions}
\label{app:Normalizations of Bessel functions}

We find it convenient to define a non-standard origin biased normalization of Bessel functions. We start with a conventionally normalized solution of the Bessel equation \be
\[ \del_x^2 + \frac{1}{x} \del_x + 1 - \frac{\nu^2}{x^2} \] B_\nu(x)=0 \, \, ,
\ee
where $B \equiv \{J,Y,H^\pm\}$, namely $B$ represents both Bessel functions of the first or second kind and Hankel functions, and $\nu$ denotes their order.
For integer $\ell$ we define \be
\tilde{b}_\ell := (2\ell+1)!! \frac{b_\ell(x)}{x^\ell} := \sqrt{\frac{\pi}{2}}(2\ell+1)!!\frac{B_{\ell+1/2}(x)}{x^{\ell+1/2}} \, \, , \ee
where $b_\ell(x)$ is a spherical Bessel function and can be expressed in terms of trigonometric functions.

The origin normalized Bessel functions $\tilde{b}_\ell$ satisfy the nice equation  \be
\[ \del_x^2 + \frac{2(\ell+1)}{x}\del_x + 1\] \tilde{b}_\ell(x) =0 ~.
\label{Modified Bessel equation2}
\ee
The purpose of the definition is to have a simple behavior of $\tilde{j}$ around the origin $x=0$ \be
\tilde{j}_\ell (x) = 1 + \co\(x^2\) ~.\ee
More precisely the Taylor expansion for $\tilde{j}_\ell (x)$ at $x=0$ is given by \be
\tilde{j}_\ell (x)  = \sum_{p=0}^\infty \frac{(-)^p (2\ell+1)!!}{(2p)!!(2\ell+2p+1)!!} x^{2p} ~.
\label{Bessel J series2}
\ee
The asymptotic form is best stated in terms of the Hankel functions $\tilde{h}^\pm:= \tilde{j} \pm i \tilde{y}$
\be
\tilde{h}^\pm_\ell(x) \sim (2\ell+1)!! \frac{(\mp i)^{\ell+1} e^{\pm i x}}{x^{\ell+1}} \, \, ,
\label{Bessel H asymptotic2}
\ee
for $x \to \infty$.

\subsection{Normalization of gravito-magnetic vector potential}
\label{app:gravito-magnetic}

The gravitational field in the system zone is parameterized as Non-Relativistic Gravitational (NRG) fields $(\phi, \vec{A}, \sigma_{ij})$ as follows \cite{CLEFT-caged,NRG} \be
 ds^2 = e^{2 \phi} \( dt - 2 \vec{A} \cdot d\vec{x} \)^2 - e^{-2\phi} \( \delta_{ij}+\sigma_{ij} \) dx^i\, dx^j \, \, , \label{def:NRG-fields} \ee
where
$\phi$ is called the Newtonian potential, $\vec{A}$ is the gravito-magnetic vector potential and $\sigma_{ij}$ is the spatial metric.
This definition redefines $A_i$ relative to \cite{CLEFT-caged,NRG} (and all previous literature) by \be
A_i^{new} := \half A_i^{old} \, \, , \ee
which turns out to clean up certain factors of $2$ in the theory as we proceed to explain.

The new definition avoids a factor of $4$ in the propagator for $A_i$ which becomes (in Feynman gauge)
\bea
 \parbox{25mm}{\includegraphics[scale=0.5]{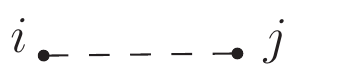}}	
 = -\frac{G}{r} \delta_{ij} \delta(t_1-t_2) ~.
\eea
 This means that the kinetic terms for the new $A$ is normalized just like in the Maxwell action.
 The price to pay is an added factor of 2 in the basic source coupling and the associated vertex \bea
 S &\supset& 2\, m \int dt\, \vec{v} \cdot \vec{A} \, \, , \non
 \parbox{25mm}{\includegraphics[scale=0.5]{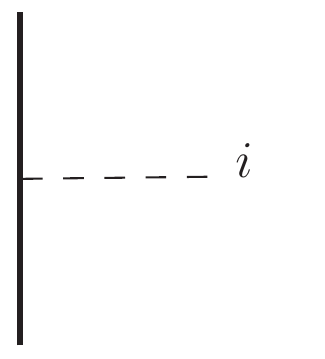}}
  &=&  2\, m\, \int dt\, v^i ~, \eea	
 namely we define $\vec{J}$ the gravitational source current to be twice the mass current.
  However, when considering both spin interaction and gravitational waves this normalization actually appears advantageous. For a small non-relativistic loop of mass current the magnetic dipole moment is now nothing but the intrinsic angular momentum (or spin) of  the system \be
  Q^M_i = S^i ~ \ee and the spin coupling and the associated vertex become \bea
 S \supset \int dt\, \vec{S} \cdot \vec{B}~, && \qquad \vec{B} := \vec{\nabla} \times \vec{A} \, \, ,  \non
 \parbox{25mm}{\includegraphics[scale=0.5]{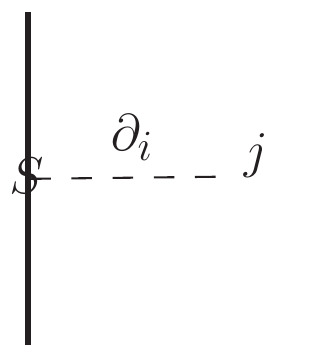}}
  &=&  \int dt\, \eps^{ijk}\, S^k \, \, .
\eea
As a result the derivation of the spin-spin interaction avoids unnecessary cancellations which appeared in the old variables $S_{s1s2} \sim \(\half S_1\) (4\, G) \(\half S_2\)
= G S_1 S_2$ where the $r$-dependence and spatial indices were suppressed.
Moreover the current multipoles defined in terms of the new gravitational source current $\vec{J}$ are such that they eliminate a relative factor of $4$ which used to appear in the power formula (see for example \cite{RossMultipoles}).

\section{Extant literature on the classical limit of CTP}
\label{sec:app-Galley}

The utility of the CTP formalism in the EFT approach to GR was already recognized and it was applied in several papers including \cite{GalleyTiglio,GalleyLeibovich,GalleyNonConservative,Collins:2012nq}. Moreover, in \cite{GalleyNonConservative} Galley recognized the importance of an intrinsically classical formulation of the classical limit of CTP and put forward such a formulation. Here we wish to point out differences between that approach and ours.
\bi
 \item \cite{GalleyNonConservative} takes its starting point to be the problem of Hamilton's least action principle with initial value, and uses it to motivate the definition of $\hS$ as an integral from $t_i$ to $t_f$ and back, namely (\ref{CTP-action}). Thus the starting point is in the $(\phi_1,\phi_2)$ basis which must be supplemented by an extra requirement that the action is expanded to the first order in $\phi_1-\phi_2$ (or equivalently by imposing a physical condition). In the current  formulation the starting point is in the Keldysh representation $(\phi,\hphi)$ without imposing a physical condition and by construction the double field action is first order in the $\hphi$ .

 \item Eq. (5) of \cite{GalleyNonConservative} defines $\hS$ and contains a function $K$ which is not specified in terms of the problem's data, namely $S=S[\phi]$. The current formulation does not contain such a function.
\ei

\section{ALD Self-force from local propagators}
\label{app:ALD local propagator}

As mentioned in section \ref{subsec:zoom}, the dissipative part of the action can be calculated using $G_{odd}$. For example, we present the calculation for an electromagnetic field $A^{\mu}=(\Phi,\vec{A})$. The field equation in the Lorentz gauge is
\bea
\Box A^{\mu}=4\pi j^\mu,
\label{app EM EOM}
\eea
with the source given by the 4-current, created by a charge $q$ with trajectory $x_p(t)$
\bea
j^{\mu}=(\rho,\vec{j}),   \,\,\,\,\,\,\,\,\,\,\,\,\,\,\,\,\, \rho&=&q\delta^{(3)}(x-x_p(t)),\,\,\,\,\,\,\,\,\,\,
\label{app EM source}
\vec{j}=\rho \vec{v_p}.
\eea
The retarded/advanced propagators are known to be (with $\Delta t=t-t'$, $\Delta r = \left|\vec{r}-\vec{r} \, ' \right|$)
\bea
G^\Phi_{ret/adv}(\vec{r}, t; \vec{r} \, ' , t')=-\frac{\delta(\Delta t \pm \Delta r)}{\Delta r},\,\,\,\,\,\,\,\,\,\,
G^A_{ret/adv}(\vec{r}, t; \vec{r} \, ' , t')=+\delta_{ij}\frac{\delta(\Delta t \pm \Delta r)}{\Delta r},\,\,\,\,\,\,\,\,\,\,
\label{app EM retarded propagator}
\eea
for the scalar $\Phi$ and vector $A^i$ potential components, respectively. By expanding the $\delta$-functions as Taylor series in $\Delta r$ around $\Delta t$ and using (\ref{even/odd prop}), we find the time-odd propagators, which contain only the terms with odd time derivatives
\bea
G^\Phi_{odd}(\vec{r}, t; \vec{r} \, ' , t')&=&-\sum_{n=0}^\infty\frac{\delta^{(2n+1)}(\Delta t)(\Delta r)^{2n}}{(2n+1)!},\nonumber\\
G^A_{odd}(\vec{r}, t; \vec{r} \, ' , t')&=&+\delta_{ij}\sum_{n=0}^\infty\frac{\delta^{(2n+1)}(\Delta t)(\Delta r)^{2n}}{(2n+1)!}.
\label{app EM odd propagator}
\eea
Now we can write the action as
\begin{align}
\label{Feynman Rule Vertex2}
S=&
 \parbox{20mm} {\includegraphics[scale=0.7]{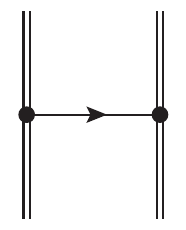}}
\,\,\,+
 \parbox{20mm} {\includegraphics[scale=0.7]{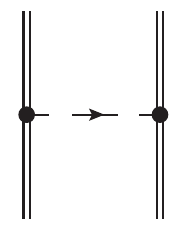}}
\nonumber\\
=&-\!\!\int\!\!qdt\!\!\int\!\!q\,dt' \sum_{n=0}^\infty\frac{\delta^{(2n+1)}(\Delta t)(\Delta r_p)^{2n}}{(2n+1)!!}
+\!\!\int\!\!q v_p^i dt\!\!\int\!\!q\, {v'}_p^j dt' \sum_{n=0}^\infty\frac{\delta_{ij}\delta^{(2n+1)}(\Delta t)(\Delta r_p)^{2n}}{(2n+1)!}
\nonumber\\
=&\sum_{n=0}^\infty \! \frac{q^2}{(2n+1)!} \left. \!\int\!\!dt \!
\left[\d_t^{2n+1}\left( (\Delta r_p)^{2n} \right)
	- \vec{v} \, '_p(t')\cdot\d_t^{2n+1}\left(\vec{v}_p(t)(\Delta r_p)^{2n}\right)
\right] \right|_{t'=t},
\end{align}
where $\Delta r_p = \left|\vec{r}_p(t)-\vec{r} \, '_p(t')\right|$, and we remark that all the $t$-derivatives are to be computed before setting $t'=t$.

This presents a natural expansion of the 2-body action in PN orders. Using the technique of field-doubling \cite{GalleyLeibovich}, we can derive the Euler-Lagrange equations for $\vec{r} \, '_p$, then substitute $\vec{r} \, '_p \to \vec{r}_p$, and obtain the self-force on q, order-by-order. We note that for each $n$, the term coming from the vector potential is always one order higher than the scalar potential term of the same $n$. We thus see that only the terms linear in $\Delta r_p$ in the Lagrangian can contribute to the self-force, and easily find the leading contributions ($n=1$ from the scalar potential, $n=0$ from the vector potential). We have also found the next-to-leading contributions in the same manner, and they match the ALD results (\ref{F LO EM ALD},\ref{F NLO EM ALD}) as expected.

\end{document}